\documentclass[preprint]{emulateapj}
\usepackage{natbib}
\usepackage{amsmath}
\usepackage{graphicx}

\begin{document} 

\title{The depletion of water during dispersal of planet-forming disk regions}

\author{A. Banzatti\altaffilmark{1,2}, K. M. Pontoppidan\altaffilmark{1}, C. Salyk\altaffilmark{3}, G. Herczeg\altaffilmark{4}, E. F. van Dishoeck\altaffilmark{5}, G. A. Blake\altaffilmark{6}} 
\altaffiltext{1}{Space Telescope Science Institute, 3700 San Martin Drive, Baltimore, MD 21218, USA} 
\altaffiltext{2}{Lunar and Planetary Laboratory, The University of Arizona, Tucson, AZ 85721, USA} 
\altaffiltext{3}{Vassar College, 124 Raymond Avenue, Poughkeepsie, NY 12604, USA}
\altaffiltext{4}{Kavli Institute for Astronomy and Astrophysics, Peking University, Yi He Yuan Lu 5, Haidian Qu, 100871 Beijing, China}
\altaffiltext{5}{Leiden Observatory, Leiden University, P.O. Box 9513, 2300RA Leiden, The Netherlands}
\altaffiltext{6}{Division of Geological \& Planetary Sciences, MC 150-21, California Institute of Technology, Pasadena, CA 91125, USA}

\email{banzatti@lpl.arizona.edu}

\begin{abstract} 
We present a new velocity-resolved survey of 2.9\,$\mu$m spectra of hot H$_2$O and OH gas emission from protoplanetary disks, obtained with CRIRES at the VLT ($\Delta v$\,$\sim$\,3\,km\,s$^{-1}$). With the addition of archival \textit{Spitzer}-IRS spectra, this is the most comprehensive spectral dataset of water vapor emission from disks ever assembled. We provide line fluxes at 2.9--33\,$\mu$m that probe from disk radii of $\sim0.05$\,au out to the region across the water snow line. With a combined dataset for 55 disks, we find a new correlation between H$_2$O line fluxes and the radius of CO gas emission as measured in velocity-resolved 4.7\,$\mu$m spectra (R$_{\rm co}$), which probes molecular gaps in inner disks. We find that H$_2$O emission disappears from 2.9\,$\mu$m (hotter water) to 33\,$\mu$m (colder water) as R$_{\rm co}$ increases and expands out to the snow line radius. These results suggest that the infrared water spectrum is a tracer of inside-out water depletion within the snow line. It also helps clarifying an unsolved discrepancy between water observations and models, by finding that disks around stars of M$_{\star}>1.5$ M$_\odot$ generally have inner gaps with depleted molecular gas content. We measure radial trends in H$_2$O, OH, and CO line fluxes that can be used as benchmarks for models to study the chemical composition and evolution of planet-forming disk regions at 0.05--20\,au. We propose that JWST spectroscopy of molecular gas may be used as a probe of inner disk gas depletion, complementary to the larger gaps and holes detected by direct imaging and by ALMA.
\end{abstract}

\keywords{circumstellar matter --- molecular processes --- planets and satellites: formation --- protoplanetary disks --- stars: pre-main sequence}

\section{INTRODUCTION} \label{sec:intro}
Protoplanetary disks are composed of the circumstellar material that feeds stars and planets. These systems evolve from early gas- and dust-rich phases to phases of dispersal, eventually unveiling planetary systems that form within them \citep[e.g.][]{alex14}. Evidence is growing that there is at least one mode of disk dispersal, working from the inside out. 
Observations of a wavelength-dependent decrease in infrared excesses \citep[e.g.][]{strut90,hai01,naji07a,esp12,ribas15}, or spatially resolved cavities in continuum dust emission \citep[e.g.][]{and11,vdm16}, indicate that dust grains in the inner disk regions are depleted before those at larger radii.
Disks that show these properties have been proposed to be in a transition phase (``transitional disks") toward their dispersal \citep[e.g.][]{strom89,strut90,takart01}. The concept of ``transitional" disk has been developing ever since its early definitions, under the evidence provided by larger samples, more detailed observations, and improved models \citep[e.g.][]{esp14,owen16}. Understanding the properties, origin, and evolution of inner disk dispersal and of gaps in protoplanetary disks bears the potential to inform on when, where, and in what physical and chemical environments planets form.

Models that produce an inside-out dispersal of disks include the effects of photoevaporation by stellar X-ray and UV radiation \citep[e.g.][]{alex06,gorti09,owen10}. In these models, disk gaps are opened where thermal heating launches the gas into a photoevaporative wind by providing enough energy to exceed the gravitational potential of the star. After a gap is opened, the inner disk is expected to be drained onto the central star by viscous accretion, and the inner hole would rapidly grow to larger radii under the effect of photoevaporation. Disks may also disperse under removal of gas by magnetohydrodynamic (MHD) winds \citep[e.g.][]{ferr06,bai16}. Another process that has been shown to open gaps in disks is the formation of giant planets \citep[e.g.][among several others]{lp86,kn12}. Where and when gaps may be opened depends on the efficiency of giant planet core formation \citep[usually assumed to increase close to the water snow line, at several to $\sim10$\,au, e.g.][]{il05,kk08}, on their migration, and on the number of planets \citep[e.g.][]{zhu11,dong15}. In this scenario, the inner region within the gap opened by the planet will also be depleted by viscous accretion onto the central star, eventually forming an inner disk hole. While there is no single model that can entirely match the currently observed properties of ``transitional" disks, there is evidence that the photoevaporative and planet scenarios are both happening in disks \citep[][]{owen16}. Models are now exploring also the combination of multiple processes, toward a more comprehensive understanding of the evolution of planet-forming disk regions. Recent models by \citet{gorti15}, for instance, explore the combination of dust evolution and disk photoevaporation, showing how the gas-to-dust ratio decreases in photoevaporating inner disks at radii $\lesssim10$\,au, producing favorable conditions for rocky planetesimal formation.

The detailed study of the properties of inner disk gaps and holes is greatly advancing thanks to the increase in spatial and spectral resolution of observing techniques. Direct imaging is able to reveal gap sizes (if their angular size is large enough to be spatially resolved) and the relative distribution of dust grain populations \citep[small versus large grains, ][]{gar13,aki16}, providing tracers of physical processes that act in inner disks. Gas spectroscopy holds unique potential to unveil the chemical evolution of planet-forming regions. Recent developments in sensitivity and spectral resolution of infrared spectrographs allowed the first large surveys of molecular gas in inner disks able to spectrally- (and sometimes also spatially-) resolve gas within dust gaps \citep[e.g.][]{pont08,sal11,vdplas15}. The combined analysis of a large sample of disks recently established rovibrational CO (carbon monoxide) emission at 4.7\,$\mu$m as a tracer of inner gaps in molecular gas. \citet{bp15} found evidence for an inside-out depletion sequence where CO gas is progressively removed from $\sim0.1$ out to $\sim20$\,au, providing a new probe of gaps over a radial disk region that extends the region probed by millimeter interferometers (typically limited to $\gtrsim5-10$\,au depending on the distance from Earth). By tracing H$_{2}$ emission at ultraviolet wavelengths, \citet{keri15} found similar evidence for a removal of molecular gas at small radii in transitional disks. The combination of dust and gas studies of small to large gaps bears the potential of a comprehensive understanding of the physical and chemical evolution of inner disks during planet formation. As a further contribution to this rapidly growing field, we report on the depletion of H$_{2}$O (water) and OH (hydroxyl) during formation of inner disk gaps.

\begin{deluxetable*}{c c c c c c c c c c c}
\tabletypesize{\small}
\tablewidth{0pt}
\tablecaption{\label{tab:water_surveys} Summary of major water vapor surveys in protoplanetary disks.}
\tablehead{\colhead{$\lambda$} & \colhead{Instr.} & \colhead{Resol.} & \colhead{\# lines} & \colhead{E$_u$} & \colhead{T$_{\rm{ex}}$} & \colhead{\# disks} & \colhead{M$_{\star} < 0.2$} & \colhead{$0.2 <$ M$_{\star} < 1.5$} & \colhead{M$_{\star} > 1.5$} & \colhead{Refs} \\ 
\colhead{[\,$\mu$m]} & \colhead{} & \colhead{} & \colhead{} & \colhead{[K]} & \colhead{[K]} & \colhead{} & \multicolumn{3}{c}{Detection fractions (and sample sizes)} & \colhead{}}
\tablecolumns{11}
\startdata

2.9--3 & VLT- & 100,000 & $\approx40$ & 8000--10,000 & 900 & $\approx40$ & N/A & 58\% (24) & 11\% (18) & this work, 1 \\

 & CRIRES &  &  &  & \multicolumn{2}{c}{\textit{Sensitivity}:}  & -- & -15 & -14.6 &  \\

\\

10--37 & \textit{Spitzer}- & 700 & $\approx200$ & 700--6000 & 300-700 & $\approx100$ & 0\% (5) & 63--85\% (64) & 0--18\% (27) & this work, 2 \\

 & IRS &  &  &  & \multicolumn{2}{c}{\textit{Sensitivity}:}  & -15.7 & -14.8 & -13.7 & \\

\\

55--200 & \textit{Herschel}- & 1000--5000 & $\approx20$ & 100--1400 & 100-300 & $\approx120$ & 0\% (10) & 15\% (80) & 15\% (27) & 3\\

 & PACS &  &  &  & \multicolumn{2}{c}{\textit{Sensitivity}:}  & -14.6 & -14.5 & -14 &  \\

\enddata
\tablecomments{Detection fractions are reported in percentage, individual sample sizes in brackets, and log sensitivities in erg cm$^{-2}$ s$^{-1}$, for three stellar mass bins as indicated.}
\tablerefs{\small{ $^{1}$ \citet{fed11,mand12}; $^{2}$ \citet{pont10a,cn11,sal11,pasc13}; $^{3}$ \citet{rivmar12,rivmar15,meeus12,fed13,blevins}}}
\end{deluxetable*}

\begin{figure*}
\includegraphics[width=1\textwidth]{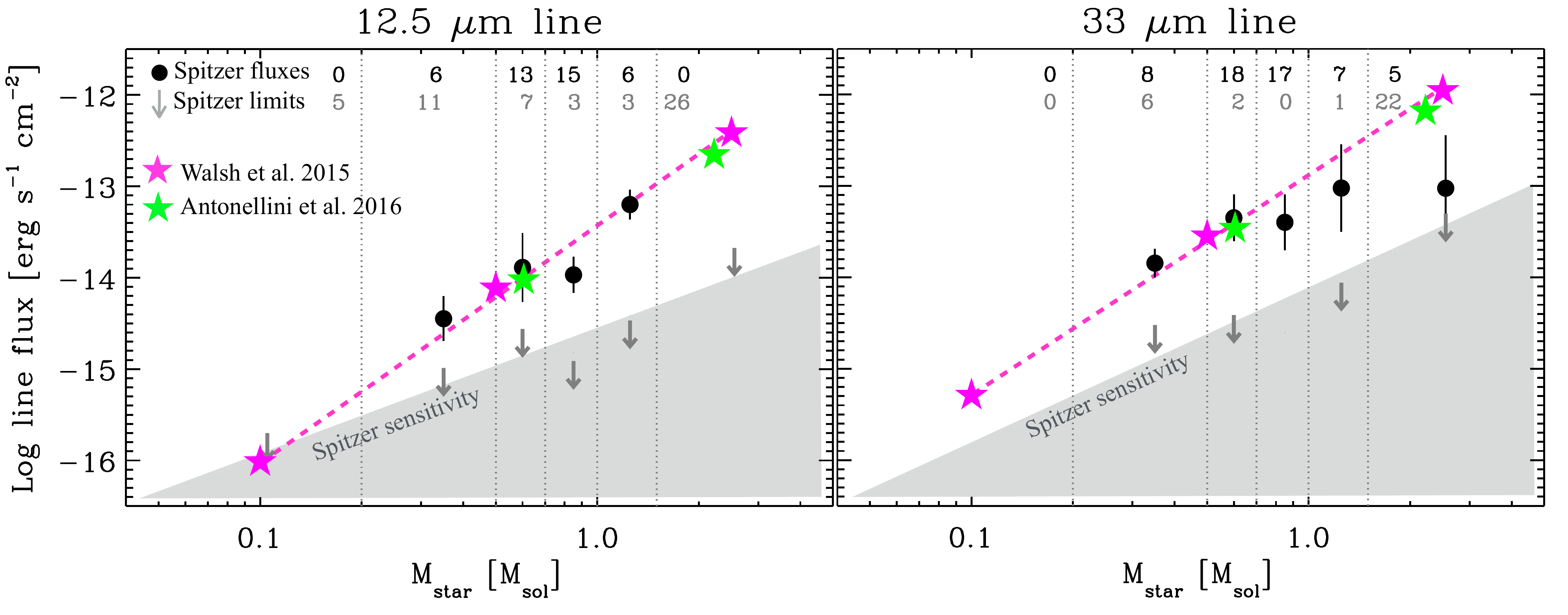} 
\caption{Mid-IR water emission dependence on stellar mass. Black dots and error-bars show median values and the median absolute deviation of water line fluxes measured in \textit{Spitzer}-IRS spectra. Median $1\sigma$ flux upper limits are shown with grey arrows. The number of disks where water emission is detected/undetected in each stellar mass bin is shown at the top. Models by \citet{walsh15} and \citet{anto16} are scaled to the measured line fluxes in disks around stars of 0.5 M$_{\odot}$. All fluxes are normalized to a common distance of 140 pc.}
\label{fig: Spitzer_spt}
\end{figure*}

\subsection{An unsolved mystery from water surveys: low detections in disks around intermediate-mass stars} \label{sec: water_surveys}
Water vapor emission has been observed in protoplanetary disks between 3 and 540\,$\mu$m, with significant differences in detection rates depending on the observed wavelength and on the mass of the central star. The most comprehensive survey of water emission in disks by a single instrument has been obtained by the IRS spectrograph on the \textit{Spitzer Space Telescope} \citep{houck04,spitzer}, which measured hundreds of water emission lines between 10 and 37\,$\mu$m in over 60 disks. \textit{Spitzer} observations showed that water emission is strongly dependent on the stellar spectral type and tentatively dependent on the evolutionary stage, with high detection rates in disks around solar-mass stars, low in disks around intermediate-mass stars, and similarly low in a small sample of ``transitional" disks \citep[1/6, see Table \ref{tab:water_surveys} and][]{pont10a,cn11,sal15}. 
Water emission at \textit{Spitzer} wavelengths has been interpreted as coming in most disks from an optically thick surface layer within the snow line radius, at temperatures of 300--700 K \citep{cn11,sal11}. Modeling of the velocity-unresolved \textit{Spitzer} spectra ($\Delta v$\,$\sim$\,400\,km\,s$^{-1}$) provided estimates of the disk emitting region to the inner few au in disks \citep[e.g.][]{naji14,anto15,walsh15}. 
At 12.4\,$\mu$m, within the \textit{Spitzer}-IRS range, a few water lines have been resolved in velocity in four disks with VLT-VISIR and Gemini-TEXES \citep[$\Delta v$\,$\sim$\,15 and $\sim$\,3.5\,km\,s$^{-1}$ respectively,][]{visir,texes}, providing support to the emitting regions estimated by models for the \textit{Spitzer} spectra \citep{pont10b,banz14,sal15}.
At shorter wavelengths ($\sim$\,2--3\,$\mu$m), hotter water (T $\approx900-1500$\,K) has been spectrally-resolved and analyzed only in four disks so far, by \citet{sal08} and \citet{mand12} with Keck-NIRSPEC and VLT-CRIRES, and by \citet{carr04} with IRTF-CSHELL. 
At longer wavelengths ($>55$\,$\mu$m), colder water has been surveyed in disks by \textit{Herschel}-PACS, finding as low detection rates in disks around solar-mass as in disks around intermediate-mass stars \citep[$\sim 15$\,\% as based on the 63.3\,$\mu$m line, see references in Table \ref{tab:water_surveys} and][]{rivmar16}. 

Table \ref{tab:water_surveys} summarizes properties, samples, and detection rates for all major surveys of water vapor in protoplanetary disks, including this work. The general trends identified to date are shown in Figure \ref{fig: Spitzer_spt}, and are compared to expectations produced by recent models of water in disks \citep{walsh15,anto15}\footnote{We reproduce the model spectra of \citet{walsh15} by using a slab model in LTE, and adopting the temperature, column density, and emitting area reported for the observable layer of gas (down to $\tau = 1$); in the case of \citet{anto16}, we use the model spectra kindly provided by the authors. We scale the models to the line fluxes measured in disks around stars of 0.5 M$_{\odot}$, to highlight global trends with M$_{\star}$.}. We investigate the measured line fluxes against the masses of the central stars M$_{\star}$, as an alternative way to look at the spectral-type dependencies found by \citet{pont10a}. Water line fluxes generally increase with stellar luminosity \citep{sal11}, and with stellar mass. Models explain this with the radial extent of the warm ($>300$\,K) disk surface where water vapor is abundant due to ice evaporation and efficient gas-phase formation, which increases to larger disk radii around more luminous stars. 
Model predictions agree well with the increase in \textit{Spitzer} line fluxes measured in disks around stars of masses between 0.2 and 1.5 M$_{\odot}$, although 20--40\% of disks show significantly lower fluxes, relative to what is expected from a pure luminosity effect. For M$_{\star} > 1.5$\,M$_{\odot}$, models produce water line fluxes $\gtrsim10$ times stronger than observed for 80-100\% of the disk sample. 
The low water detection frequency at \textit{Spitzer} and CRIRES wavelengths in disks around stars of M$_{\star} > 1.5$\,M$_{\odot}$ is still not understood \citep[e.g.][]{walsh15}.

In this work, we test the hypothesis that inside-out inner disk depletion may produce the progressive disappearance of water vapor emission at infrared wavelengths. If water vapor in inner disks is depleted from smaller to larger disk radii, an essential spectral region to include in the investigation of this process is where the hot water emits, in high-energy transitions at 2.9-3\,$\mu$m \citep[e.g.][]{mand12}. In Sections \ref{sec:obs} and \ref{sec:ana} we describe the new survey of hot water emission as observed with CRIRES at 2.9\,$\mu$m, and the comparative analysis with the CO dataset at 4.7\,$\mu$m. In Section \ref{sec:rad_trends}, we report the discovery of a correlation between water line fluxes at 2.9--33\,$\mu$m and the size of a gas-depleted region in the inner disk as probed by rovibrational CO emission. We discuss the properties of these molecular gaps, and propose their potential for studies of the evolution of planet-forming disk regions, in Section \ref{sec:disc}.

\subsection{A note on molecular gaps/holes in inner disks} \label{sec:nomencl}
The terms disk ``gap" and ``hole" have been referred (sometimes interchangeably) to radial regions where the dust in the disk shows some degree of depletion as compared to a continuous distribution adopted by models for disk considered ``primordial" \citep[e.g.][]{esp12}. In this work, we adopt these terms to identify inner disk regions of various sizes (namely $\approx$\,0.1--20 au) empirically defined to where molecular gas is depleted enough that its infrared emission (at wavelengths of 2.9--35\,$\mu$m) is not detected.
The term ``gap" was utilized in \citet{bp15} with this meaning, with the discovery of a strong correlation between velocity-resolved line widths and the $v2/v1$ vibrational ratios in rovibrational CO emission at 4.7\,$\mu$m in a large survey of protoplanetary disks. By converting these measurements respectively into a characteristic CO emitting radius R$_{\rm{co}}$ and the vibrational excitation temperature, this correlation revealed cooler CO gas as larger gaps are formed in disks (i.e. as R$_{\rm{co}}$ increases). 
This observational finding suggested an inside-out depletion scenario, which is now being increasingly supported by independent modeling explorations \citep[][]{woitke16,hbert16}. To distinguish them from dust gaps or holes detected by other techniques, we call them``molecular gaps/holes", especially after the discovery in this work that they are shared by three of the most abundant molecules in disks: CO, H$_{2}$O, and OH (Sections \ref{sec:ana} and \ref{sec:rad_trends}). In absence of constraints on the radial distribution of any undetected residual CO gas within the CO-depleted inner region, we use the term gap/hole interchangeably. An upper limit to the CO column density of $\lesssim10^{15}$\,cm$^{-2}$ has been estimated by \citet{carm16} in one of these gaps/holes, implying that very little residual CO gas is left (if any) in the inner CO-depleted region.

\begin{deluxetable*}{l c c c c c c c c c c c c}
\tabletypesize{\small}
\tablewidth{0pt}
\tablecaption{\label{tab: sample} Sample properties adopted or calculated from the literature.}
\tablehead{\colhead{Name} & \colhead{M$_{\star}$}  & \colhead{L$_{\star}$} & \colhead{log M$_{acc}$} & \colhead{log L$_{acc}$} & \colhead{incl.} & \colhead{$v2/v1$} & \colhead{R$_{\rm{co}}$} & \colhead{R$_{\rm snow}$} & \colhead{dist.} & \colhead{cont. (3.3\,$\mu$m)} & \colhead{cont. (4.6\,$\mu$m)} & \colhead{Refs}
\\ 
 &  [$M_{\odot}$] & [$L_{\odot}$] & [$M_{\odot}$ yr$^{-1}$] & [$L_{\odot}$] & [$^{\circ}$] &  & [au] & [au] & [pc] & [Jy] & [Jy] & }
\tablecolumns{13}
\startdata

AATau & 0.85 & 0.71 & -8.31 & -1.43 & 71 & 0.14 & 0.20 & 1.45 & 140 & 0.322 $\pm$ 0.011 & 0.337 $\pm$ 0.006 &   \\
AS205 N & 1.10 & 7.10 & -7.10 & -0.18 & 20 & 0.38 & 0.17 & 5.45 & 125 & 4.390 $\pm$ 0.370 & 5.220 $\pm$ 0.400 &   \\
AS209 & 1.40 & 2.50 & -7.52 & -0.35 & 39 & 0.91 & 0.08 & 3.84 & 125 & 0.829 $\pm$ 0.055 & 0.806 $\pm$ 0.032 &   \\
CVCha & 2.10 & 8.00 & -6.81 & 0.41 & 30 & 0.25 & 0.18 & 9.09 & 215 & 0.816 $\pm$ 0.020 & 1.170 $\pm$ 0.030 & 1 \\
CWTau & 1.20 & 0.76 & -7.80 & -0.49 & 28 & 0.32 & 0.32 & 2.74 & 140 & 1.470 $\pm$ 0.180 & 1.760 $\pm$ 0.160 &   \\
DFTau & 0.53 & 1.97 & -6.93 & -0.63 & 65 & 0.43 & 0.20 & 5.08 & 140 & 1.190 $\pm$ 0.130 & 1.380 $\pm$ 0.090 &   \\
DGTau & 0.30 & 1.70 & -6.39 & 0.70 & 18 & 0.59 & 0.01 & 7.30 & 140 & 1.030 $\pm$ 0.070 & 1.500 $\pm$ 0.080 & 2 \\
DoAr24E S & 0.70 & 8.80 & -7.44 & -0.92 & 20 & 0.30 & 0.04 & 3.31 & 120 & 2.360 $\pm$ 0.380 & 3.130 $\pm$ 0.340 &   \\
DoAr44 & 1.40 & 1.40 & -8.43 & -1.16 & 25 & 0.03 & 0.33 & 1.51 & 125 & 0.610 $\pm$ 0.031 & 0.593 $\pm$ 0.014 &   \\
DOTau & 0.72 & 1.05 & -7.28 & -0.67 & 31 & 0.49 & 0.12 & 3.93 & 140 & 0.892 $\pm$ 0.080 & 1.150 $\pm$ 0.060 &   \\
DRTau & 1.00 & 0.90 & -6.82 & -0.24 & 9 & 0.39 & 0.06 & 7.03 & 140 & 1.420 $\pm$ 0.150 & 1.910 $\pm$ 0.160 &   \\
EC82 & 0.75 & 3.21 & -- & 0.03 & 59 & 0.03 & 0.81 & -- & 415 & 0.348 $\pm$ 0.008 & 0.498 $\pm$ 0.010 & 3 \\
EXLup08 & 0.80 & 0.60 & -7.50 & 0.30 & 44 & 0.61 & 0.06 & 3.25 & 155 & 1.300 $\pm$ 0.300 & 4.200 $\pm$ 0.600 & 4 \\
EXLup14 & 0.80 & 0.60 & -10.00 & -1.30 & 48 & 0.51 & 0.07 & 0.25 & 155 & 0.290 $\pm$ 0.040 & 0.340 $\pm$ 0.030 & 5 \\
FNTau & 0.33 & 0.80 & -8.25 & -1.77 & 20 & $<$0.12 & 1.69 & 1.12 & 140 & 0.294 $\pm$ 0.009 & 0.308 $\pm$ 0.005 & 6 \\
FZTau & 0.70 & 0.51 & -7.33 & -0.27 & 38 & 0.20 & 0.98 & 3.70 & 140 & 1.060 $\pm$ 0.070 & 1.340 $\pm$ 0.070 &   \\
GQLup & 0.80 & 0.80 & -8.15 & -1.10 & 65 & 0.28 & 0.20 & 1.67 & 150 & 1.090 $\pm$ 0.090 & 1.030 $\pm$ 0.050 &   \\
HD36112 & 2.00 & 30.00 & -7.35 & -0.10 & 22 & 0.03 & 2.26 & 5.15 & 200 & 4.100 $\pm$ 0.320 & 4.850 $\pm$ 0.390 &   \\
HD95881 & 2.00 & 2.98 & $<$-5.65 & $<$1.63 & 55 & 0.06 & 4.65 & $<$29.31 & 170 & 4.830 $\pm$ 0.900 & 6.990 $\pm$ 0.970 &   \\
HD97048 & 2.20 & 33.00 & $<$-8.16 & $<$-0.55 & 43 & 0.33 & 12.57 & $<$2.32 & 180 & 3.010 $\pm$ 0.200 & 3.250 $\pm$ 0.130 &   \\
HD98922 & 2.20 & 891.00 & $<$-6.97 & $<$0.41 & 45 & 0.29 & 8.85 & $<$7.84 & 350 & 16.500 $\pm$ 2.400 & 30.700 $\pm$ 0.400 &   \\
HD101412 & 2.30 & 1.36 & $<$-7.61 & $<$-0.04 & 80 & -- & 0.79 & $<$4.13 & 160 & 1.090 $\pm$ 0.060 & 1.720 $\pm$ 0.090 &   \\
HD135344B & 1.60 & 8.00 & -8.35 & -1.11 & 14 & 0.05 & 1.70 & 1.72 & 140 & 3.170 $\pm$ 0.390 & 3.540 $\pm$ 0.400 &   \\
HD139614 & 1.50 & 6.60 & -7.63 & -0.10 & 20 & 0.25 & 2.77 & 3.51 & 140 & 1.370 $\pm$ 0.070 & 1.560 $\pm$ 0.040 &   \\
HD141569 & 1.90 & 19.10 & -7.65 & -0.05 & 53 & 0.56 & 17.02 & 3.72 & 100 & 0.652 $\pm$ 0.026 & 0.441 $\pm$ 0.008 &   \\
HD142527 & 3.50 & 69.00 & -7.02 & 0.07 & 20 & 0.04 & 2.33 & 8.69 & 140 & 6.090 $\pm$ 0.510 & 7.850 $\pm$ 0.500 &   \\
HD144432S & 1.70 & 14.80 & -7.69 & -0.35 & 27 & 0.09 & 0.78 & 3.44 & 160 & 2.200 $\pm$ 0.360 & 2.530 $\pm$ 0.260 &   \\
HD150193 & 1.90 & 16.20 & -7.45 & 0.10 & 38 & 0.24 & 0.88 & 4.57 & 150 & 5.130 $\pm$ 1.260 & 6.470 $\pm$ 1.220 &   \\
HD163296 & 2.30 & 36.00 & -7.13 & 0.28 & 46 & 0.14 & 1.40 & 6.75 & 122 & 9.810 $\pm$ 2.630 & 11.600 $\pm$ 3.600 &   \\
HD179218 & 2.70 & 75.90 & -6.76 & 0.42 & 57 & 0.37 & 20.80 & 10.40 & 240 & 3.550 $\pm$ 0.700 & 4.820 $\pm$ 0.590 &   \\
HD190073 & 2.85 & 83.20 & -5.82 & 1.38 & 23 & 0.27 & 6.03 & 27.72 & 767 & 3.950 $\pm$ 0.790 & 5.620 $\pm$ 1.170 &   \\
HD244604 & 3.05 & 97.70 & -7.19 & 0.14 & 50 & 0.11 & 2.76 & 6.98 & 336 & 0.742 $\pm$ 0.025 & 0.840 $\pm$ 0.019 &   \\
HD250550 & 3.40 & 190.00 & -5.63 & 1.82 & 10 & 0.16 & 0.91 & 35.70 & 280 & 1.690 $\pm$ 0.210 & 2.030 $\pm$ 0.160 &   \\
HTLup & 2.50 & 14.50 & -- & -0.71 & 28 & 0.65 & 0.05 & -- & 150 & 1.500 $\pm$ 0.170 & 1.590 $\pm$ 0.110 &   \\
IMLup & 0.52 & 1.30 & -11.00 & -1.08 & 49 & 1.16 & 0.03 & 0.08 & 190 & 0.500 $\pm$ 0.023 & 0.423 $\pm$ 0.009 &   \\
IRS48 & 2.00 & 14.30 & -8.40 & -0.89 & 42 & 0.38 & 16.22 & 1.76 & 125 & 1.490 $\pm$ 0.140 & 2.220 $\pm$ 0.230 &   \\
LkHa330 & 2.50 & 16.00 & -7.66 & -0.46 & 12 & 0.09 & 4.74 & 4.04 & 250 & 1.130 $\pm$ 0.110 & 1.360 $\pm$ 0.090 &   \\
RNO90 & 1.50 & 5.70 & -7.40 & -0.19 & 37 & 0.21 & 0.23 & 4.44 & 125 & 2.070 $\pm$ 0.130 & 2.460 $\pm$ 0.160 &   \\
RULup & 0.70 & 0.42 & -7.75 & -0.47 & 35 & 0.37 & 0.14 & 2.41 & 150 & 1.220 $\pm$ 0.140 & 1.700 $\pm$ 0.110 &   \\
RYLup & 1.50 & 2.60 & -7.50 & 0.08 & 68 & $<$0.04 & 14.13 & 4.01 & 150 & 1.020 $\pm$ 0.080 & 1.340 $\pm$ 0.050 & 7 \\
RWAur & 1.34 & 1.70 & -7.50 & -0.20 & 55 & 0.62 & 0.05 & 3.86 & 140 & 1.000 $\pm$ 0.040 & 1.380 $\pm$ 0.040 & 8 \\
SCrA S & 0.60 & 0.76 & -7.53 & -0.65 & 26 & 0.40 & 0.04 & 2.87 & 130 & 2.710 $\pm$ 0.310 & 3.040 $\pm$ 0.220 &   \\
SCrA N & 1.50 & 2.30 & -7.36 & -0.13 & 10 & 0.30 & 0.05 & 4.63 & 130 & 2.710 $\pm$ 0.310 & 3.040 $\pm$ 0.220 &   \\
SR9 & 1.35 & 1.41 & -9.20 & -- & 60 & $<$0.07 & 0.47 & 0.68 & 120 & 0.729 $\pm$ 0.049 & 0.660 $\pm$ 0.018 & 9 \\
SR21 & 2.20 & 15.00 & -7.90 & -0.70 & 15 & 0.64 & 6.46 & 3.03 & 125 & 1.110 $\pm$ 0.090 & 1.080 $\pm$ 0.040 & 10 \\
TTau N & 2.10 & 7.30 & -7.16 & 0.05 & 20 & 0.57 & 0.07 & 6.35 & 147 & 8.800 $\pm$ 0.800 & 12.000 $\pm$ 0.800 &   \\
TTau S & 2.50 & -- & -- & 0.55 & 25 & 0.20 & 1.10 & -- & 147 & 8.800 $\pm$ 0.800 & 12.000 $\pm$ 0.800 &   \\
TWCha & 1.00 & 0.38 & -8.13 & -1.66 & 67 & 0.16 & 0.71 & 1.84 & 180 & 0.178 $\pm$ 0.004 & 0.174 $\pm$ 0.003 & 6 \\
TWHya & 0.70 & 0.23 & -8.70 & -1.30 & 4 & 0.09 & 0.34 & 0.91 & 54 & 0.482 $\pm$ 0.023 & 0.303 $\pm$ 0.006 &   \\
VSSG1 & 0.52 & 1.50 & -7.56 & -0.87 & 53 & 0.36 & 0.27 & 2.65 & 120 & 0.757 $\pm$ 0.045 & 1.010 $\pm$ 0.060 &   \\
VVSer & 3.00 & 85.00 & -7.13 & 0.33 & 65 & 0.16 & 2.89 & 7.38 & 415 & 2.600 $\pm$ 0.480 & 2.530 $\pm$ 0.180 &   \\
VWCha & 0.60 & 3.37 & -7.50 & -0.77 & 44 & 0.32 & 0.16 & 2.95 & 180 & 1.060 $\pm$ 0.050 & 0.847 $\pm$ 0.047 &   \\
VZCha & 0.80 & 0.46 & -7.91 & -0.73 & 25 & 0.44 & 0.18 & 2.14 & 180 & 0.354 $\pm$ 0.009 & 0.397 $\pm$ 0.007 &   \\
WaOph6 & 0.90 & 0.67 & -6.64 & -0.73 & 39 & 0.33 & 0.07 & 8.15 & 120 & 1.010 $\pm$ 0.070 & 1.000 $\pm$ 0.050 &   \\
WXCha & 0.60 & 0.77 & -8.47 & -- & 87 & 0.45 & 0.11 & 1.09 & 180 & 0.399 $\pm$ 0.011 & 0.392 $\pm$ 0.007 & 11 \\

\enddata

\tablecomments{
Stellar masses, luminosities, distances, and accretion rates and luminosities are taken from \citet{sal13} and references therein for solar-mass stars and from \citet{fair15} for intermediate-mass stars, unless noted otherwise. Disk inclinations are taken from spatially-resolved dust observations or velocity-resolved CO observations as compiled in \citet{bp15} and \citet{vdplas15}. Values for $v2/v1$ and R$_{\rm{co}}$ are measured as explained in \citet{bp15} and in Eqn.\,\ref{eqn: R_co} below. R$_{\rm{snow}}$ is derived using Eqn.\,\ref{eqn: R_snow}. Flux continuum measurements at 3.3 and 4.6\,$\mu$m are taken from the WISE survey as reported in \citet{wise12,wise14}, and are used to calibrate the molecular line fluxes reported in Table \ref{tab: fluxes} as explained in the Appendix, Section A\ref{app:line_flux}. References for stellar masses, luminosities, distances, and accretion rates and luminosities, when different from \citet{fair15} or the compilation in \citet{sal13}: (1) \citet{ingl13}; (2) \citet{isel10}; (3) \citet{rigl15}; (4) \citet{asp10}; (5) \citet{sicagu12}; (6) \citet{manara16}; (7) \citet{jokr00}; (8) \citet{whgh01}; (9) \citet{don11,ale12}; (10) \citet{manara14}; (11) \citet{sal11}. 
}

\end{deluxetable*}

\section{OBSERVATIONS and SAMPLE} \label{sec:obs}
CO emission at 4.7\,$\mu$m has been observed with the CRyogenic Infrared Echelle Spectrometer \citep[CRIRES,][]{crires} on the Very Large Telescope (VLT) of the European Southern Observatory (ESO). Most spectra have been taken as part of the ESO Large Program 179.C-0151 \citep{pont_msgr}; some measurements have been added from programs 079.C-0349 and 081.C-0833 as published in \citet{vdplas15}, from program 091.C-0671 as published in \citet{carm16}, and from program 088.C-0898 as published in \citet{hbert16}. CRIRES has a resolving power of $\sim96,000$ or 3.2\,km\,s$^{-1}$, which provides detailed velocity-resolved line profiles of CO gas from inner disks and allows us to radially locate the emission through Keplerian line broadening \citep[e.g.,][]{brown13,bp15,vdplas15}. In total, we include in this work 52 disks that have velocity-resolved CO lines from CRIRES observations. 

Within the ESO Large Program 179.C-0151, 28 disks were observed at 2.9\,$\mu$m to measure their H$_{2}$O and OH emission spectra. Three of these disks were previously published in \citet{mand12}. Three main-sequence stars were observed to be used as photospheric standards, sampling stellar spectral types from K to M \citep[spectral types are taken from][]{gray,koen}: NLTT29176 (K3), HD111631 (K7), and HIP49986 (M1.5). These spectra are used to correct the 2.9\,$\mu$m H$_{2}$O and OH emission spectra, as explained in Section \ref{sec:ana3}. We also include a second epoch of the 2.9 and 4.7\,$\mu$m spectra of a strongly variable accreting source, EX Lupi, obtained as part of program 093.C-0432 \citep{banz15}, and measurements at 2.9\,$\mu$m for HD\,250550 and HD\,98922 from program 082.C-0491 as published in \citet{fed11}.
The CRIRES data from programs 179.C-0151 and 093.C-0432 have been reduced using procedures developed for the ESO Large Program 179.C-0151 as described in \citet{pont11a}.
To complement the CRIRES dataset, we include in this work the Keck-NIRSPEC \citep[R $\sim$ 25,000,][]{nirspec} 4.7\,$\mu$m CO spectra of DO\,Tau, HD\,244604, and HD\,36112, previously published in \citet{blake04} and \citet{sal11b}. These targets have H$_{2}$O spectra available at 2.9\,$\mu$m and/or at 10--35\,$\mu$m, but have not been observed at 4.7\,$\mu$m with CRIRES.

The dataset of H$_{2}$O and CO emission at 2.9 and 4.7\,$\mu$m is analyzed in this work in combination with \textit{Spitzer}-IRS spectra covering H$_{2}$O emission at 10--35\,$\mu$m at the resolution of R $\sim$ 700. The \textit{Spitzer} dataset has been presented and analyzed in previous work \citep{pont10a,sal11,cn11}. The advantage of a combined analysis of water emission over such a large wavelength range is to include lines that probe upper level energies between 1500 and 9000 K and the whole disk region out to the water snow line (see references in Section \ref{sec: water_surveys}).
The combined sample includes 55 disks around stars with estimated masses between 0.3 and 3.5 M$_{\odot}$ (Table \ref{tab: sample}), observed in different star-forming regions (mainly Taurus, Ophiucus, Lupus, Chamaeleon for solar-mass stars, and several isolated intermediate-mass stars), and spanning evolutionary phases from primordial to transitional disks. The sample has been assembled from previous work as driven by the availability of CO spectra at 4.7\,$\mu$m (available for all 55 disks) and H$_{2}$O spectra at 2.9\,$\mu$m (available for 31 disks in this sample) and/or at 10--35\,$\mu$m (available for 44 disks in this sample).

\begin{deluxetable}{c c c c c c c}
\tabletypesize{\small}
\tablewidth{0pt}
\tablecaption{\label{tab:lines} Molecular lines properties.}
\tablehead{\colhead{Wavelength} & Transition & $A_{ul}$ & $E_{u}$ &  \\ \colhead{ ($\mu$m)} &  &  (s$^{-1}$) &  (K)}
\tablecolumns{4}
\startdata
\cutinhead{H$_{2}$O lines} 
2.9273 & $15_{\:1\:15} \rightarrow 16_{\:1\:16}$ & 49 &  8744  \\
2.9278 & $12_{\:3\:9} \rightarrow 13_{\:3\:10}$ & 52 &  8388  \\
2.9291 & $13_{\:3\:11} \rightarrow 14_{\:3\:12}$ & 48 &  8583 \\
2.9292 & $14_{\:1\:13} \rightarrow 15_{\:1\:14}$ & 49 &  8698  \\
12.396 & $17_{\:4\:13} \rightarrow 16_{\:3\:14}$ & 7.7 &  5781 \\
12.407 & $16_{\:3\:13} \rightarrow 15_{\:2\:14}$ & 4.2 &  4945 \\
12.454 & $13_{\:7\:6} \rightarrow 12_{\:4\:9}$ & 1.2 &  4213  \\
12.519 & $15_{\:12\:13} \rightarrow 14_{\:1\:14}$ & 1.5 &  4133  \\
12.596 & $13_{\:6\:7} \rightarrow 12_{\:3\:10}$ & 0.8 &  3966  \\
17.103 & $12_{\:5\:8} \rightarrow 11_{\:2\:9}$ & 3.7 &  3274  \\
17.225 & $11_{\:3\:9} \rightarrow 10_{\:0\:10}$ & 1.0 &  2439  \\
17.358 & $11_{\:2\:9} \rightarrow 10_{\:1\:10}$ & 0.9 &  2432  \\
30.525 & $7_{\:6\:1} \rightarrow 6_{\:5\:2}$ & 14 &  1750  \\
30.871 & $8_{\:5\:4} \rightarrow 7_{\:4\:3}$ & 8.7 &  1806  \\
32.991 & $7_{\:5\:2} \rightarrow 6_{\:4\:3}$ & 8.3 &  1525  \\
33.005 & $6_{\:6\:0} \rightarrow 5_{\:5\:1}$ & 13 &  1504  \\
63.323 & $8_{\:1\:8} \rightarrow 7_{\:0\:7}$ & 1.7 &  1070  \\

\cutinhead{OH lines} 
2.9343 & 3/2 P4.5e & 11.7 & 5400 \\
2.9346 & 3/2 P4.5f & 11.7 & 5400 \\
12.647 & 1/2 R22.5ef & 213 &  15100  \\
12.657 & 3/2 R23.5ef & 213 &  15100  \\
30.277 & 3/2 R8.5f & 14.3 &  2380  \\
30.346 & 3/2 R8.5e & 14.3 &  2380  \\

\cutinhead{CO lines} 
4.7545 & v=1-0 P10 & 17.3 & 3300 \\
4.7589 & v=2-1 P4 & 37.4 & 6160

\enddata
\tablecomments{Line properties are taken from the HITRAN 2012 database \citep{hitran}.}
\end{deluxetable}

\begin{figure*}[ht]
\includegraphics[width=1\textwidth]{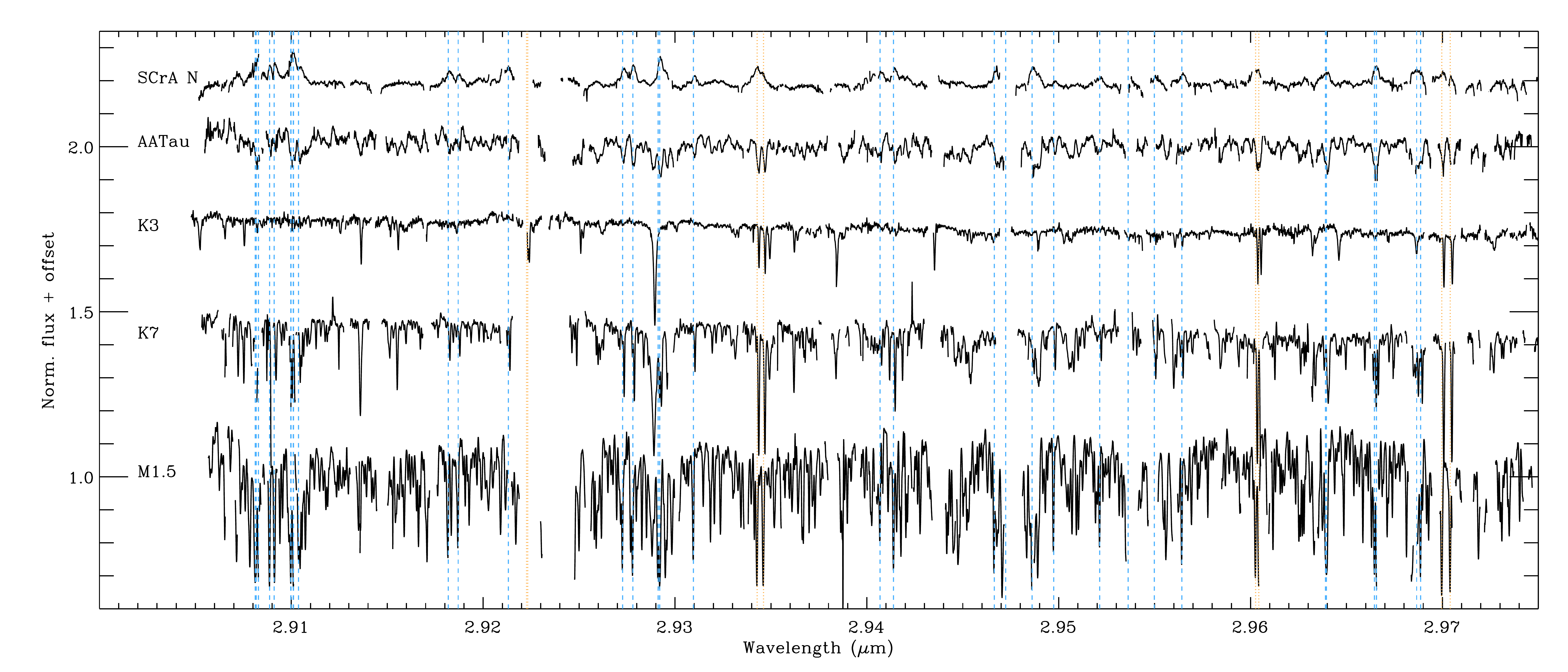} 
\caption{Examples of VLT-CRIRES spectra at 2.9\,$\mu$m. From top to bottom: a disk spectrum in emission (SCrA\,N, a K3 star), a spectrum showing moderately veiled photospheric absorption (AA\,Tau, a K7 star), and the three photospheric templates used for correction of the CRIRES spectra (K3 type: NLTT\,29176, K7 type: HD\,111631, and M1.5 type: HIP\,49986). Individual water transitions are marked with blue dashed lines, while OH transitions are marked with orange dotted lines.}
\label{fig: phot stand}
\end{figure*}

\begin{figure}[ht]
\includegraphics[width=0.5\textwidth]{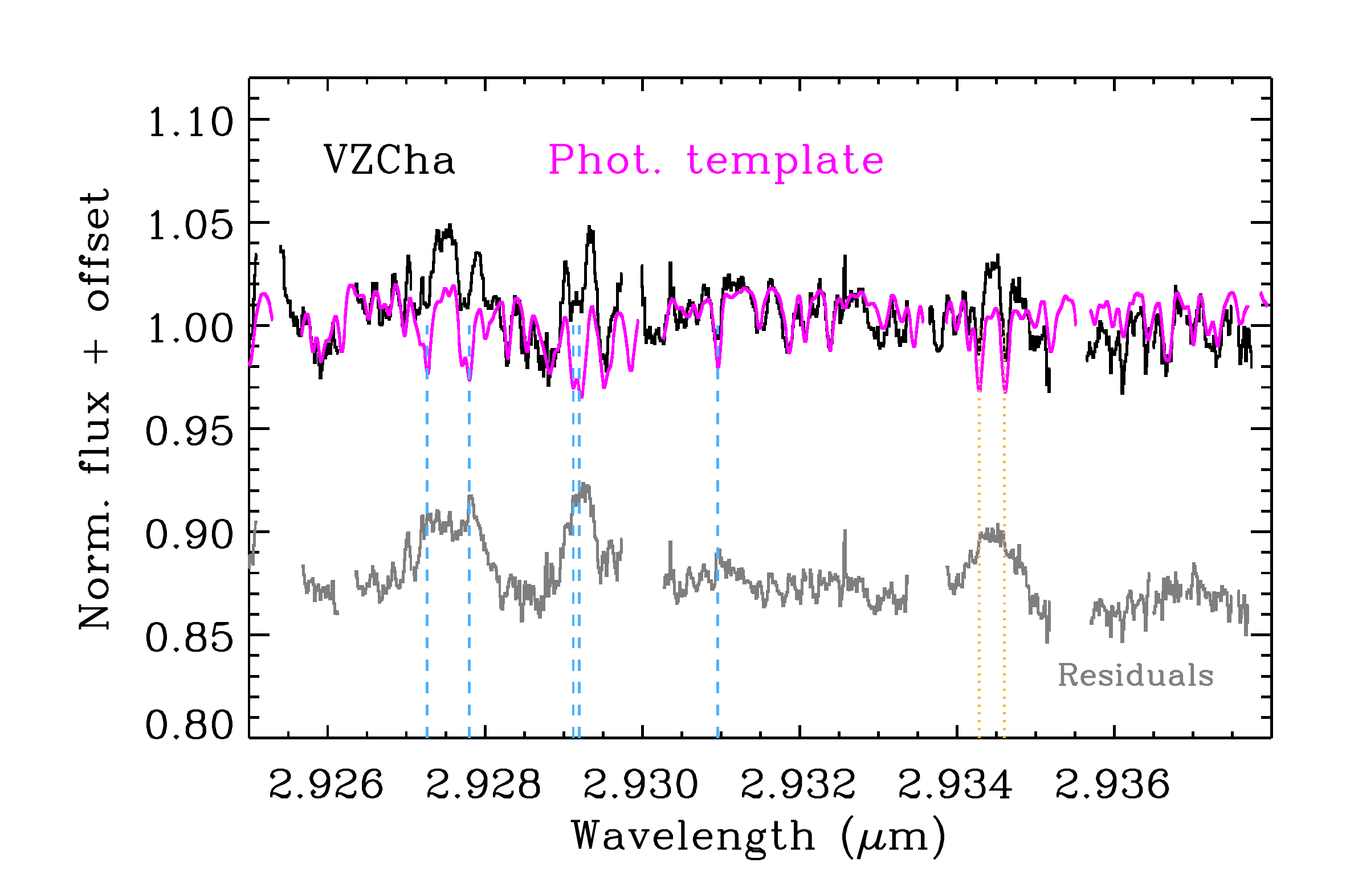} 
\caption{Example of the photospheric correction of protoplanetary disk spectra observed with CRIRES at 2.9\,$\mu$m (the spectra from the entire survey are shown in the Appendix). The spectrum is shown at the top in black, with its photospheric template in magenta (veiled and broadened). The molecular emission spectrum after correction is shown at the bottom in grey. Water and OH transitions are marked in blue and orange as in Figure \ref{fig: phot stand}.}
\label{fig: phot_corr}
\end{figure}

\section{SPECTRAL LINE ANALYSIS} \label{sec:ana}
In this Section, we present the analysis of the H$_{2}$O and OH emission spectra in the CRIRES setting at 2.905--2.99\,$\mu$m. The analysis of these spectra presents two main challenges: one due to stellar photosphere absorption and one to the blending of emission lines, which can both be addressed by exploiting the high spectral resolution provided by CRIRES. The level of photospheric absorption at 2.9\,$\mu$m depends on the spectral type of the star (e.g. a cooler M1 star has more photospheric absorption from H$_{2}$O and OH than a hotter K3 star, see Figure \ref{fig: phot stand}) and on the amount of veiling from hot dust emission (the ratio between continuum and photospheric flux $F_{\rm cont}/F_{\rm phot}$). We use photospheric templates to correct for photospheric absorption and retrieve the signal from disk gas emission (Section \ref{sec:ana3}). Most of the disk emission lines are blended due to their large velocity widths produced by Keplerian rotation in the inner disk (with full widths at half maximum FWHM up to 200 km/s). We use the velocity-resolved CO line profiles observed at 4.7\,$\mu$m to characterize the H$_{2}$O and OH emission at 2.9\,$\mu$m (Section \ref{sec:ana4}). The reason to use CO emission at 4.7\,$\mu$m is that it includes several lines spanning 3000--8000 K in upper level energy $E_u$, which are close to and overlap with the upper level energies of H$_{2}$O ($E_u=$ 8000--10000 K) and OH ($E_u\sim$ 5400--5700 K) at 2.9\,$\mu$m (Table \ref{tab:lines}). These lines therefore may share similar excitation conditions and emit from a similar disk region, as supported by previous observations of similar line profiles for 4.7\,$\mu$m CO and 2.9\,$\mu$m H$_{2}$O and OH emission in a handful of disks \citep{sal08,pont10b,banz15,brit16}.

\begin{deluxetable}{l l c c c}[h!]
\tabletypesize{\small}
\tablewidth{0pt}
\tablecaption{\label{tab:phot_corr} Photospheric correction and veiling of 2.9\,$\mu$m disk spectra.}
\tablehead{\colhead{Name} & \colhead{SpT} & \colhead{Phot. templ.} & \colhead{2.9\,$\mu$m veil.} & \colhead{Note} }
\tablecolumns{5}
\startdata

AATau & K7 & M1.5 & 2.2  & $^{1}$  \\
CWTau & K3 & (M1.5) & (14.0)  & $^{1}$ \\
DFTau & M1 & M1.5 & 2.0  &  \\
DOTau & M0 & M1.5 & 7.0   & \\
EXLup14 & M0 & M1.5 & 2.7  & \\
FNTau & M5 & M1.5 & 0.24  & $^{2}$ \\
GQLup & K7 & K7 & 4.8  &   \\
HTLup & K2 & (K7) & (13.4)  & $^{1}$ \\
IMLup & M0 & (M1.5) & (1.7)  & $^{1}$ \\
RULup & K7 & M1.5 & 13.0  &  \\
SCrA S & M0 & M1.5 & 13.3   &  \\
TWCha & K7 & M1.5 & 4.4  &  \\
TWHya & K7 & M1.5 & 0.8  & $^{2}$ \\
VWCha & K5 & K7 & 5.4  &  \\
VZCha & K6 & M1.5 & 6.8  &  \\
WaOph6 & K6 & K7 & 2.2  &  \\
WXCha & M0 & M1.5 & 11.4  & 

\enddata
\tablecomments{$^{1}$\,Residual photospheric absorption. $^{2}$\,Possible emission as narrow as photospheric lines. See the Appendix for details.}
\end{deluxetable}

\subsection{Photospheric correction of 2.9\,$\mu$m spectra} \label{sec:ana3}
We take each 2.9\,$\mu$m spectrum that has substantial stellar photospheric absorption and we correct it by subtraction of the photospheric template that best matches its absorption spectrum. Given the coarse grid of spectral type templates obtained in this survey (Figure \ref{fig: phot stand}), the best matching template is usually easily identified for each science spectrum. We apply the following procedure to the spectral range between 2.925 and 2.935\,$\mu$m, where the S/N is higher, the spectrum is less affected by detector or telluric gaps, and the H$_{2}$O and OH lines are more isolated and stronger (Figure \ref{fig: phot stand}). The template spectrum is first shifted to compensate for velocity differences with the science target, by cross-correlating the H$_{2}$O and OH photospheric features in the science and template spectra. To match the science spectrum, the template is then artificially veiled to account for circumstellar disk emission, and its absorption lines broadened. This is achieved by adding an excess flat continuum and by convolving the template spectrum with a Gaussian of variable width \citep[for the technique see also][]{hartigan89,baba90,hartigan91}. The best fit is found by minimizing the residuals after subtraction of the veiled and broadened template from the science spectrum, over a grid of veiling and broadening values. Residuals are measured as the standard deviation of pixel values from the median-smoothed de-veiled spectrum, in a spectral range free of disk molecular emission (2.930--2.934\,$\mu$m). Figure \ref{fig: phot_corr} shows an example of the photospheric correction procedure as applied to the spectrum of VZ\,Cha; plots for all the other targets are reported in the Appendix. Table \ref{tab:phot_corr} lists, for each target, the best photospheric standard match and the estimated veiling at 2.9\,$\mu$m. Veiling measurements may be refined in future work by using a denser sampling of template spectral types.

\subsection{Characterization of CO line profiles} \label{sec:ana1}
The analysis of CO lines has been presented in \citet{bp15}, and we provide here a short summary.
We built high S/N CO line profiles for each disk by stacking rovibrational transitions observed in the 4.7\,$\mu$m spectra. The stacked line profiles preserve the line flux as well as the line shape, through a homogeneous analysis that we adopted for the entire disk sample. We stacked those lines that are most commonly available across the sample (due to the observing strategies and the detector characteristics), that have similar flux, and that are not blended with higher-level transitions. These requirements led us to typically use the $v=1-0$ lines between P7 and P12, and $v=2-1$ lines between P3 and P6. Each line was normalized to its local continuum, re-sampled on a common velocity grid, and stacked by weighted average. 

\begin{figure}
\includegraphics[width=0.5\textwidth]{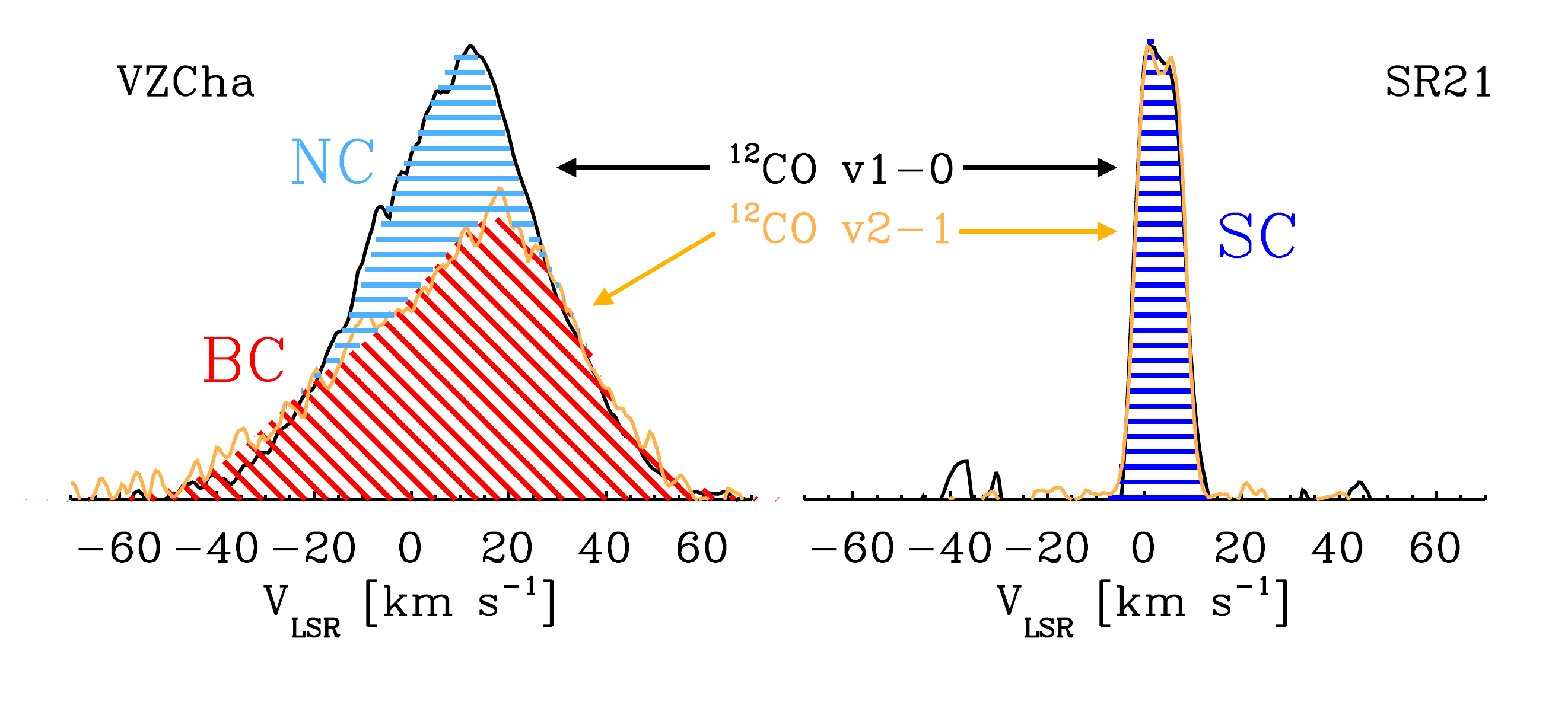} 
\includegraphics[width=0.5\textwidth]{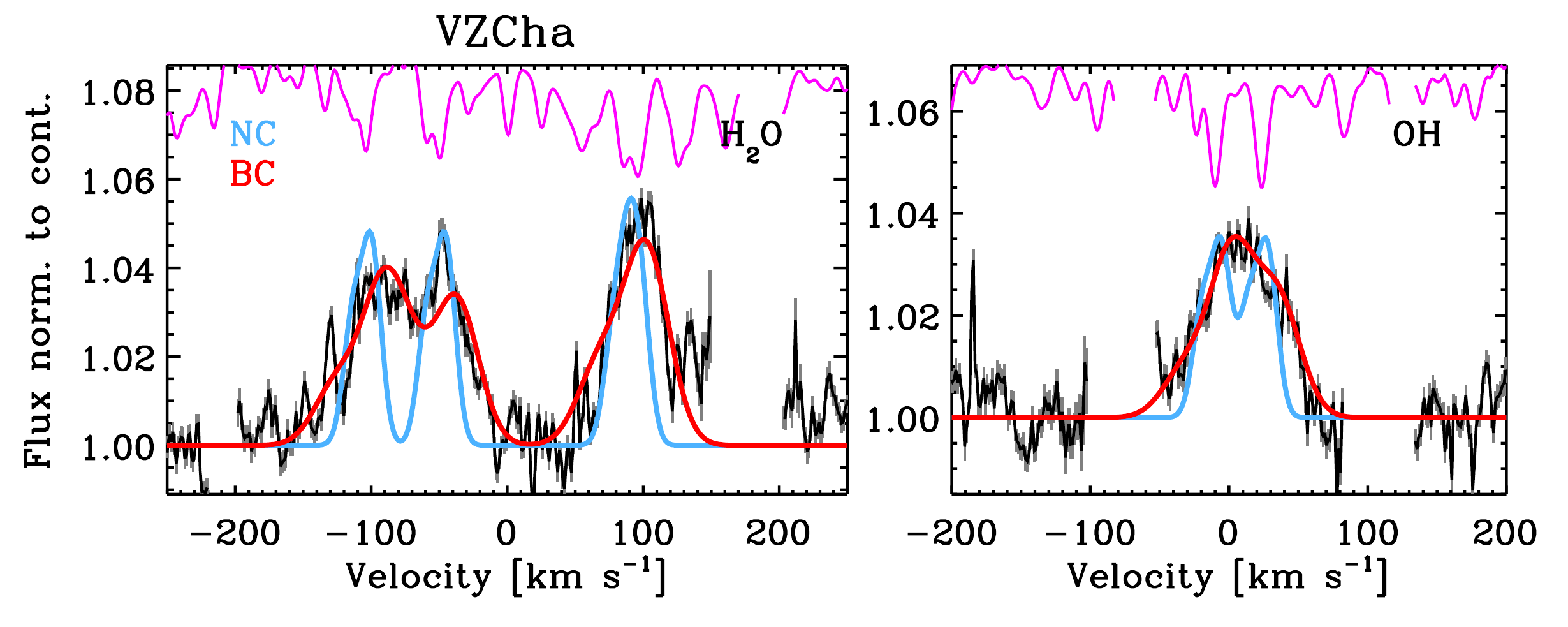} 
\includegraphics[width=0.5\textwidth]{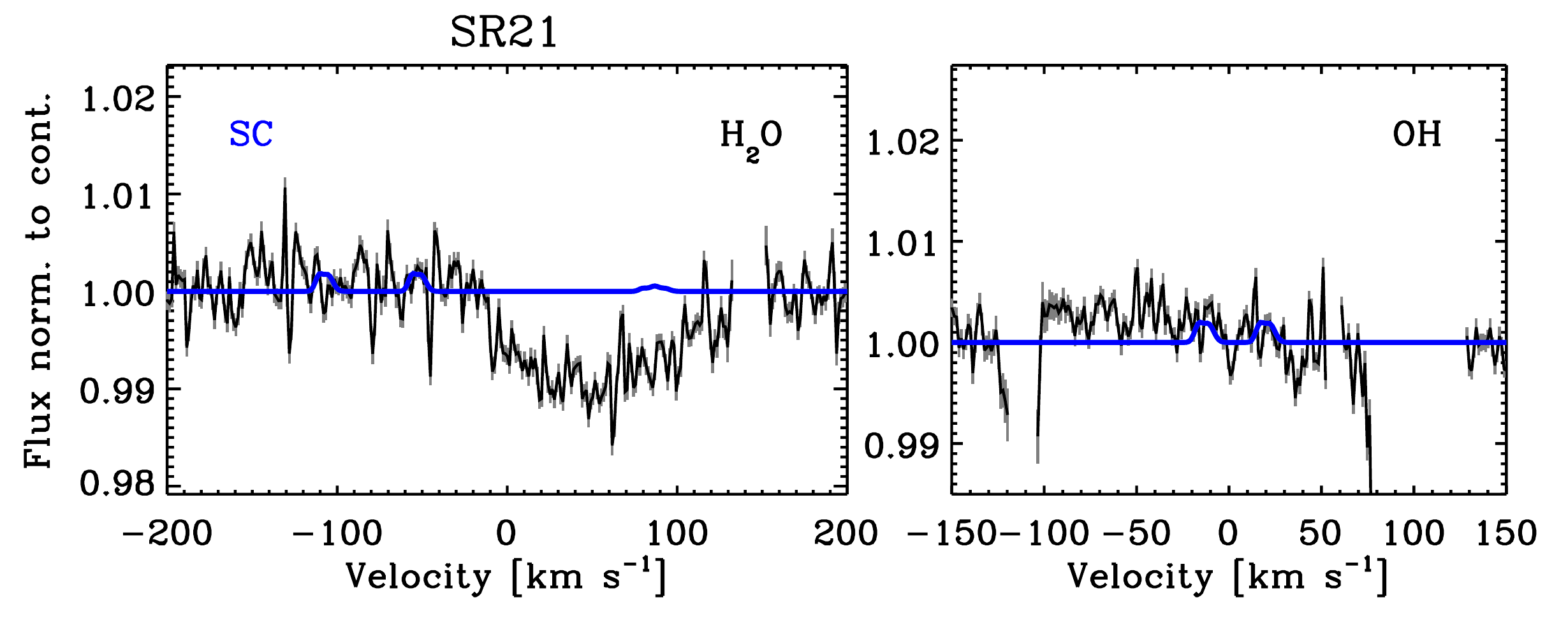} 
\caption{\textit{Top:} Spectral decomposition of CO lines profiles, as presented in \citet{bp15}. The spectrum of VZCha is an example of a ``double-component disk", where CO rovibrational emission at 4.7\,$\mu$m has two velocity components (the broad component BC marked in red, the narrow component NC marked in cyan). The spectrum of SR21 shows an example of a ``single-component disk" disk (SC), where only a narrow CO component is found. The gallery of CO line profiles is shown in \citet{bp15}. \textit{Center \& Bottom:} Examples of fits to H$_{2}$O and OH line profiles at 2.93\,$\mu$m by using the CO emission components (fits to other disks are shown in the Appendix). The spectrum is shown in black with grey error-bars, centered in velocity on (for OH) or in between (for H$_{2}$O) the lines for visualization. The best-fit line profiles obtained by combining CO lines are shown in cyan (using the NC), red (using the BC), and in blue for the single-component disk. The photospheric template, where used to remove the stellar photosphere, is shown at the top in magenta for reference.}
\label{fig: mol comp}
\end{figure}

The $v=1-0$ and $v=2-1$ stacked line profiles were then used to identify and separate multiple emission components, where present. The $v=2-1$ lines, with higher $E_u$ (Table \ref{tab:lines}), are mostly excited in the hotter innermost disk region, while the $v=1-0$, with lower $E_u$, often show additional flux contributions from colder gas at larger disk radii. This has been empirically demonstrated through the systematic decomposition of CO lines, where a broad (inner) component ``BC" is identified by the $v=2-1$ lines that match the wings of $v=1-0$ lines, and a narrow (outer) component ``NC" is obtained from the residuals by subtraction of the $v=2-1$ profile (which typically includes only BC) from the $v=1-0$ profile (which includes both BC and NC). An example of this procedure is included in Figure \ref{fig: mol comp}. In several disks, the two velocity components are clearly seen by visual inspection of the $v=1-0$ lines, which show a narrower peak on top of broader wings \citep[][]{hercz11,bast11,bp15}. Those disks that, following this analysis, have been found to have only one CO component (i.e., those where the $v=1-0$ and $v=2-1$ stacked line profiles match), have been identified as ``single-component" disks (``SC", e.g. SR21 in Figure \ref{fig: mol comp}).

 . 

 .

\begin{deluxetable}{l c c c c c}
\tabletypesize{\small}
\tablewidth{0pt}
\tablecaption{\label{tab:molec summary} Summary of 2.9--4.7\,$\mu$m emission detections and properties.}
\tablehead{ \colhead{Component} & \colhead{FWHM} & \colhead{R$_{\rm{co}}$}  & \colhead{CO} & \colhead{H$_{2}$O} & \colhead{OH} \\
  & [km\,s$^{-1}$] &   [au] & \multicolumn{3}{c}{(detection fractions)}  }
\tablecolumns{6}
\startdata
Broad (BC)  & 50--200  & 0.04--0.3 & 25/25 & 15/20 & 17/20 \\
Narrow (NC)  & 10--70   & 0.2--3 & 23/25 & 3/20 & 3/20 \\
Single (SC) & 6--70  & 0.3--20 & 28/28 & 1/10 & 4/10
\enddata
\tablecomments{The definition of components in the first column is based on the analysis of 4.7\,$\mu$m CO emission in \citet{bp15}, as summarized in Section \ref{sec:ana1}. Details on individual sources are given in the Appendix.}
\end{deluxetable} 

\begin{figure}
\centering
\includegraphics[width=0.4\textwidth]{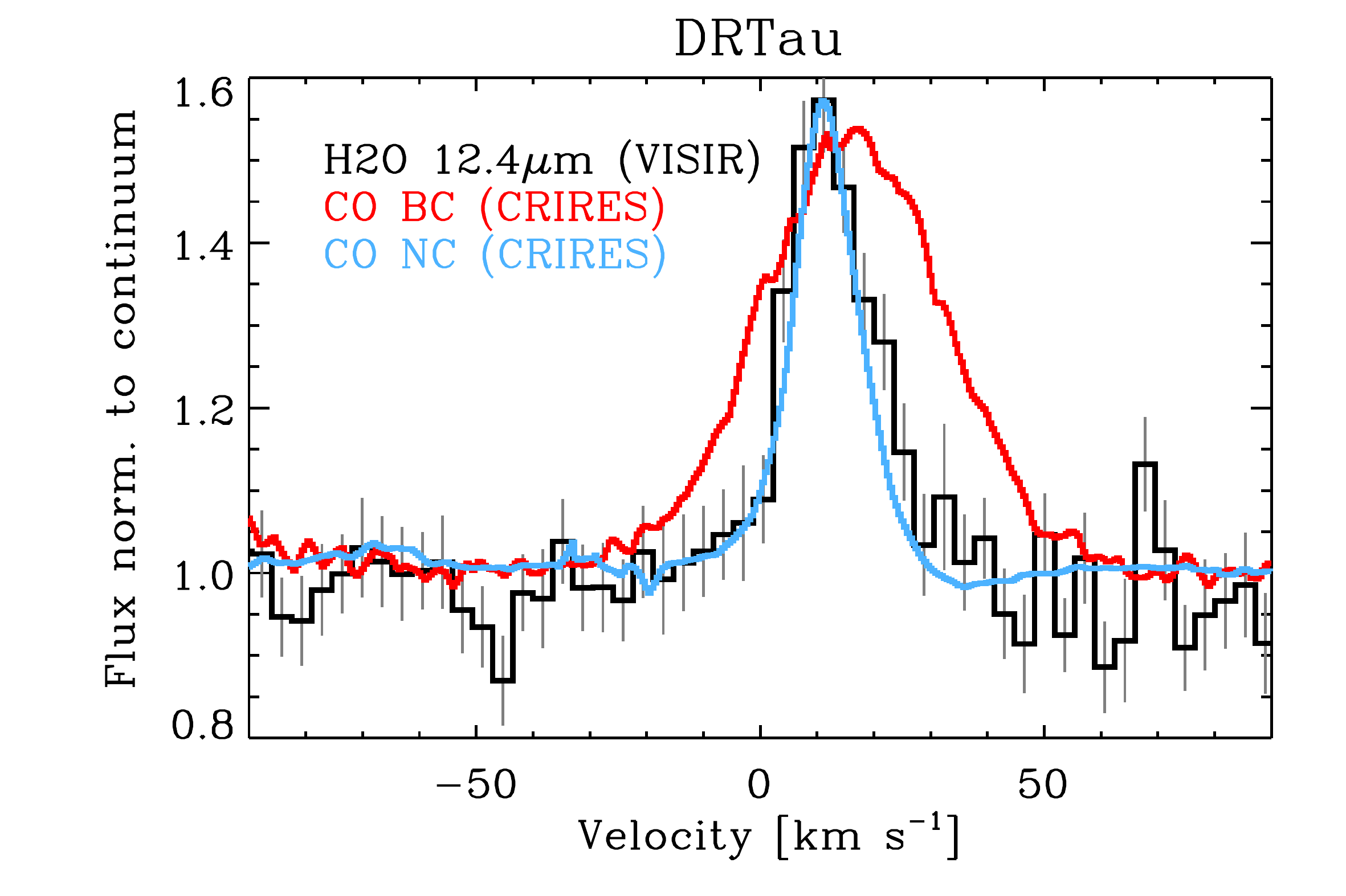} 
\caption{H$_{2}$O line at 12.396\,$\mu$m as observed in DRTau with VLT-VISIR at a resolution of $\sim15$\,km\,s$^{-1}$ \citep{banz14}. The rovibrational CO line profiles of the broad and narrow components are shown in red and cyan, scaled to the VISIR line peak for comparison.}
\label{fig: DRTau_VISIRcomp}
\end{figure}

\begin{figure*}[ht]
\includegraphics[width=1\textwidth]{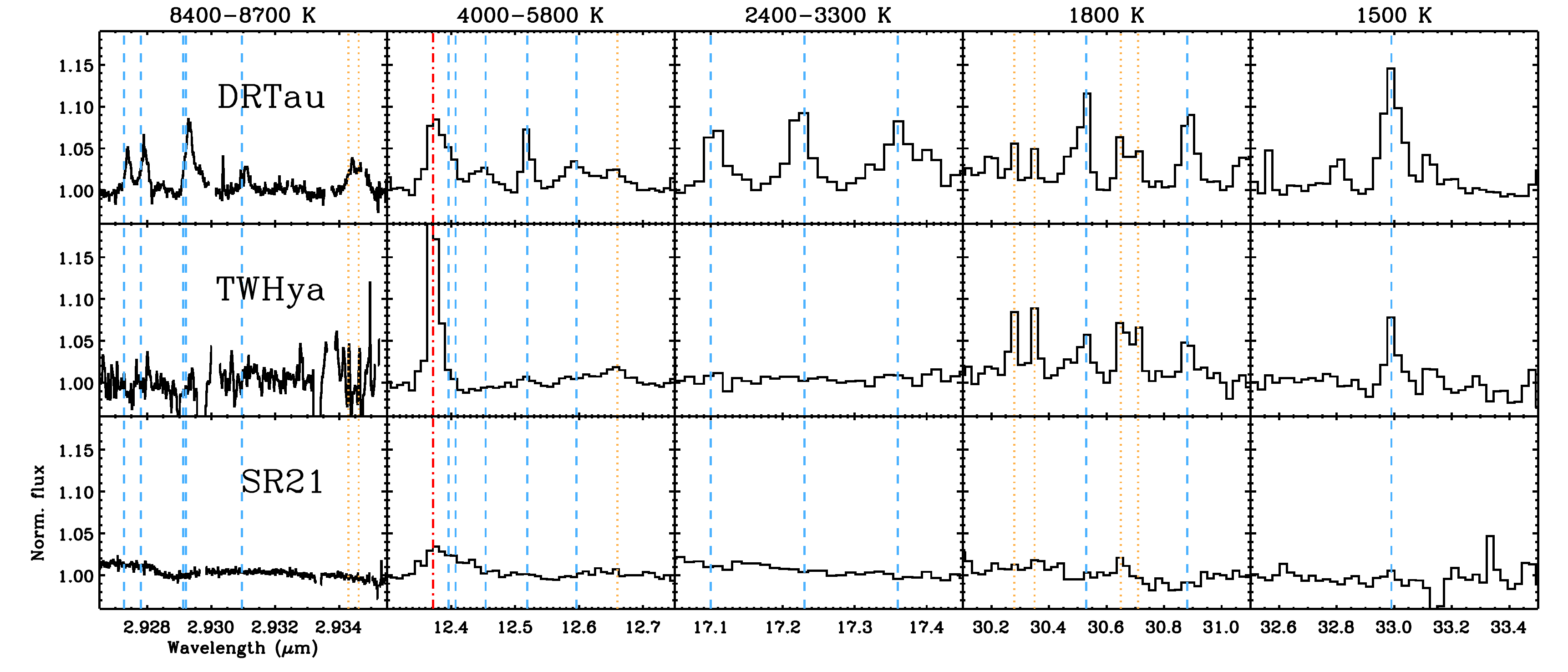} 
\caption{Representative samples of water vapor emission from 2.9 to 35\,$\mu$m, combining observations from VLT-CRIRES and \textit{Spitzer}-IRS. Individual water transitions are marked with blue dashed lines, while OH transitions are marked with orange dotted lines. The HI 7-6 line is marked in red dashed-dotted line to avoid confusion with water emission. Increasing wavelengths probe colder water transitions (the range in upper level energy in each spectral sample is displayed at the top of the figure; see also Table \ref{tab:lines}).}
\label{fig: multiwl_lines}
\end{figure*}

\begin{figure}[ht]
\includegraphics[width=0.5\textwidth]{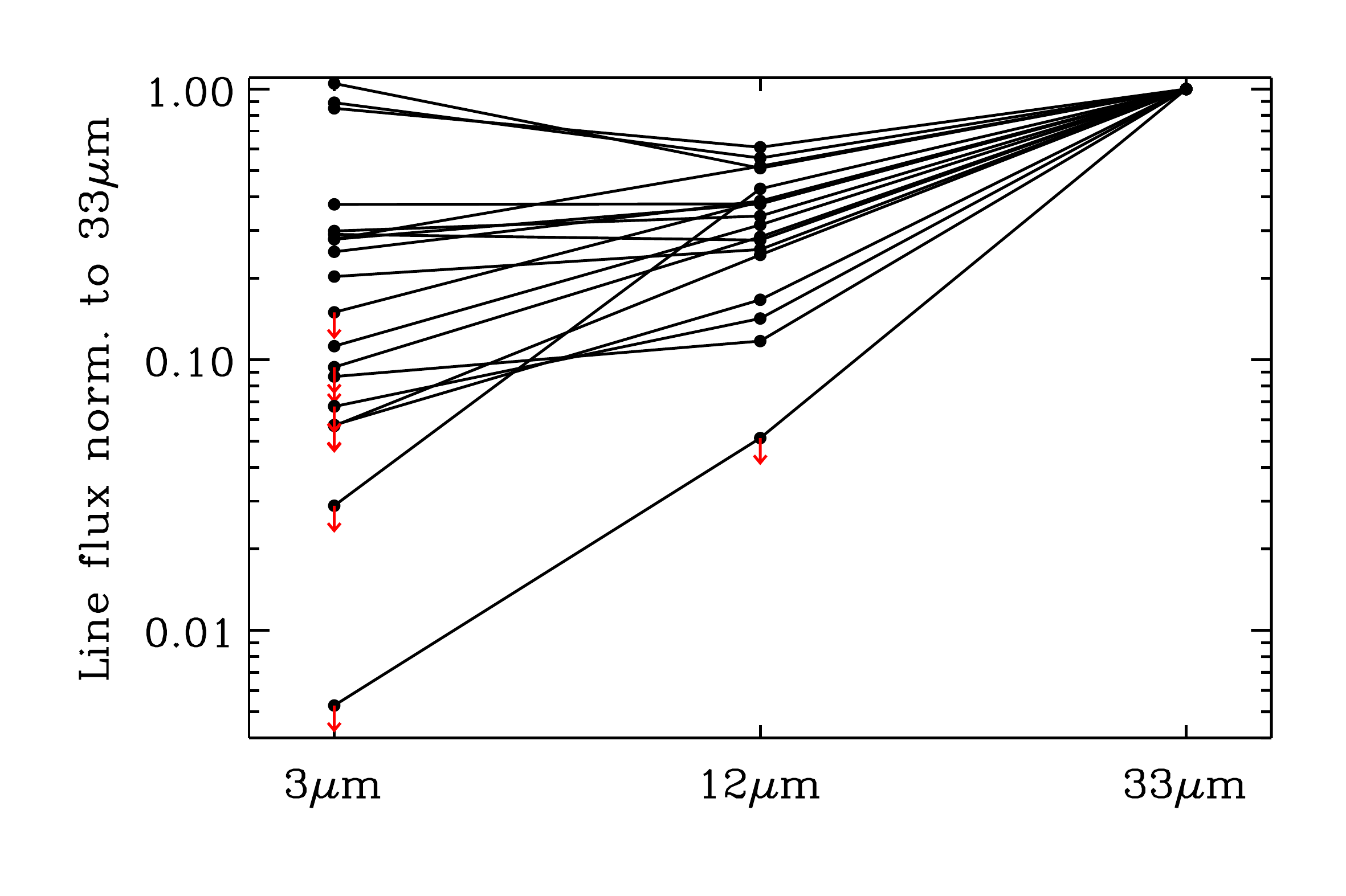} 
\caption{Measured H$_{2}$O line fluxes in disks that have both CRIRES and \textit{Spitzer} spectra available. Line fluxes from each disk are linked with a line, and are normalized to their 33\,$\mu$m line flux (the lowest-energy line that we include in the analysis, see Table \ref{tab:lines}). Upper limits are shown with red arrows. The wavelength dependence of the decrease in the line fluxes is found both in the line-to-continuum strength, illustrated for three disks in Figure \ref{fig: multiwl_lines}, and in the integrated line fluxes, as shown here.}
\label{fig: multiwl_lines2}
\end{figure}

\subsection{Characterization of H$_{2}$O and OH line profiles} \label{sec:ana4}
We use the stacked CO line profiles to reproduce the observed H$_{2}$O and OH emission features, to test if their shapes and width are equal to those of CO implying that they emit from a similar disk region. We adopt the CO line profile at each rest wavelength of H$_{2}$O and OH transitions, combine them into a single feature if the lines are broad and overlap (this typically happens for the broad component), and then vary the peak flux of the combined line profile to fit the 2.9\,$\mu$m spectrum (see Figure \ref{fig: mol comp} and the Appendix).

H$_{2}$O and OH lines at 2.9\,$\mu$m are typically detected in double-component disks only (i.e. when the BC is not detected in CO, H$_{2}$O and OH at 2.9\,$\mu$m are typically not detected either). Where detected, H$_{2}$O and OH lines at 2.9\,$\mu$m are dominated by a broad component that matches the CO BC profile (Figure \ref{fig: mol comp} and Table \ref{tab:molec summary}); we therefore conclude that these lines are emitted at similar disk radii. This was first recognized in the 3--5\,$\mu$m spectrum of EX Lupi by \citet{banz15}, and applies to the majority of disks in the sample studied here (see Table \ref{tab:molec summary}). 
Disks that lack the CO broad component (the ``single-component" disks) do not show H$_{2}$O emission at 2.9\,$\mu$m either.  This is a first hint that disks with evidence for an inner CO-depleted region are also depleted in water, which will be further detailed and discussed in the next Sections. 
In these disks, OH is detected in a few cases. This finding agrees with previous work that detected OH but not H$_{2}$O in some disks around intermediate-mass stars \citep{mand08,fed11}; disks in this stellar mass bin are in fact found to typically have only a single, narrow CO component \citep{bp15}.

While H$_{2}$O and OH emission at 2.9\,$\mu$m comes from similar disk radii as the CO broad component ($\sim 0.04-0.3$\,au), a survey of velocity-resolved H$_{2}$O and OH emission at $>$\,2.9\,$\mu$m is still missing, and is essential to determine what disk radii are probed at longer wavelengths. H$_{2}$O emission at 12.4\,$\mu$m may come from a region at larger disk radii that matches the emitting region of the narrow CO component, as shown in Figure \ref{fig: DRTau_VISIRcomp} for one disk. In fact, the 12.4\,$\mu$m H$_{2}$O lines have lower energy than the 2.93\,$\mu$m lines (Table \ref{tab:lines}), and may therefore match the colder/outer component of CO emission rather than the hotter/inner one.

\section{LINE FLUXES ACROSS DISK RADII} \label{sec:rad_trends}
By combining H$_{2}$O and OH spectra between 2.9 and 35\,$\mu$m, we find a wavelength dependence in the decrease of line fluxes. When the emission is detected at 2.9\,$\mu$m, it is always detected at longer wavelengths up to 33\,$\mu$m. On the contrary, when lines at 30--33\,$\mu$m are detected, those at 12 or 2.9\,$\mu$m may not necessarily be (see Figures \ref{fig: multiwl_lines} and \ref{fig: multiwl_lines2}, and Table \ref{tab: fluxes}). The decrease of water emission is found as a sequential intrinsic decrease of the observed line fluxes and of the line-to-continuum strength, where water line fluxes at short wavelengths decrease when water line fluxes at longer wavelengths are still strong. Since the emission probes hotter to colder gas from short to long wavelengths \citep[following a global trend in $E_{u}$ energies that decrease with increasing wavelength -- see Table \ref{tab:lines}, Figure \ref{fig: multiwl_lines}, and][]{blevins}, this wavelength dependence of the line flux decrease may reflect the physical removal of molecular gas in the disk from the inside-out \citep[as found in the disk of TW\,Hya,][]{zhang13}. 

In Section \ref{sec:ana}, we reported that H$_{2}$O and OH emission at 2.9\,$\mu$m comes from a disk region that is consistent with that of the broad CO emission at 4.7\,$\mu$m, so that we can adopt R$_{\rm{co}}$ as descriptive of H$_{2}$O and OH emission too. The fact that the high-energy H$_{2}$O lines at 2.9\,$\mu$m disappear when the broad component of CO emission disappears from the observed spectra, suggests that the depletion of the hotter/inner CO gas and the depletion of the hotter/inner H$_{2}$O gas proceed together. We now combine measurements from \citet{bp15} and this work to test the dependence of H$_{2}$O and OH line fluxes on R$_{\rm{co}}$ as a tracer of molecular gaps in inner disks.

\begin{figure*}[ht]
\includegraphics[width=1\textwidth]{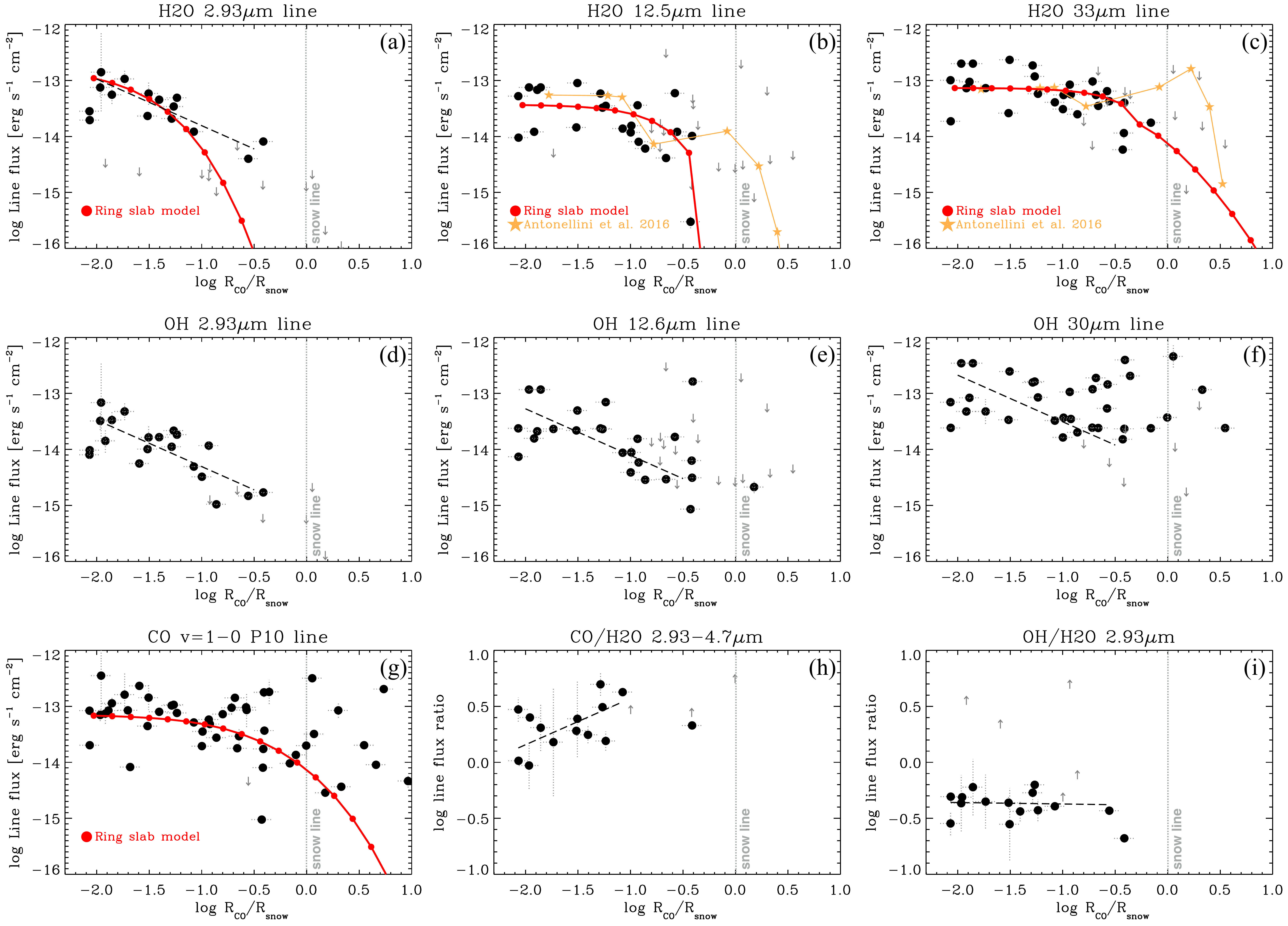} 
\caption{Measured line fluxes versus R$_{\rm{co}}$/R$_{\rm{snow}}$. Panel (a) shows the H$_{2}$O line fluxes at 2.93\,$\mu$m; a linear fit is shown in black dashed line, and is reported in panels (d) to (f) for comparison (scaled to the OH line fluxes). The ring slab model described in the text is shown in red in this and other panels. Panels (b) and (c) show the H$_{2}$O line fluxes at 12.5 and 33\,$\mu$m; model results for increasing gap size by \citet{anto16} are shown in orange. Panels (d), (e), and (f) show OH line fluxes, together with the linear fit from the H$_{2}$O line fluxes at 2.93\,$\mu$m for comparison of the radial trend. Panel (g) shows the CO $v=1-0$ P10 line fluxes for the broad component in double-component disks and for the single-component disks. Panels (h) and (i) show line flux ratios as measured in the CRIRES dataset, with linear fits (the rightmost datapoint in each panel is excluded from the fits as possibly part of a different trend, see Section \ref{sec:disc_ratios}). All fluxes are normalized to a common distance of 140 pc.}
\label{fig: R_fl_corr}
\end{figure*}

\subsection{Relations between line fluxes and gap size}
R$_{\rm{co}}$ measures the location of CO emission in the disk by taking into account the stellar mass and disk inclination, which are both linked to the measured line FWHM through Kepler's law, as 
\begin{equation} \label{eqn: R_co}
\rm{R}_{co} = (2\, \rm sin \,\textit{i}\, /\, FWHM)^2 \, G\,M_{\star} \\
\end{equation}
where M$_{\star}$ is the stellar mass, $i$ the disk inclination, G the Gravitational constant, and we take the line velocity at the half width at half maximum (FWHM/2) as in \citet{bp15}.
As a reference comparison for R$_{\rm{co}}$, we take the midplane water snow line radius R$_{\rm{snow}}$, defined as:
\begin{equation} \label{eqn: R_snow}
\begin{aligned}
R_{\rm{snow}} = & \; 2.1 \; \rm{au} \, (M_{\star} / M_{\odot}) ^{1/3} \, (M_{acc} / 10^{-8}\,M_{\odot}yr^{-1}) ^{4/9}  \\ 
& (T_{\rm snow}/160 \rm \,K) ^{-10/9}
\end{aligned}
\end{equation}
where M$_{acc}$ the accretion rate, and $T_{snow}$ the water freeze-out temperature. Equation \ref{eqn: R_snow} is the midplane snow line radius for a viscously heated disk as derived by \citet{muld15}\footnote{We assume a gas-to-dust ratio of 100, an $\alpha$ viscosity parameter of 0.01, and a Rosseland mean dust opacity of 770 cm$^{2}$ g$^{-1}$ that normalize the extra terms in Eqn. \ref{eqn: R_snow} as formulated in \citet{muld15}.}, who used $T_{\rm snow} = 160$ K as in \citet{min11}. This formula assumes that the disk midplane heating by stellar irradiation is negligible, and this has been shown to be valid for disks around stars younger than $\sim3$\,Myr with mass up to $\sim2$\,M$_{\odot}$ and accretion rates between $10^{-9}$ and $10^{-6}$\,M$_{\odot}$\,yr$^{-1}$ \citep{kk08,min11,muld15}. Some disks in our sample exceed these limits in stellar mass and age, while some have inner dust gaps that may change the disk temperature structure. For simplicity we adopt the same equation for the entire sample.
By normalizing to R$_{\rm{snow}}$, we put R$_{\rm{co}}$ from individual disks in reference to a disk radius of equal temperature across the entire disk sample, as set by water ice evaporation in the disk midplane. R$_{\rm{co}}$/R$_{\rm{snow}}$ therefore enables a comparison of results from individual disks within a common reference framework.

Figure \ref{fig: R_fl_corr} shows that line fluxes of H$_{2}$O, OH, and CO decrease with increasing R$_{\rm{co}}$/R$_{\rm{snow}}$, i.e. as the inner gap radius increases and reaches the snow line radius\footnote{Snow line radii following \citet{muld15} are on average $\lesssim1.5$ times larger than those formulated by \citet{kk08} at stellar masses between 1 and 3\,M$_{\odot}$; while the R$_{\rm{co}}$/R$_{\rm{snow}}$ values slightly shift depending on the formula used for the snow line radius, the global trends presented here are preserved.}.
H$_{2}$O line fluxes at 2.9\,$\mu$m show a decreasing linear correlation at radii $-2\lesssim$ log\,R$_{\rm{co}}$/R$_{\rm{snow}} \lesssim -0.5$, beyond which they fall under the detection limit ($\sim 2 \times 10^{-15}$ erg\,s$^{-1}$\,cm$^{-2}$ distance/140\,pc). A linear fit to the detected lines provides \citep[using the Bayesian method by][]{kelly}:
\begin{equation}
\rm{Flux_{140}(H_2O)} = -14.6 \pm 0.2 + (R_{\rm{co}}/R_{\rm{snow}})^{-0.8 \pm 0.2}
\end{equation}
where Flux$_{140}$ is the measured line flux normalized to a distance of 140 pc.
At 12.5 and 33\,$\mu$m, H$_{2}$O lines are detected out to log\,R$_{\rm{co}}$/R$_{\rm{snow}} \sim -0.2$, and decrease only at log\,R$_{\rm{co}}$/R$_{\rm{snow}} \gtrsim -1$. OH emission follows similar trends, but OH lines are tentatively detected up to larger disk radii (especially the cold lines at 30\,$\mu$m). 
Variability in accretion luminosity may contribute to the order-of-magnitude scatter in line fluxes along these trends; however, a factor up to $\approx10$ change in line fluxes has been observed so far only in the extreme accretion outburst of EX Lupi, and not during the typically lower variability of other young stars included in this sample \citep{banz12,banz14,banz15}.
Similarly to H$_{2}$O and OH, the CO $v=1-0$ line flux decreases with R$_{\rm{co}}$/R$_{\rm{snow}}$, but it is detected in disks well beyond the snow line radius.
We interpret this as evidence that what regulates the observed H$_{2}$O and OH emission is the molecular gas left within the snow line radius, or in other words that H$_{2}$O and OH emission disappears as inner disks are depleted from the inside out to the snow line (i.e. as R$_{\rm{co}}$ grows larger and approaches R$_{\rm{snow}}$), beyond which the reservoir of water vapor is strongly limited due to freeze-out at low temperatures. CO gas is instead still detected  beyond the snow line because it freezes out at lower temperatures ($\sim20$\,K), found at larger disk radii.

As a basic test of this inside-out depletion scenario, we use a simple disk model composed of concentric rings of increasing radius, where each ring is a slab of gas described by one temperature and one column density as in \citet{banz12}. We adopt power-law profiles to set how the temperature and column density of each ring decrease with increasing disk radius, as $T$(R) = $T_{0}$ (R/R$_{0}$)$^{-q}$ and $N$(R) = $N_{0}$ (R/R$_{0}$)$^{-p}$. We calibrate the model using the observed CO line fluxes, by adopting model parameter values from the radial temperature profile measured by \citet{bp15} in rovibrational CO emission, as $T_{0} = 1300$\,K, $q=0.3$, $N_{0} = 10^{20}$\,cm$^{-2}$, and $p=1$. In Figure \ref{fig: R_fl_corr} (panel g) we show how the integrated CO line flux from the model decreases with increasing gap size, by removing one ring at a time from R$_0=0.03$\,au to R$_{\rm out}=100$\,au to mimic an inside-out removal of hotter to colder gas (the radial grid of disk rings is shown by the dots). A radially decreasing power-law temperature profile reproduces the decrease in the CO line fluxes as a curve, where as the inner hole radius increases, colder gas emission beyond the hole produces weaker CO lines at 4.7\,$\mu$m. CO line fluxes that exceed the model at log\,R$_{\rm{co}}$/R$_{\rm{snow}} \gtrsim -0.5$ come from disks where the emission is pumped by UV radiation \citep{brit03,bp15}. 

The same ring slab model is then used for H$_{2}$O emission. Such a model produces line fluxes stronger than observed at 12.5 and 33\,$\mu$m. We therefore include a step function for the column density, to simulate the presence of a radius in the surface of the disk beyond which water vapor is reduced by freeze-out or by chemistry in cold regions \citep[e.g.][]{vdish13}. We keep $p=1$ as used for CO at disk radii smaller than $\approx2.5$\,au, and adopt $p=2.5$ to produce a steeper decrease in water column density at larger disk radii (where the gas temperature decreases to $\lesssim300$\,K). This model, shown in panels a, b, and c of Figure \ref{fig: R_fl_corr}, approximately reproduces both the absolute fluxes and the wavelength-dependent radial decrease in water emission at 2.93--33\,$\mu$m. A step function in the column density produces a decrease/cutoff in the 12.5 and 33\,$\mu$m line fluxes at log\,R$_{\rm{co}}$/R$_{\rm{snow}} \sim -0.5$, which is consistent with the observed water line fluxes in the \textit{Spitzer} spectra.

\begin{figure*}[ht]
\includegraphics[width=1\textwidth]{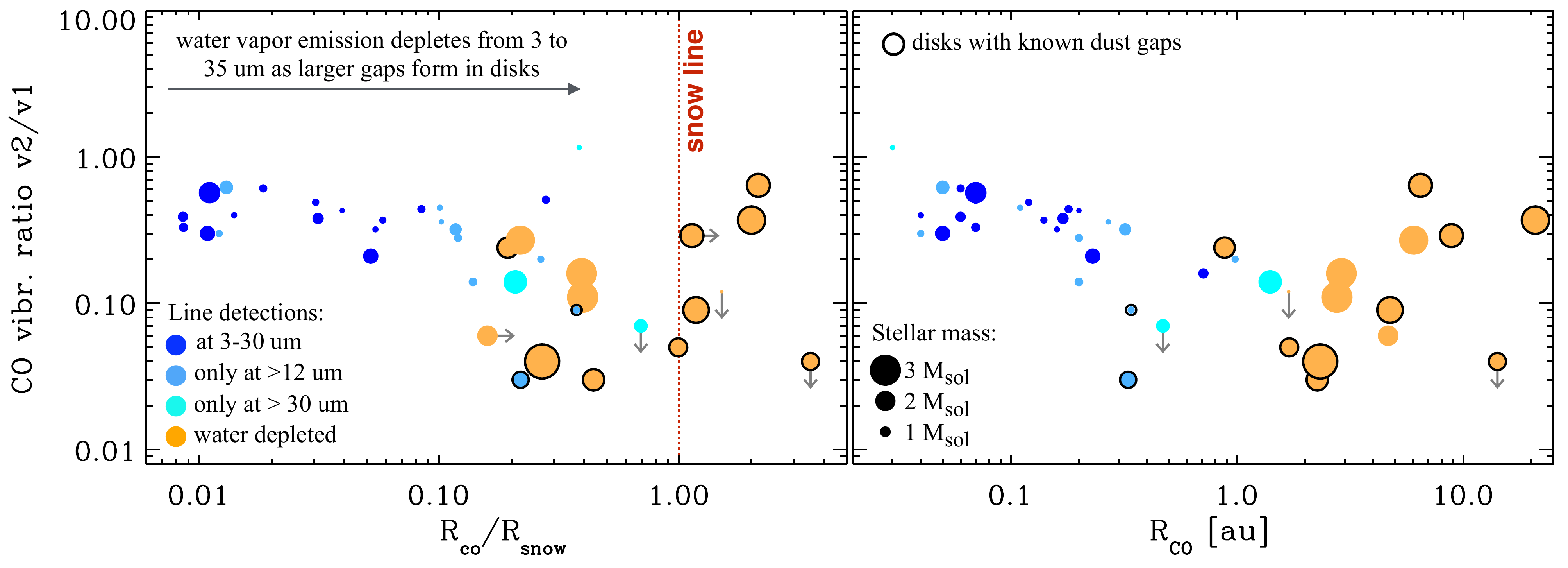} 
\includegraphics[width=1\textwidth]{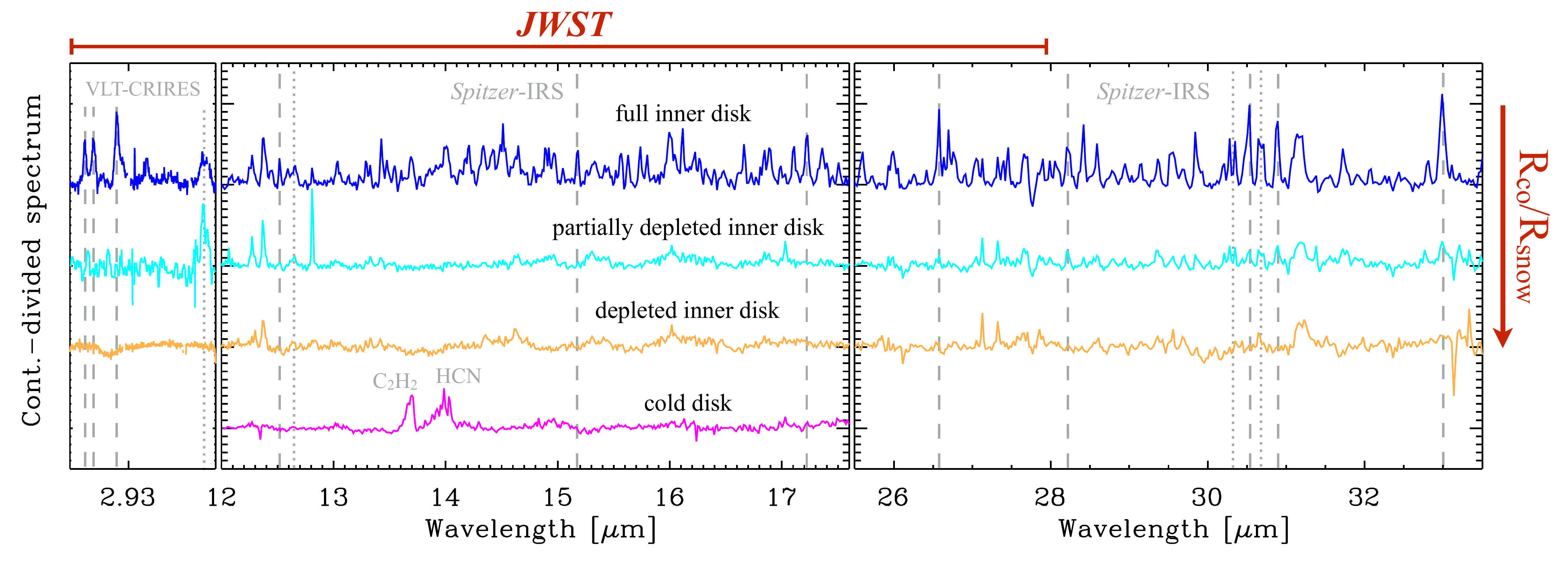} 
\caption{Inside-out water vapor depletion from 2.9--33\,$\mu$m spectra of protoplanetary disks. \textit{Top:} The datapoints show the CO sequence in the temperature-radius diagram of CO emission at 4.7\,$\mu$m from \citet{bp15},  normalized to the midplane snow line radius given by Eqn. \ref{eqn: R_snow} in the left plot. The data are color-coded according to water detections (line fluxes are included in Table \ref{tab: fluxes}). The size of datapoints is proportional to the stellar mass as shown, and disks where dust gaps have been found in previous work are marked with a black circle. \textit{Bottom:} Three steps along the CO and H$_{2}$O depletion sequence are visualized using averaged \textit{Spitzer} spectra: \textit{i)} water emission is strong at 2.9--33\,$\mu$m in disks when R$_{\rm{co}}$/R$_{\rm{snow}} \lesssim 0.1$, \textit{ii)} water emission is weaker but still detected at 25--33\,$\mu$m in disks when $0.1 \lesssim$\,R$_{\rm{co}}$/R$_{\rm{snow}} \lesssim 1$, and \textit{iii)} water emission is undetected at all infrared wavelengths 2.9--33\,$\mu$m when R$_{\rm{co}}$/R$_{\rm{snow}} \gtrsim 1$. The averaged \textit{Spitzer} spectrum of disks around low-mass stars (M$_{\star} < 0.2$ M$_{\odot}$) is shown at the bottom for comparison. Some prominent water and OH emission lines are marked respectively with dashed and dotted lines for reference.}
\label{fig: TR_water}
\end{figure*}

The purpose of the ring slab model is to test in the simplest terms if an inside-out removal of disk gas can explain the observed line flux decrease within the snow line. Clearly, this simple model ignores several important factors that should be considered, such as a realistic physical disk structure, disk chemistry, and their adaptation to the formation and development of disk gaps. However, the model seems to have the key ingredients needed to qualitatively reproduce the observed data, by including a radially decreasing temperature and the inside-out removal of molecular gas. 
Detailed modeling of the water spectrum and of its dependence on disk regions of water vapor depletion at and beyond the snow line has been presented by \citet[][]{meij09,zhang13,blevins,notsu16}. 
Recent explorations by \citet{anto16} for a disk around an intermediate-mass star found a wavelength-dependent decrease in the \textit{Spitzer} water line fluxes for a grid of increasing disk hole sizes, in qualitative agreement with what we measure in the data out to the snow line radius (Figure \ref{fig: R_fl_corr}, panels b and c). These models produce an excess flux when the gap reaches the snow line radius, contrarily to the observations (see Section \ref{sec:disc_ratios}).

The wavelength-dependent decrease in water vapor emission between 2.9 and 33\,$\mu$m is summarized in Figure \ref{fig: TR_water}, where we show the temperature-radius (T-R) diagram originally defined for CO rovibrational emission by \citet{bp15} and we color-code the individual disks by their H$_{2}$O detections. For R$_{\rm{co}}$/R$_{\rm{snow}}$ between 0.1 and 1, all disks have lost their detectable H$_{2}$O emission at 2.9\,$\mu$m (their line fluxes are $\lesssim 10^{-15}$ erg\,s$^{-1}$\,cm$^{-2}$ distance/140\,pc), while some still retain detectable emission at 12 or 33\,$\mu$m. Beyond the snow line (R$_{\rm{co}}$/R$_{\rm{snow}} = 1$), H$_{2}$O emission between 2.9 and 35\,$\mu$m is not detected in any disk. As a diagnostic of R$_{\rm{co}}$/R$_{\rm{snow}}$, we illustrate in Figure \ref{fig: TR_water} the wavelength-dependent H$_{2}$O emission decrease shown earlier in Figure \ref{fig: multiwl_lines}, here in reference to the increasing disk gap size.

\subsection{Molecular line flux ratios in inner disks}
Figure \ref{fig: R_fl_corr} (panels h and i) shows molecular line flux ratios as measured in the CRIRES dataset (2.9--4.7\,$\mu$m). We find a constant OH/H$_{2}$O ratio of $\sim0.4$ in the range $-2\lesssim$ log\,R$_{\rm{co}}$/R$_{\rm{snow}} \lesssim -0.5$.
As for the CO/H$_{2}$O ratio, we find an increasing trend with increasing radius in the range $-2\lesssim$ log\,R$_{\rm{co}}$/R$_{\rm{snow}} \lesssim -1$. This trend is due to a decreasing H$_{2}$O line flux, while the CO line flux is less steeply decreasing over these disk radii (Figure \ref{fig: R_fl_corr}, panels a and g).
By fitting the measured line ratios, we find the following linear relation \citep[using][]{kelly}:
\begin{equation}
\rm{Flux(CO)/Flux(H_2O)} = 1.0 \pm 0.4 + (R_{\rm{co}}/R_{\rm{snow}})^{0.4 \pm 0.2}
\end{equation}
These trends are further analyzed and discussed in Section \ref{sec:disc_ratios}.

\section{Discussion} \label{sec:disc}

\begin{figure*}
\includegraphics[width=1\textwidth]{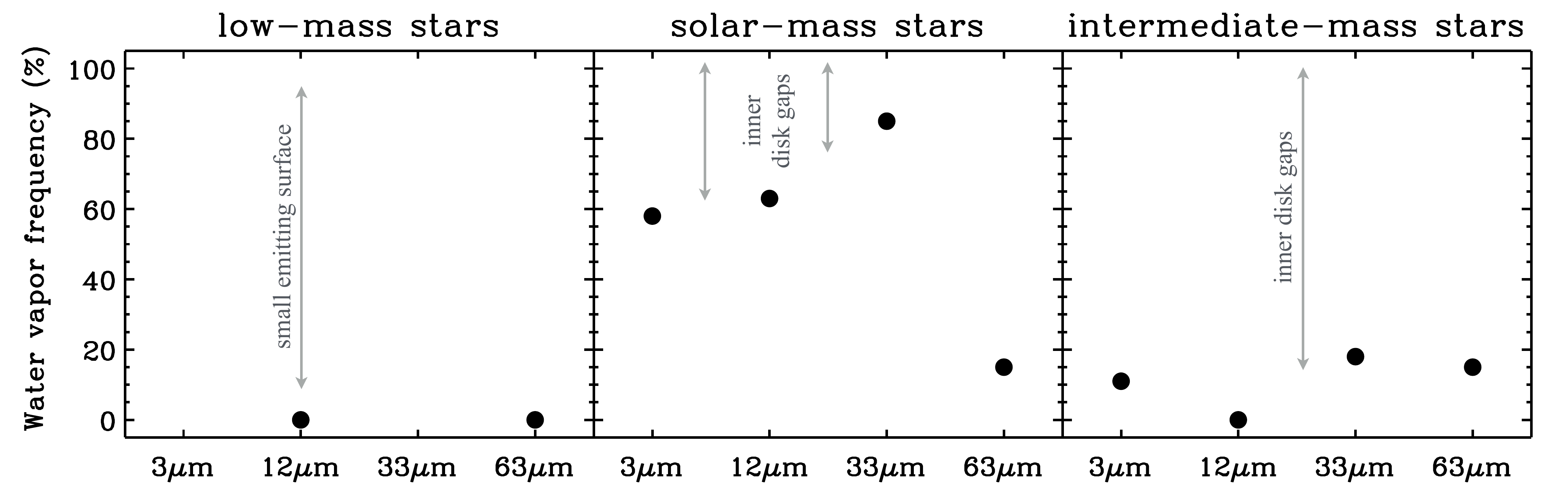} 
\caption{Detection rates of infrared water vapor emission for three stellar mass bins: low-mass stars and Brown Dwarfs ($<0.2$\,M$_{\odot}$), solar-mass stars (0.2--1.5 M$_{\odot}$), and intermediate mass stars (1.5--4 M$_{\odot}$). Results from VLT-CRIRES, \textit{Spitzer}-IRS, and \textit{Herschel}-PACS surveys as reported in Table \ref{tab:water_surveys} are shown as black dots.}
\label{fig: Spitzer_det_rates}
\end{figure*}

\subsection{Water vapor and the depletion of inner disks} \label{sec:disc_water}
Water emission from protoplanetary disks has been detected by the \textit{Spitzer}-IRS in over 60 disks (see Section \ref{sec: water_surveys}). This sample revealed two major trends: 1) water line fluxes increase with increasing stellar luminosity \citep[and mass, see Figures \ref{fig: Spitzer_spt} and \ref{fig: Lum_trends}, and][]{sal11}, and, in contrast with the first trend, 2) the water vapor frequency is much higher in disks around stars of $0.2 <$ M$_{\star} < 1.5$\,M$_{\odot}$ (with a frequency of 60--80\%) than in disks around stars of M$_{\star} > 1.5$\,M$_{\odot}$ \citep[with a frequency of $<20$\%; see Table \ref{tab:water_surveys}, Figure \ref{fig: Spitzer_det_rates}, and][]{pont10a}. Recent chemical modeling of the molecular content of inner disks by \citet{walsh15}, while able to reproduce the luminosity trend, was still unable to reproduce the absence of water emission in disks around intermediate-mass stars. 
We have demonstrated above that the strength of the infrared water spectrum is linked to an inside-out depletion scenario in disks, by finding: 1) a similar emitting region and detection frequency for the high-energy H$_{2}$O lines at 2.9\,$\mu$m and the innermost CO gas (the broad component), 2) a wavelength-dependent decrease of water line fluxes, where hotter emission (as probed at 2.9\,$\mu$m) decreases before colder emission (as probed at 33\,$\mu$m), and 3) a correlation between the decrease of water line fluxes at 2.9--33\,$\mu$m and the increase in size of an inner molecular hole measured from CO emission. 

One major implication of this analysis is that the low frequency of water vapor detections in disks around M$_{\star} > 1.5$\,M$_{\odot}$ is found to be linked to inner holes that have depleted the molecular content of their disks out to close to or beyond the snow line, contrary to most disks around solar-mass stars. Other independent techniques find signatures of dust inner holes in a growing number of disks around intermediate-mass stars \citep[e.g.][]{maask13,menu15}. 
With the analysis of disks that have CO emission observed at 4.7\,$\mu$m and H$_{2}$O at 2.9\,$\mu$m and/or 10--33\,$\mu$m (Table \ref{tab: fluxes}), we also find that the depletion scenario is not limited to the known transitional disks or to disks around intermediate-mass. As shown in Figure \ref{fig: TR_water}, several disks around solar-mass stars show at least some degree of water-depletion (those that we find to have $0.1 \lesssim$\,R$_{\rm{co}}$/R$_{\rm{snow}} \lesssim 1$). By interpreting results from \textit{Spitzer} and VLT-CRIRES surveys in the context of this analysis (Figure \ref{fig: Spitzer_det_rates}), we could conclude that $\approx$\,20--40\% of disks around solar-mass stars may have molecular gaps in inner disks, while this fraction may be $\approx$\,80\% for disks around intermediate-stars. 

The analysis of velocity-resolved molecular line spectra suggests a path of disk evolution where gas-phase CO and water are depleted from the planet-forming region in an inside-out fashion, a path that may be shared by disks around stellar masses of $0.3\lesssim$\,M$_{\star}$/M$_{\odot}$\,$\lesssim 3.5$ (as probed in this sample). By detecting H$_{2}$O emission only at long wavelengths in SR\,9, IM\,Lup, HD\,35929, EC\,82, TW\,Hya, DoAr\,44 and HD\,163296 \citep[the last three of this list have been previously studied by][]{zhang13,sal15,fed12}, we propose that these disks may be in an intermediate phase of depletion ($0.1 \lesssim$\,R$_{\rm{co}}$/R$_{\rm{snow}} \lesssim 1$) and still retain some water vapor in their planet-forming region. Disks that have 2.9--33\,$\mu$m water line fluxes lower than $\approx 5 \times 10^{-15}$\,erg\,s$^{-1}$\,cm$^{-2}$ (distance/140\,pc), have R$_{\rm{co}}$/R$_{\rm{snow}} \gtrsim 0.5$ and have already depleted their molecular gas in the planet-forming region within the snow line. This line flux limit may be used to identify molecular holes/gaps in inner disks when only the line flux, and not the line profile or a direct image, can be obtained.

In Figure \ref{fig: TR_water} we include for comparison also the case of disks around low-mass stars (M$_{\star} < 0.2$ M$_{\odot}$), observed with Spitzer at 10--19\,$\mu$m \citep{pasc13}, divided by 10 to re-scale the strong emission observed from organic molecules (HCN and C$_2$H$_2$). Rovibrational CO emission at 4.7\,$\mu$m has not been obtained for any of these disks so far, so we cannot include them in the T-R diagram. However, their spectra at \textit{Spitzer} wavelengths could be distinguished from that of disks with large gaps by considering the emission from organics. \citet{walsh15} suggested that disks around low-mass stars may be too cold to produce strong water emission. The luminosity dependence shown in Figure \ref{fig: Spitzer_spt} may provide an explanation for the non-detection of water in these disks with \textit{Spitzer} \citep[only a tentative detection was reported by][]{pasc13}: models produce fluxes at or below the measured sensitivity limits, due to a radial extent of the warm water-emitting layer that is $\sim10$ times smaller than in a disk around a solar-mass star \citep[$\sim0.1$ au, see][]{walsh15}. However, according to models, these disks would still show prominent organic emission lines from a disk atmosphere at 0.1--10\,au only mildly UV-irradiated as compared to higher-mass stars. Therefore, a continuous disk around low-mass stars would produce emission at least from organics at 13--15\,$\mu$m, while no molecular emission at 10--19\,$\mu$m may be the signature of inner disk depletion with R$_{\rm{co}}$/R$_{\rm{snow}} > 0.1$ also in disks around low-mass stars.

\begin{figure*}
\includegraphics[width=1\textwidth]{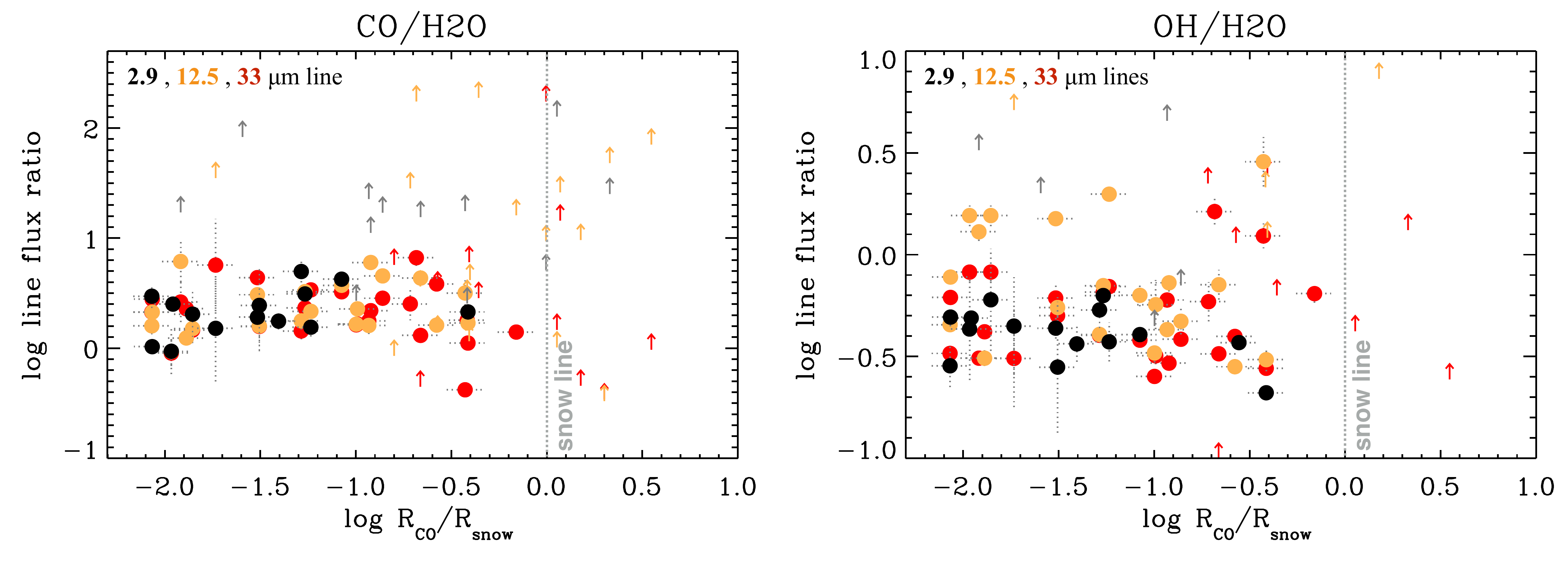} 
\caption{\textit{Left:} CO/H$_{2}$O line flux ratio as measured using the P10 line for CO and the 2.93\,$\mu$m (black), 12.5\,$\mu$m (orange), and 33\,$\mu$m (red) lines for H$_{2}$O. \textit{Right:} OH/H$_{2}$O ratio as measured at 2.9--33\,$\mu$m.}
\label{fig: mol_ratios}
\end{figure*}

\subsection{Origin of molecular holes/gaps in inner disks} \label{sec: gaps_origin}
Previous studies proposed the idea that UV radiation may be responsible for the depletion of water vapor in inner disk surfaces, through an efficient photo-dissociation of H$_{2}$O into OH \citep{mand08,pont10a,fed11,brit16}. Current thermochemical disk models, however, show that the balance at thermochemical equilibrium between the production and destruction of the water molecules produces large columns of water vapor even in highly-UV-irradiated disk surfaces \citep[e.g.][]{woitke09,walsh15}. In fact, models produce more water emission than in less UV-luminous stars (or less massive stars, see Figure \ref{fig: Spitzer_spt}), because H$_{2}$O and OH are able to self-shield to UV radiation at the column densities typically measured in inner disks \citep[$\gtrsim10^{17}$\,cm$^{-2}$,][]{bb09}. Therefore, it is still unclear whether UV photo-dissociation alone may be responsible for a global depletion of water vapor in inner disks. Further, CO and H$_{2}$O emission at 2.9--4.7\,$\mu$m recede at larger disk radii together (Section \ref{sec:ana}), and the CO molecule is able to self-shield to UV photo-dissociation at even lower columns \citep[$\approx10^{15}$\,cm$^{-2}$,][]{brud13}. As a consequence, a depletion of the molecular column densities to below the self-shielding limits may be required, before UV photo-dissociation may deplete any residual CO and H$_{2}$O gas.

The destruction of H$_{2}$O by UV radiation has been observationally tested by measuring if OH emission is consistent with prompt emission from excited OH, produced by photo-dissociation of H$_{2}$O \citep{cn14}. However, H$_{2}$O line fluxes increase (instead of decreasing) with increasing UV radiation, by considering the accretion luminosity as a proxy \citep[at least up to moderate accretion luminosities; see][and Figure \ref{fig: accretion_figure}]{banz12,banz14}. 
The measured $v2/v1$ vibrational ratios of CO lines, which are sensitive to UV pumping and to the UV irradiation that reaches the disk \citep[e.g.][]{thi13}, provide another way to test if and how H$_{2}$O line fluxes are linked to UV irradiation. The ratio $v2/v1$ decreases as water emission decreases from shorter to longer wavelengths, and $v2/v1$ is not anomalously high in several disks around intermediate-mass stars that show no water emission (Figure \ref{fig: TR_water}). 
However, the increase in $v2/v1$ with inner gap size for R$_{\rm co} \gtrsim 2$\,au demonstrates that enough UV radiation is present to pump the CO at large radial distances. In this regime of disk gaps around intermediate-mass stars, UV photo-dissociation of water could contribute more significantly to the depletion of residual water vapor. We find that this regime is activated after the inner disk region at $\lesssim 2$\,au has already been depleted from CO and water. 

If not through UV photo-dissociation, the depletion of CO and H$_{2}$O may be initiated in inner disks by direct removal of disk gas.
Disk photo-evaporation by X-ray and UV photons potentially provides an inside-out dispersal mechanism, although this is proposed to be efficient only beyond the radius where the gas is gravitationally bound to the star ($\gtrsim$\,1--2\,au in disks around solar-mass stars). Another removal mechanism could be through portions of MHD winds \citep[e.g.][]{ferr06,bai16} that may be probed at optical wavelengths through a low-velocity component in the forbidden oxygen lines \citep{simon16}. A slow disk wind has been proposed to contribute also to the CO narrow component in double-component disks \citep{pont11a,bast11}, and may play a role in the depletion of molecular gas in inner disks.
A promising direction for future work is to test if wind models, potentially combined to grain growth and planetesimal formation, can reproduce the formation and development of the molecular holes revealed by the analysis of CO and H$_{2}$O velocity-resolved infrared spectroscopy.

\subsection{Molecular composition and evolution of inner disks} \label{sec:disc_ratios}
In Figure \ref{fig: R_fl_corr}, we showed linear trends in the CO/H$_{2}$O and OH/H$_{2}$O line flux ratios, as measured in the 2.9--4.7\,$\mu$m lines observed with CRIRES. Both ratios were characterized only out to log R$_{\rm{co}}$/R$_{\rm{snow}} \sim -1$, due to the absence of 2.93\,$\mu$m H$_{2}$O detections at larger radii. To explore these trends over a larger disk region, we include in Figure \ref{fig: mol_ratios} the line flux ratios as measured using H$_{2}$O line detections at longer wavelengths from the \textit{Spitzer} dataset, which is sensitive to emission at larger disk radii (Sections \ref{sec:ana} and \ref{sec:rad_trends}). We find that the CO/H$_{2}$O line flux ratio of all three water lines show similar radial trends. The increase in CO/H$_{2}$O is confirmed out to log R$_{\rm{co}}$/R$_{\rm{snow}} \sim -0.7$ (Figure \ref{fig: mol_ratios}, left plot) by all three water lines. At larger radii, the \textit{Spitzer} lines suggest a decreasing CO/H$_{2}$O out to the snow line radius. Beyond the snow line, lower limits show that the ratio may increase again. 

Following a similar procedure, we show the measured OH/H$_{2}$O line flux ratios in Figure \ref{fig: mol_ratios}, right plot. The constant $\sim0.4$ ratio with disk radius is confirmed also in this case by all three OH/H$_{2}$O ratios as measured at different wavelengths (2.93\,$\mu$m using the CRIRES data, 12.5-12.6\,$\mu$m and 30-33\,$\mu$m using the \textit{Spitzer} data). Some disks show a higher OH/H$_{2}$O ratio of $\sim1.7$ at 12.5--33\,$\mu$m, perhaps linked to phases of stronger production of OH through increased UV irradiation \citep[among these disks are in fact EX\,Lupi and DG\,Tau, where enhanced production of OH through UV photodissociation has been found by][]{banz12,cn14}. A decreasing trend of OH/H$_{2}$O may be present at log R$_{\rm{co}}$/R$_{\rm{snow}} \gtrsim -0.7$. Beyond the snow line, lower limits suggest that the ratio may increase. Models propose that the OH/H$_{2}$O ratio in the observable column density of gas should increase significantly at the snow line, due to a drop in the column density of H$_{2}$O \citep[e.g.][]{walsh15}. Future velocity-resolved observations must test if the 12.5--33\,$\mu$m water lines have velocity profiles that match the 12.6--30\,$\mu$m OH lines, indicating a similar emitting region (so far, no OH lines have been velocity-resolved at wavelengths $\gtrsim10$\,$\mu$m). 

An open question is if these observed trends are due to the disk physical/chemical radial structure in a static ``primordial" disk, rather than by the formation and evolution of disk gaps that may change it. The decrease of CO/H$_{2}$O and OH/H$_{2}$O at log R$_{\rm{co}}$/R$_{\rm{snow}} \gtrsim -0.7$ out to the snow line radius may be due to changes in the disk chemistry as molecular holes expand out to the snow line radius. A promising direction for future modeling is to calibrate the disk chemistry with the measured molecular line flux ratios. The explorations suggested above within the framework of models that include disk dispersal processes may provide important constraints on the evolution of planet-forming regions. Interesting in this context is, for instance, the notable discrepancy between the data and the model shown in Figure \ref{fig: R_fl_corr} (panel b and c), where \citet{anto16} predict an increase in water emission when disk gaps reach the snow line radius, as due to a new reservoir of water emission released by photo-desorption of icy grains (I. Kamp, private communication). The fact that this is not observed in the data may suggest that the desorption efficiency is over-estimated, either by the adopted desorption rates or because dust grains are larger (i.e. the total surface available for ice desorption smaller) or are removed from disk surfaces. In the evolved disk of TW\,Hya, to reproduce weak water lines observed at $>250$\,$\mu$m with \textit{Herschel}, \citet{hog11} proposed that icy dust grains may have been removed from the disk surface by settling towards the midplane, hiding a large reservoir of icy solids in a region shielded from UV irradiation and photo-desorption.

\section{SUMMARY and CONCLUSIONS} \label{sec:end}
1) This analysis combines a sample of 55 protoplanetary disks that have CO rovibrational emission spectra observed at 4.7\,$\mu$m as well as H$_{2}$O and OH emission spectra observed at 2.9 and/or 10--33\,$\mu$m. H$_{2}$O and OH spectra at 2.9\,$\mu$m (from VLT-CRIRES) are published here for the first time for 25 disks; CO 4.7\,$\mu$m spectra (from VLT-CRIRES) and H$_{2}$O and OH spectra at 10--33\,$\mu$m (from \textit{Spitzer}-IRS) have been published previously. This combined sample probes the molecular content of inner disks at radii of $\approx$0.05--20\,au. 

2) The velocity-resolved line profiles of H$_{2}$O and OH emission at 2.9\,$\mu$m are matched by the broader/inner component of 4.7\,$\mu$m CO emission as identified by \citet{bp15}. These emission lines probe a similar inner disk region at $\approx$\,0.04--0.3\,au, as shown by their similar profiles, and when CO is depleted in this region, H$_{2}$O is depleted too. H$_{2}$O lines at longer wavelengths (12--33\,$\mu$m), probing progressively colder water, may be co-spatial with the colder component of CO emission at disk radii of $\gtrsim0.3$\,au.

3) We propose the infrared water spectrum as a tracer of inside-out clearing of inner disks around stars of masses between 0.3 and 3.5 M$_{\odot}$. We find a wavelength-dependent decrease of water line fluxes where water emission is absent at 2.9\,$\mu$m but still present at longer wavelengths (12--33\,$\mu$m) in some disks. We discover a correlation between water line fluxes at 2.9--33\,$\mu$m and the size of inner disk gaps/holes as probed by CO emission, where water emission decreases as $R_{\rm{co}}$ approaches the snow line radius. All taken together, we interpret this as due to the progressive depletion of inner disks in an inside-out fashion, where inner/hotter gas is depleted before outer/colder gas.

4) This analysis helps clarifying the origin of the absence of infrared H$_{2}$O emission in disks around intermediate-mass stars: we find that these disks have inner gaps or holes in molecular gas (i.e. they have large R$_{\rm co}$). This adds to growing evidence from other studies of gaps and holes in dust emission \citep[e.g.][]{maask13,menu15}, as well as to recent model explorations \citep{anto16}.

5) We measure the radial profiles of CO/H$_{2}$O and OH/H$_{2}$O line flux ratios in inner disks out to the water snow line. We find an increasing CO/H$_{2}$O with disk radius within the snow line, possibly followed by an inversion in the trend with a CO/H$_{2}$O ratio decreasing with disk radius as the molecular hole expands out the snow line radius. For OH/H$_{2}$O, we find a constant ratio within the snow line, again possibly followed by an inversion in the trend.

To conclude, our work finds that the decrease in CO and H$_{2}$O line fluxes is linked to the depletion of molecular gas through inside-out formation of inner disk gaps or holes. Line flux measurements and trends from this work provide an opportunity to model the properties of evolving inner disks, to investigate how physical changes (e.g. in the gas-to-dust ratio, dust grain growth, and the removal of inner disk gas) may affect the composition of planet-forming disk regions (e.g. through an enhanced UV penetration at large disk radii, and the depletion of molecular gas from inner disks). Measured radial trends in the chemistry and depletion of water during the evolution of protoplanetary disks may offer new ground to investigate the origins of radial gradients in composition measured in Solar System bodies \citep[e.g.][]{morbi12} and potentially also in exoplanet populations.

\subsection{Future prospects with JWST} \label{sec: JWST}
The \textit{James Webb Space Telescope} (\textit{JWST}) will provide R $\sim3000$ (or $\sim100$\,km/s) spectroscopy of molecular spectra from the 2.9\,$\mu$m lines included in this analysis (with \textit{JWST}-NIRSpec) to the longer wavelengths covered by \textit{Spitzer} (up to 28\,$\mu$m, with \textit{JWST}-MIRI). The observed molecular emission spectra will include CO, H$_{2}$O, OH, and CO$_{2}$. The broad spectral coverage will provide the data to study more molecular trends than those included in this work, and to better characterize the chemistry and its evolution when gaps form in disks.
\textit{JWST} has also the potential to study infrared molecular emission in disks around low-mass stars, and disk samples not limited to the closest star-forming regions at 120-180\,pc.
A \textit{JWST} spectrum at 3--28\,$\mu$m may distinguish the evolutionary phase of inner disks from the measured line fluxes and the molecular species detected (Figure \ref{fig: TR_water}), spanning a range of disk radii and gap sizes \citep[$\sim$\,0.05--20\,au,][]{bp15} that is highly complementary to the larger scales probed by millimeter interferometry with the Atacama Large Millimeter Array ALMA \citep[$\gtrsim5$\,au at 140\,pc, e.g.][]{vdm16} and by direct imaging techniques. 
The joint use of these observatories can transform our understanding of the development of disk gaps from the innermost disk regions to the larger-scale disk structures, and of the physical and chemical evolution of planet-forming environments.  \\

We thank S.Antonellini and I.Kamp for providing water emission spectra from their ProDiMo model explorations, and for insightful discussions on water emission in disks and on its dependence on disk model parameters. We also thank A.Carmona for providing the CRIRES spectrum of HD139614.
A.B. wishes to thank all colleagues who made this work possible through their pioneering work in infrared spectroscopy of disks.
A.B. thanks Simon Bruderer for the seminal work he did in 2011-2012, which provided knowledge and tools to enable this work.
A.B. acknowledges financial support by a NASA Origins of the Solar System grant No. OSS 11-OSS11-0120, a NASA Planetary Geology and Geophysics Program under grant NAG 5-10201.
E.v.D. acknowledges European Union A-ERC grant 291141 CHEMPLAN.
This work is partly based on observations made with ESO telescopes at the Paranal Observatory under programs 179.C-0151, 093.C-0432, 079.C-0349, 081.C-0833, 091.C-0671, 082.C-0491, and 088.C-0898.
This work is partly based on observations made with the Spitzer Space Telescope, which is operated by the Jet Propulsion Laboratory, California Institute of Technology. 
This publication makes use of data products from the Wide-field Infrared Survey Explorer, which is a joint project of the University of California, Los Angeles, and the Jet Propulsion Laboratory/California Institute of Technology, funded by the National Aeronautics and Space Administration.
This research has made use of the VizieR catalogue access tool, CDS, Strasbourg, France. The original description of the VizieR service was published in A\&AS 143, 23.
This work is partly based on observations obtained at the W. M. Keck Observatory, which is operated as a scientific partnership among the California Institute of Technology, the University of California and the National Aeronautics and Space Administration. The Observatory was made possible by the generous financial support of the W. M. Keck Foundation.

\appendix

\subsection{Line flux measurements} \label{app:line_flux}
We measure emission line fluxes by fitting a first-order polynomial continuum over neighboring pixels and by summing continuum-subtracted pixel flux values over the spectral range of each line. In the \textit{Spitzer}-IRS spectra, to select pixels of continuum free from line emission, we use models of the strongest emitters at 10--35\,$\mu$m as identified in previous work \citep[water, OH, HCN, C$_2$H$_2$, CO$_2$, HI; see e.g.][]{cn11,banz12}.
The following are the ranges adopted: the 12.52\,$\mu$m line is measured on 12.50--12.537\,$\mu$m, and its continuum fitted on 12.48--12.5\,$\mu$m and 12.7--12.72\,$\mu$m (at the two sides of the line); the 33\,$\mu$m line on 32.92--33.06\,$\mu$m and its continuum on 32.53--32.63\,$\mu$m and 33.4--33.47\,$\mu$m; the 12.65\,$\mu$m OH line on 12.64--12.68\,$\mu$m and its continuum on 12.48--12.5\,$\mu$m and 12.7--12.72\,$\mu$m; the 30\,$\mu$m OH lines on 30.245--30.385\,$\mu$m and 30.62--30.73\,$\mu$m and their continua on 30.1--30.17\,$\mu$m and 30.38--30.43\,$\mu$m, and 30.57--30.62\,$\mu$m and 30.73--30.83\,$\mu$m (we sum the flux of these OH doublets to increase the S/N). OH detections at 30\,$\mu$m are only tentative in some spectra of disks around intermediate-mass stars, and will need to be confirmed by higher-spectral-resolution observations.
Line flux errors are estimated as the standard deviation of the distribution of measured line fluxes after re-sampling the spectrum using the local noise, as in \citet{banz12}.
Following this procedure, we find values consistent within 1 $\sigma$ (or a few \%) to those measured by \citet{pont10a}, adopting as a test case the 17.22\,$\mu$m water line. 
Line fluxes and errors are reported in Table \ref{tab: fluxes}; in all figures throughout this paper, line fluxes are normalized to a common distance of 140\,pc for a homogeneous comparison, using distances from Table \ref{tab: sample}.

In the CRIRES spectra, line fluxes are measured in the continuum-normalized spectra and then flux calibrated using fluxes measured by the Wide-field Infrared Survey Explorer \citep[WISE,][]{wise} at 3.35 and 4.6\,$\mu$m as released by \citet{wise12,wise14}. In some cases where the WISE 4.6\,$\mu$m measurement disagrees with previous photometry, we adopt the \textit{Spitzer}-IRAC measurement or an interpolated value between WISE measurements at 3.35\,$\mu$m and 11.6\,$\mu$m. The adopted continuum fluxes are reported in Table \ref{tab: sample}. Unlike in the Spitzer spectra, the lines observed with CRIRES are velocity-resolved and present a broad range of widths that makes it inappropriate to adopt fixed pixel ranges for their measurement. To adopt a homogeneous method, we therefore measure the line flux included within the full width at 10\% of the line peak. Table \ref{tab: fluxes} reports the fluxes and errors measured for the lines listed in Table \ref{tab:lines}. The four water lines at $\sim$2.93\,$\mu$m are measured together into a single flux value, because in most spectra they are blended together due to their large line widths.

\begin{deluxetable*}{l c c c c c c c c}
\tabletypesize{\small}
\tablewidth{0pt}
\tablecaption{\label{tab: fluxes} Line fluxes.}
\tablehead{\colhead{Name} & \multicolumn{3}{|c|}{H$_{2}$O} & \multicolumn{3}{|c|}{OH} & \multicolumn{2}{|c|}{CO v=1-0 P10}
\\ 
 & 2.93\,$\mu$m & 12.52\,$\mu$m & 33\,$\mu$m & 2.93\,$\mu$m & 12.65\,$\mu$m & 30\,$\mu$m & BC & NC }
\tablecolumns{9}
\startdata

AATau & $<$ 0.25 & 0.61  $\pm$  0.04 & 2.48  $\pm$  0.10 & 0.11  $\pm$  0.00 & 0.29  $\pm$  0.03 & 2.01  $\pm$  0.08 & -- & 2.74  $\pm$  0.02 \\
AS205 N & 7.33  $\pm$  2.71 & 11.26  $\pm$  0.61 & 29.32  $\pm$  1.88 & 2.06  $\pm$  0.76 & 6.20  $\pm$  0.61 & 30.84  $\pm$  1.34 & 17.98  $\pm$  7.19 & 13.38  $\pm$  5.35 \\
AS209 & -- & -- & -- & -- & -- & -- & 1.04  $\pm$  0.03 & 0.71  $\pm$  0.02 \\
CVCha & -- & -- & -- & -- & -- & -- & 3.55  $\pm$  0.11 & 1.97  $\pm$  0.07 \\
CWTau & $<$ 0.46 & 3.59  $\pm$  0.09 & 8.42  $\pm$  0.06 & 1.17  $\pm$  0.21 & 1.54  $\pm$  0.07 & 10.66  $\pm$  0.13 & 5.80  $\pm$  0.93 & 5.51  $\pm$  0.88 \\
DFTau & 4.55  $\pm$  0.59 & -- & -- & 1.65  $\pm$  0.21 & -- & -- & 7.99  $\pm$  0.72 & 3.25  $\pm$  0.29 \\
DGTau & -- & 2.49  $\pm$  0.20 & 4.70  $\pm$  0.70 & -- & 5.50  $\pm$  0.18 & 16.60  $\pm$  0.64 & 5.42  $\pm$  0.43 & -- \\
DoAr24E S & $<$ 1.09 & 1.66  $\pm$  0.15 & 9.85  $\pm$  0.35 & 1.92  $\pm$  0.73 & 2.13  $\pm$  0.12 & 6.44  $\pm$  0.34 & 10.15  $\pm$  3.45 & 5.51  $\pm$  1.87 \\
DoAr44 & $<$ 2.02 & 0.52  $\pm$  0.05 & 4.39  $\pm$  0.16 & $<$ 0.56 & 0.36  $\pm$  0.05 & 3.00  $\pm$  0.15 & -- & 2.24  $\pm$  0.03 \\
DOTau & 2.31  $\pm$  0.19 & 1.46  $\pm$  0.07 & 2.63  $\pm$  0.33 & 1.01  $\pm$  0.08 & 2.19  $\pm$  0.06 & 3.39  $\pm$  0.22 & 4.44  $\pm$  0.27 & 0.59  $\pm$  0.04 \\
DRTau & 2.84  $\pm$  0.43 & 5.25  $\pm$  0.12 & 10.11  $\pm$  0.13 & 0.80  $\pm$  0.12 & 2.37  $\pm$  0.13 & 6.97  $\pm$  0.28 & 8.36  $\pm$  1.34 & 6.55  $\pm$  1.05 \\
EC82 & $<$ 0.60 & $<$ 0.36 & 2.05  $\pm$  0.18 & 0.39  $\pm$  0.02 & $<$ 0.33 & 1.73  $\pm$  0.17 & -- & 2.04  $\pm$  0.02 \\
EXLup08 & 8.69  $\pm$  2.61 & $<$ 0.99 & 5.99  $\pm$  0.90 & 3.88  $\pm$  1.16 & 1.89  $\pm$  0.24 & 3.85  $\pm$  1.53 & 13.20  $\pm$  7.92 & 5.51  $\pm$  3.31 \\
EXLup14 & 0.32  $\pm$  0.02 & 1.00  $\pm$  0.16 & 3.43  $\pm$  0.48 & 0.12  $\pm$  0.01 & $<$ 0.46 & $<$ 1.14 & 0.70  $\pm$  0.02 & 0.89  $\pm$  0.03 \\
FNTau & $<$ 0.05 & $<$ 0.19 & $<$ 0.27 & $<$ 0.03 & 0.21  $\pm$  0.05 & $<$ 0.42 & -- & 0.29  $\pm$  0.01 \\
FZTau & -- & 5.94  $\pm$  0.10 & 6.44  $\pm$  0.05 & -- & 1.67  $\pm$  0.08 & 5.38  $\pm$  0.11 & -- & 9.59  $\pm$  0.68 \\
GQLup & $<$ 0.57 & 0.70  $\pm$  0.04 & 4.92  $\pm$  0.09 & $<$ 0.26 & 0.51  $\pm$  0.04 & 3.06  $\pm$  0.11 & 4.18  $\pm$  0.21 & 2.22  $\pm$  0.12 \\
HD36112 & -- & $<$ 1.85 & $<$ 6.55 & -- & $<$ 1.81 & 9.76  $\pm$  1.36 & -- & 8.85  $\pm$  3.45 \\
HD95881 $^{*}$ & -- & $<$ 2.24 & $<$ 3.00 & -- & $<$ 2.19 & $<$ 2.03 & -- & 4.91  $\pm$  0.20 \\
HD97048 $^{*}$ & -- & -- & -- & -- & -- & -- & -- & 12.30  $\pm$  0.40 \\
HD98922 $^{*}$ & $<$ 0.08 & $<$ 7.48 & $<$ 6.05 & $<$ 0.08 & $<$ 7.29 & 7.29  $\pm$  2.66 & -- & 5.08  $\pm$  0.14 \\
HD101412 & -- & $<$ 1.17 & $<$ 1.27 & -- & $<$ 1.14 & 1.89  $\pm$  0.25 & -- & -- \\
HD135344B & $<$ 0.30 & $<$ 0.65 & $<$ 1.84 & $<$ 0.14 & $<$ 0.63 & 3.66  $\pm$  0.36 & -- & 1.99  $\pm$  0.79 \\
HD139614 & -- & -- & -- & -- & -- & -- & -- & 1.36  $\pm$  0.05 \\
HD141569 $^{*}$ & -- & -- & -- & -- & -- & -- & -- & 1.77  $\pm$  0.14 \\
HD142527 & -- & $<$ 2.37 & $<$ 10.41 & -- & $<$ 2.31 & 14.30  $\pm$  2.00 & -- & 8.58  $\pm$  4.29 \\
HD144432S & -- & -- & -- & -- & -- & -- & -- & 2.73  $\pm$  0.71 \\
HD150193 $^{*}$ & -- & $<$ 3.55 & $<$ 7.00 & -- & $<$ 3.46 & 10.40  $\pm$  0.94 & -- & 8.22  $\pm$  1.94 \\
HD163296 $^{*}$ & -- & $<$ 3.91 & 7.21  $\pm$  1.02 & -- & $<$ 3.83 & 24.77  $\pm$  0.99 & -- & 18.70  $\pm$  1.60 \\
HD179218 $^{*}$ & -- & $<$ 5.27 & $<$ 9.89 & -- & $<$ 4.50 & $<$ 4.89 & -- & 2.88  $\pm$  0.74 \\
HD190073 $^{*}$ & -- & $<$ 2.43 & $<$ 1.18 & -- & $<$ 2.38 & 5.77  $\pm$  0.59 & -- & 6.99  $\pm$  0.59 \\
HD244604 & -- & $<$ 1.53 & $<$ 1.66 & -- & $<$ 1.50 & $<$ 0.95 & -- & 0.64  $\pm$  0.01 \\
HD250550 $^{*}$ & $<$ 0.14 & -- & -- & 0.14  $\pm$  0.01 & -- & -- & -- & 5.80  $\pm$  0.35 \\
HTLup & $<$ 0.73 & $<$ 1.15 & $<$ 1.19 & 0.36  $\pm$  0.06 & 1.26  $\pm$  0.31 & 2.58  $\pm$  0.48 & 2.27  $\pm$  0.25 & 0.49  $\pm$  0.06 \\
IMLup & $<$ 0.18 & $<$ 0.18 & 0.65  $\pm$  0.12 & $<$ 0.08 & 0.34  $\pm$  0.04 & $<$ 0.34 & 0.44  $\pm$  0.01 & 0.07  $\pm$  0.11 \\
IRS48 & -- & -- & -- & -- & -- & -- & -- & 0.58  $\pm$  0.13 \\
LkHa330 & -- & $<$ 0.23 & $<$ 1.77 & -- & $<$ 0.23 & $<$ 0.82 & -- & 1.01  $\pm$  0.09 \\
RNO90 & 2.61  $\pm$  0.34 & 7.38  $\pm$  0.26 & 23.29  $\pm$  0.35 & 1.40  $\pm$  0.18 & 2.98  $\pm$  0.25 & 19.66  $\pm$  0.35 & 12.93  $\pm$  2.07 & 6.24  $\pm$  1.00 \\
RULup & 4.27  $\pm$  0.60 & 3.08  $\pm$  0.16 & 5.00  $\pm$  0.23 & 1.60  $\pm$  0.22 & 6.08  $\pm$  0.16 & 7.39  $\pm$  0.21 & 6.63  $\pm$  0.73 & 4.93  $\pm$  0.54 \\
RYLup & -- & $<$ 0.96 & $<$ 1.52 & -- & $<$ 0.93 & 2.11  $\pm$  0.30 & -- & 1.76  $\pm$  0.09 \\
RWAur & -- & 6.79  $\pm$  0.20 & 9.48  $\pm$  0.17 & -- & 2.09  $\pm$  0.17 & 8.35  $\pm$  0.25 & 8.47  $\pm$  0.38 & 5.83  $\pm$  0.25 \\
SCrA S & 6.46  $\pm$  2.00 & 8.59  $\pm$  0.87 & 23.16  $\pm$  2.33 & 3.85  $\pm$  1.19 & 13.32  $\pm$  0.82 & 40.11  $\pm$  1.69 & 13.25  $\pm$  2.92 & 4.86  $\pm$  1.07 \\
SCrA N & 8.68  $\pm$  2.69 & 8.59  $\pm$  0.87 & 23.16  $\pm$  2.33 & 3.75  $\pm$  1.16 & 13.32  $\pm$  0.82 & 40.11  $\pm$  1.69 & 8.11  $\pm$  1.78 & 5.76  $\pm$  1.27 \\
SR9 & -- & $<$ 0.93 & 2.42  $\pm$  0.21 & -- & $<$ 0.91 & 3.27  $\pm$  0.13 & -- & 1.31  $\pm$  0.03 \\
SR21 & $<$ 0.03 & $<$ 1.16 & $<$ 4.60 & $<$ 0.02 & $<$ 1.13 & 14.42  $\pm$  0.98 & -- & 0.46  $\pm$  0.02 \\
TTau N & 12.77  $\pm$  10.22 & -- & -- & 6.14  $\pm$  4.91 & -- & -- & 32.12  $\pm$  25.70 & 9.65  $\pm$  7.72 \\
TTau S & 9.66  $\pm$  7.73 & -- & -- & 11.31  $\pm$  9.05 & -- & -- & 37.56  $\pm$  30.05 & -- \\
TWCha & 0.49  $\pm$  0.00 & 0.62  $\pm$  0.02 & 2.42  $\pm$  0.05 & 0.10  $\pm$  0.00 & 0.19  $\pm$  0.02 & 1.41  $\pm$  0.04 & -- & 1.05  $\pm$  0.01 \\
TWHya & $<$ 0.02 & 0.20  $\pm$  0.05 & 3.96  $\pm$  0.47 & $<$ 0.01 & 0.58  $\pm$  0.04 & 10.21  $\pm$  0.40 & -- & 0.64  $\pm$  0.00 \\
VSSG1 & -- & 2.11  $\pm$  0.26 & 7.49  $\pm$  0.45 & -- & 1.20  $\pm$  0.24 & 4.94  $\pm$  0.35 & 4.83  $\pm$  0.29 & 2.88  $\pm$  0.17 \\
VVSer & -- & $<$ 1.27 & $<$ 1.73 & -- & 1.88  $\pm$  0.32 & 4.48  $\pm$  0.25 & -- & 2.02  $\pm$  0.36 \\
VWCha & 2.07  $\pm$  0.11 & 1.97  $\pm$  0.08 & 7.14  $\pm$  0.08 & 1.32  $\pm$  0.07 & 1.38  $\pm$  0.07 & 9.85  $\pm$  0.12 & 6.46  $\pm$  0.30 & 2.96  $\pm$  0.14 \\
VZCha & 0.74  $\pm$  0.01 & 0.83  $\pm$  0.05 & 2.46  $\pm$  0.07 & 0.30  $\pm$  0.00 & 0.52  $\pm$  0.05 & 1.97  $\pm$  0.05 & 3.11  $\pm$  0.03 & 1.01  $\pm$  0.02 \\
WaOph6 & 2.66  $\pm$  0.19 & 1.29  $\pm$  0.04 & 2.56  $\pm$  0.14 & 1.32  $\pm$  0.09 & 1.00  $\pm$  0.04 & 3.30  $\pm$  0.12 & 2.75  $\pm$  0.14 & 0.96  $\pm$  0.05 \\
WXCha & $<$ 0.51 & 1.18  $\pm$  0.04 & 3.08  $\pm$  0.07 & 0.32  $\pm$  0.01 & 0.39  $\pm$  0.04 & 1.63  $\pm$  0.06 & 1.98  $\pm$  0.03 & 0.66  $\pm$  0.02 \\

\enddata

\tablecomments{
Line fluxes are reported in $10^{-14}$\,erg\,s$^{-1}$\,cm$^{-2}$, equivalent to $10^{-17}$\,W\,m$^{-2}$. Upper limits are given as $2\times \sigma$, where $\sigma$ is the sum of pixel flux errors. For targets marked with *, the CO line flux is adopted from \citet{vdplas15}, \citet{hbert16}, and \citet{brit07}, while the H$_{2}$O and OH line fluxes at 2.9\,$\mu$m, where available, are adopted from \citet{fed11}.
}

\end{deluxetable*}

\subsection{Photospheric correction plots} \label{app:phot_plots}
Figures \ref{fig: phot_corr_1} and \ref{fig: phot_corr_2} show the stellar photospheric correction of CRIRES 2.9\,$\mu$m spectra as explained in Section \ref{sec:ana3}.
This procedure proves to be effective in correcting most science spectra, those that match well with a spectral template and where the disk gas emission lines are broader than the absorption lines from the same molecules in the stellar photosphere. In some cases, however, there is no unambiguous/acceptable spectral match with a template (in CW\,Tau, HT\,Lup, and IM\,Lup), or the disk emission lines are possibly as narrow as the photospheric lines (in FN\,Tau and TW\,Hya); in these cases, the residuals after photospheric correction leave larger uncertainties on potential detections of H$_{2}$O and OH and on their line profiles.

\begin{figure*}
\includegraphics[width=0.5\textwidth]{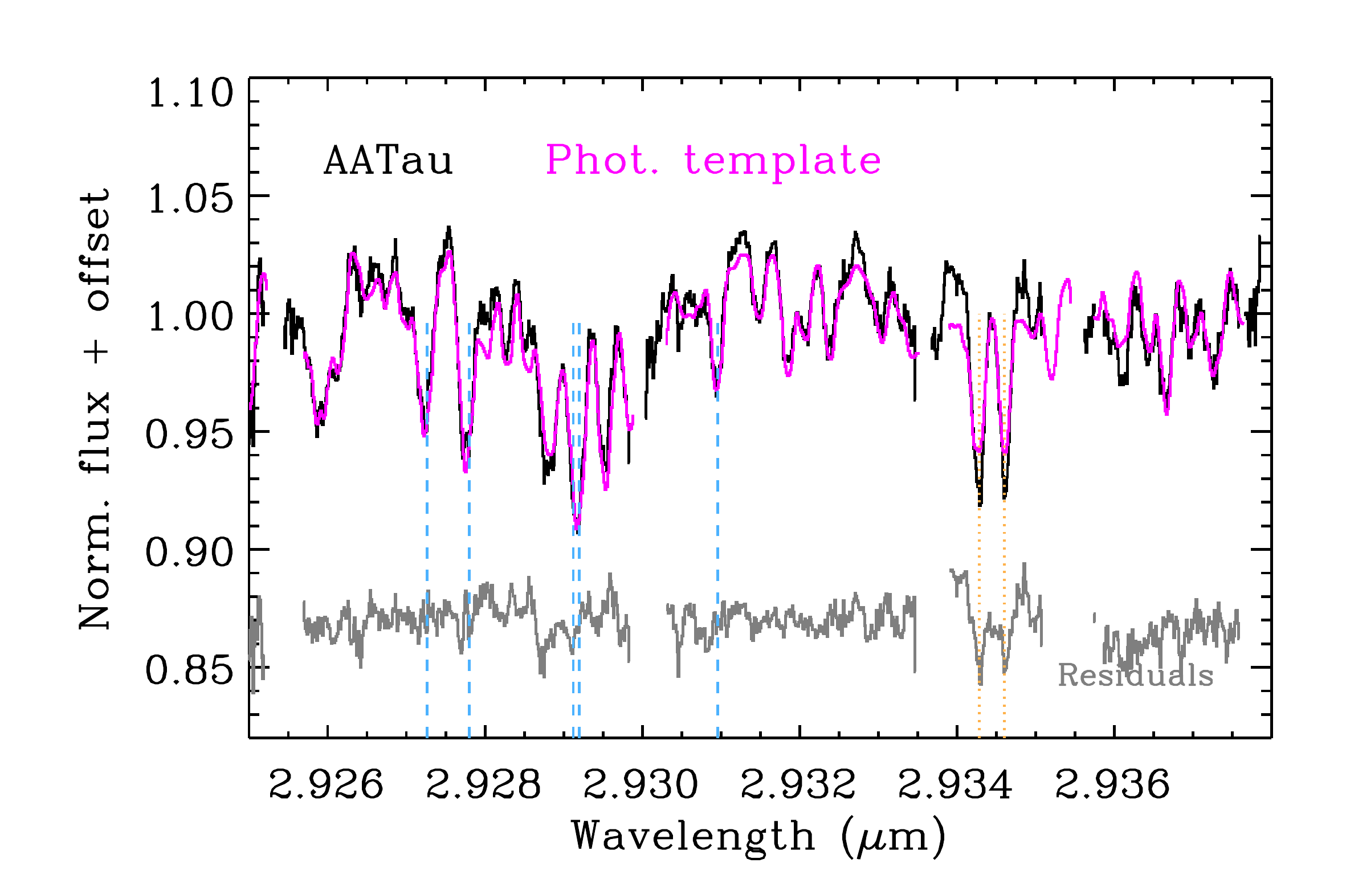} 
\includegraphics[width=0.5\textwidth]{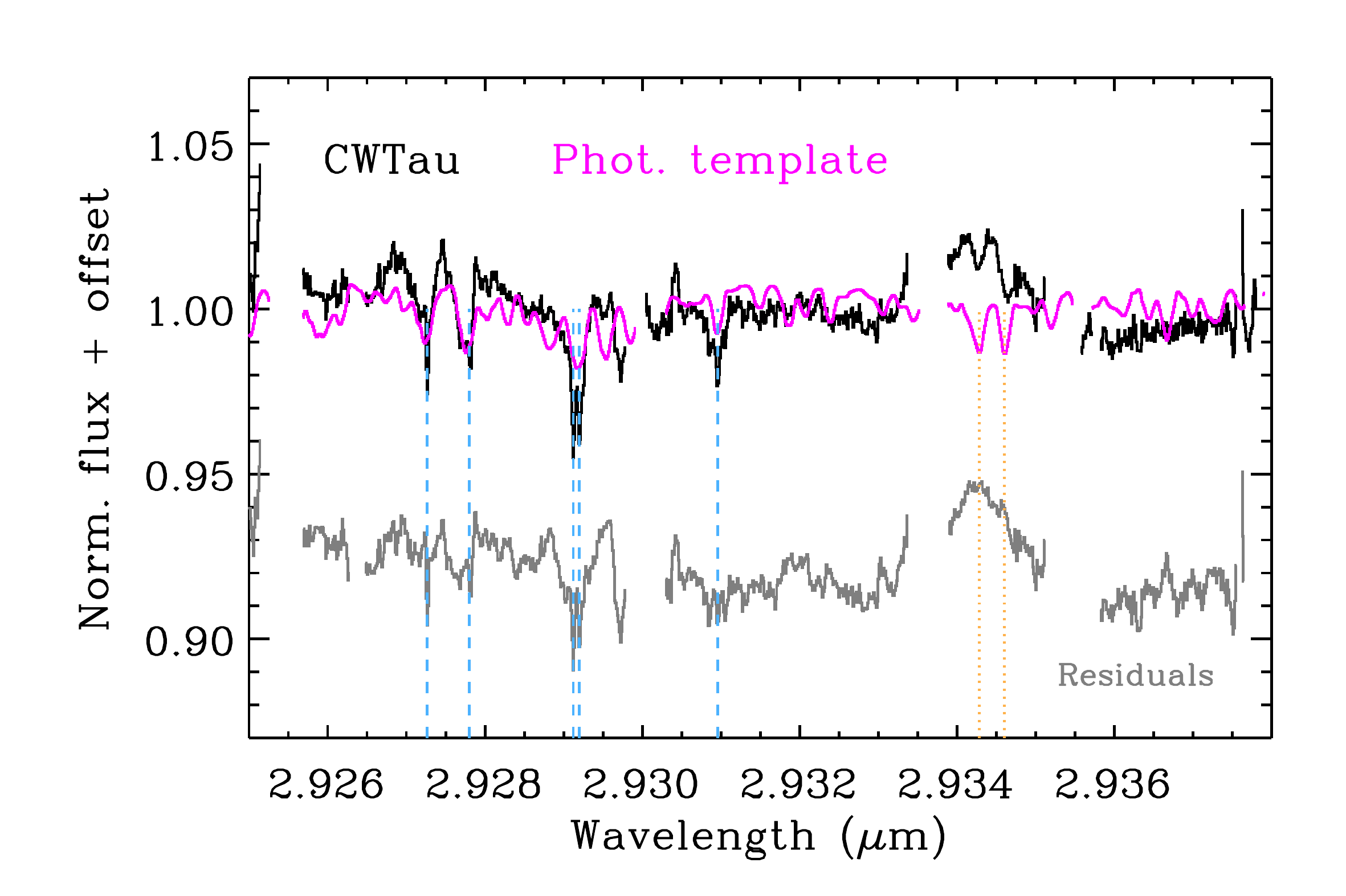} 
\includegraphics[width=0.5\textwidth]{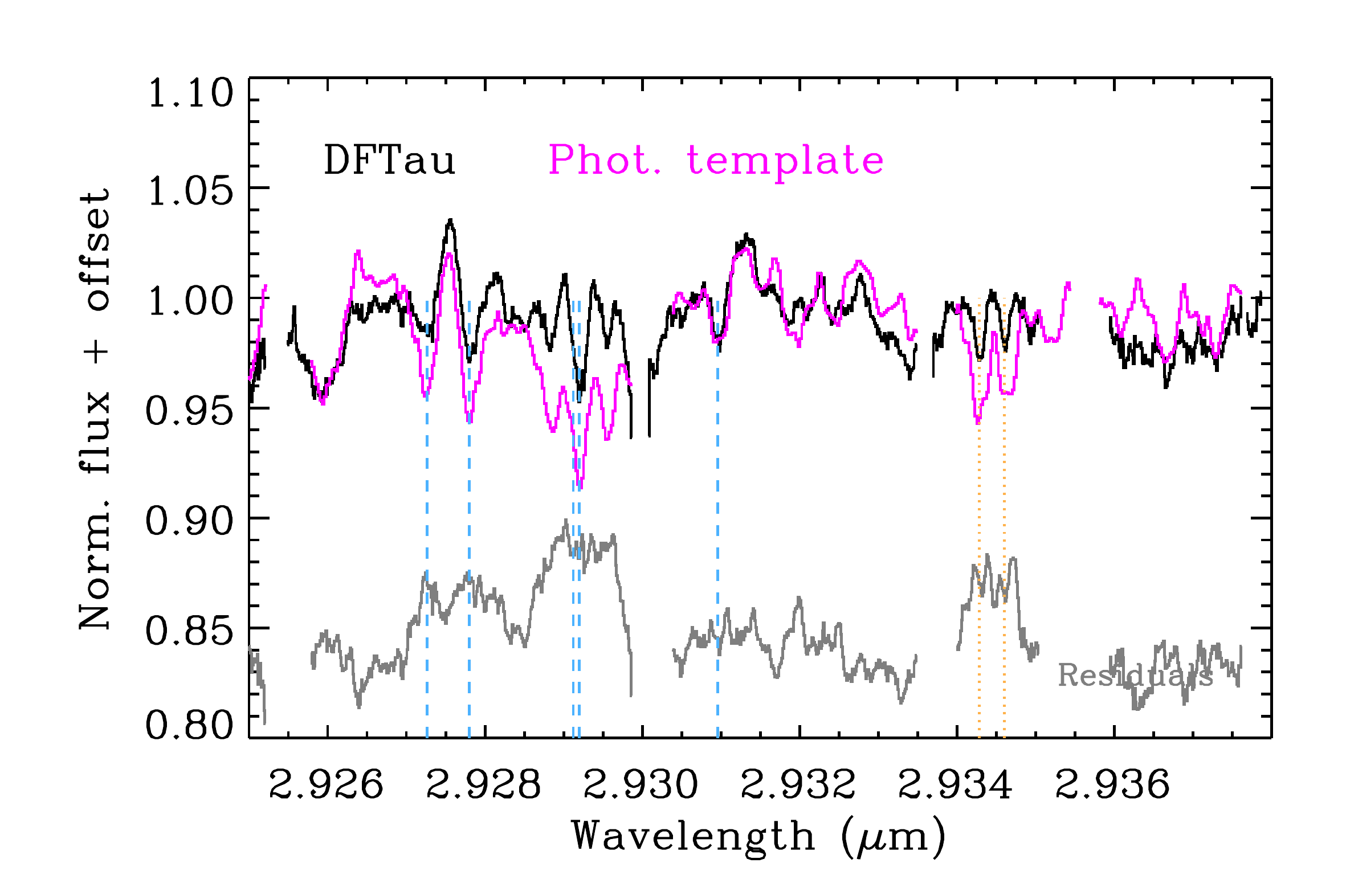} 
\includegraphics[width=0.5\textwidth]{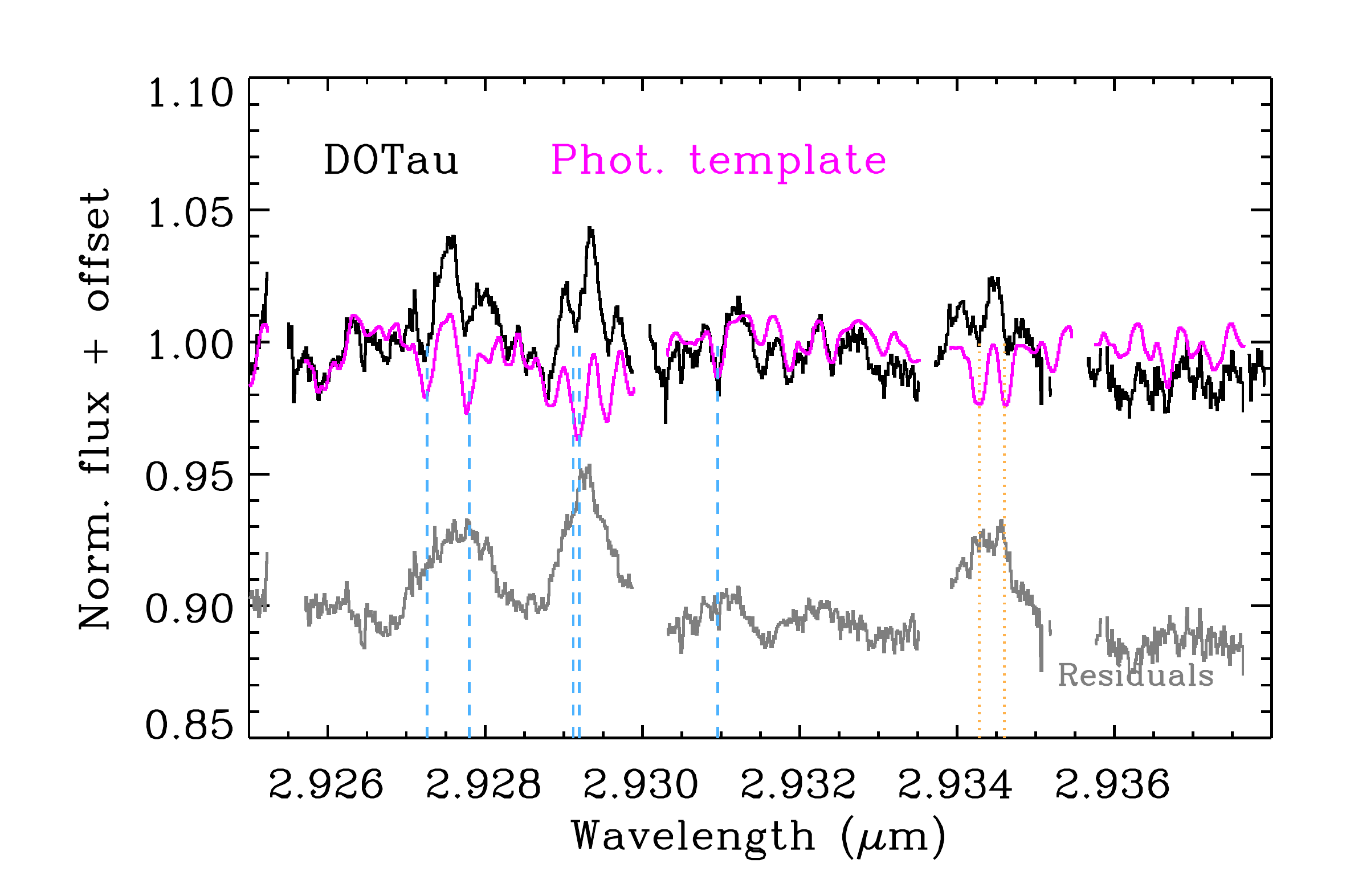} 
\includegraphics[width=0.5\textwidth]{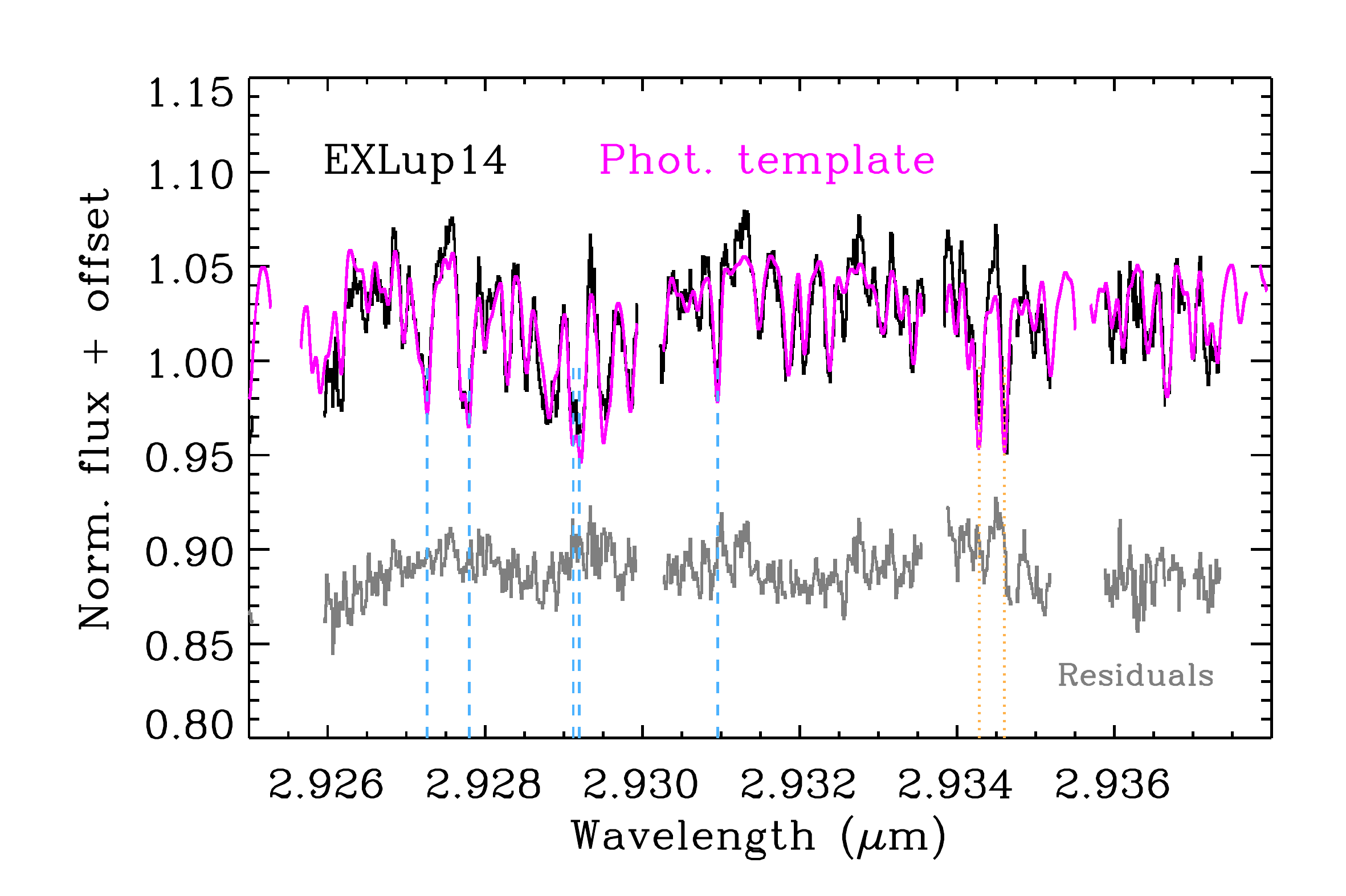} 
\includegraphics[width=0.5\textwidth]{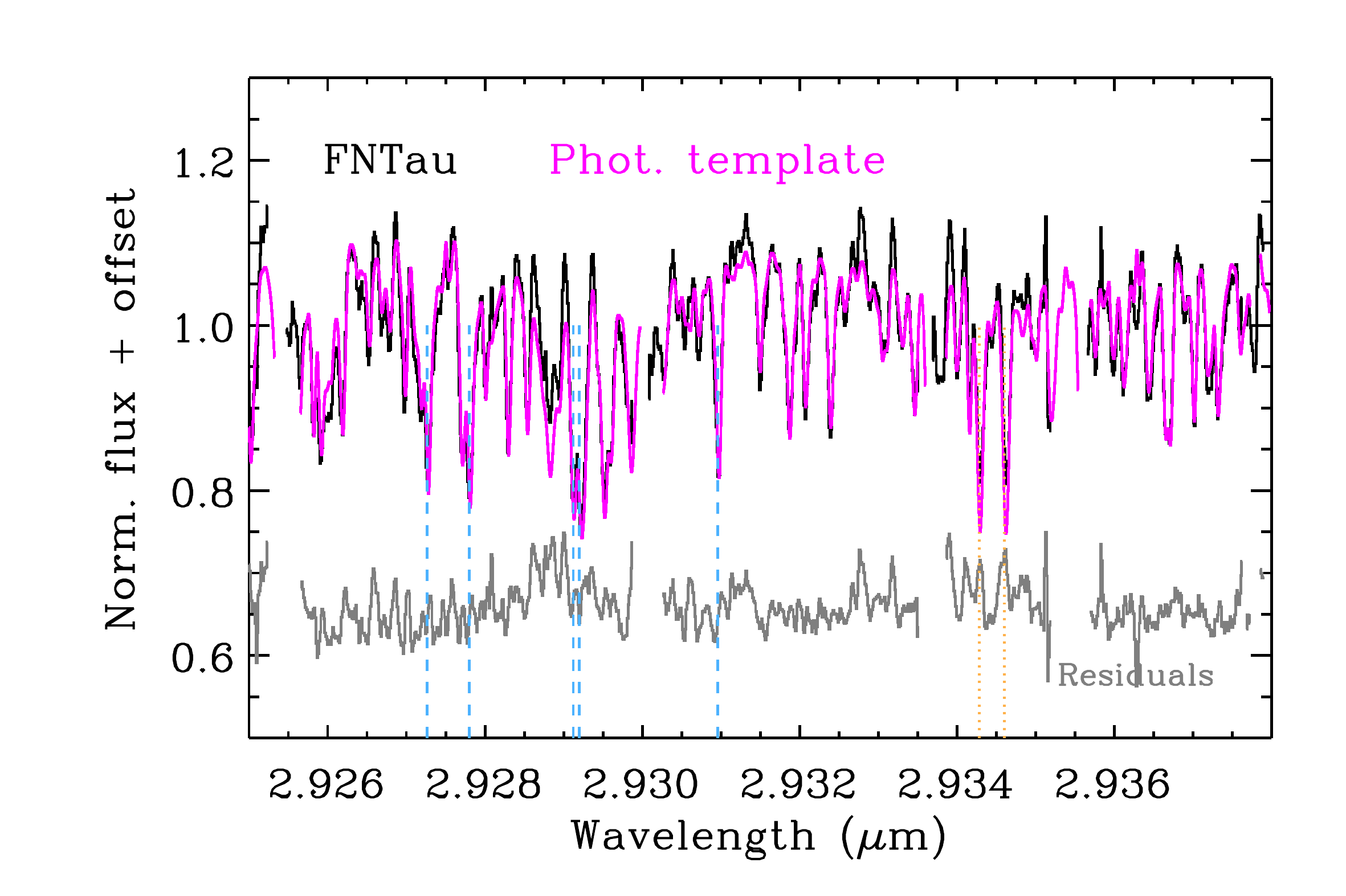} 
\includegraphics[width=0.5\textwidth]{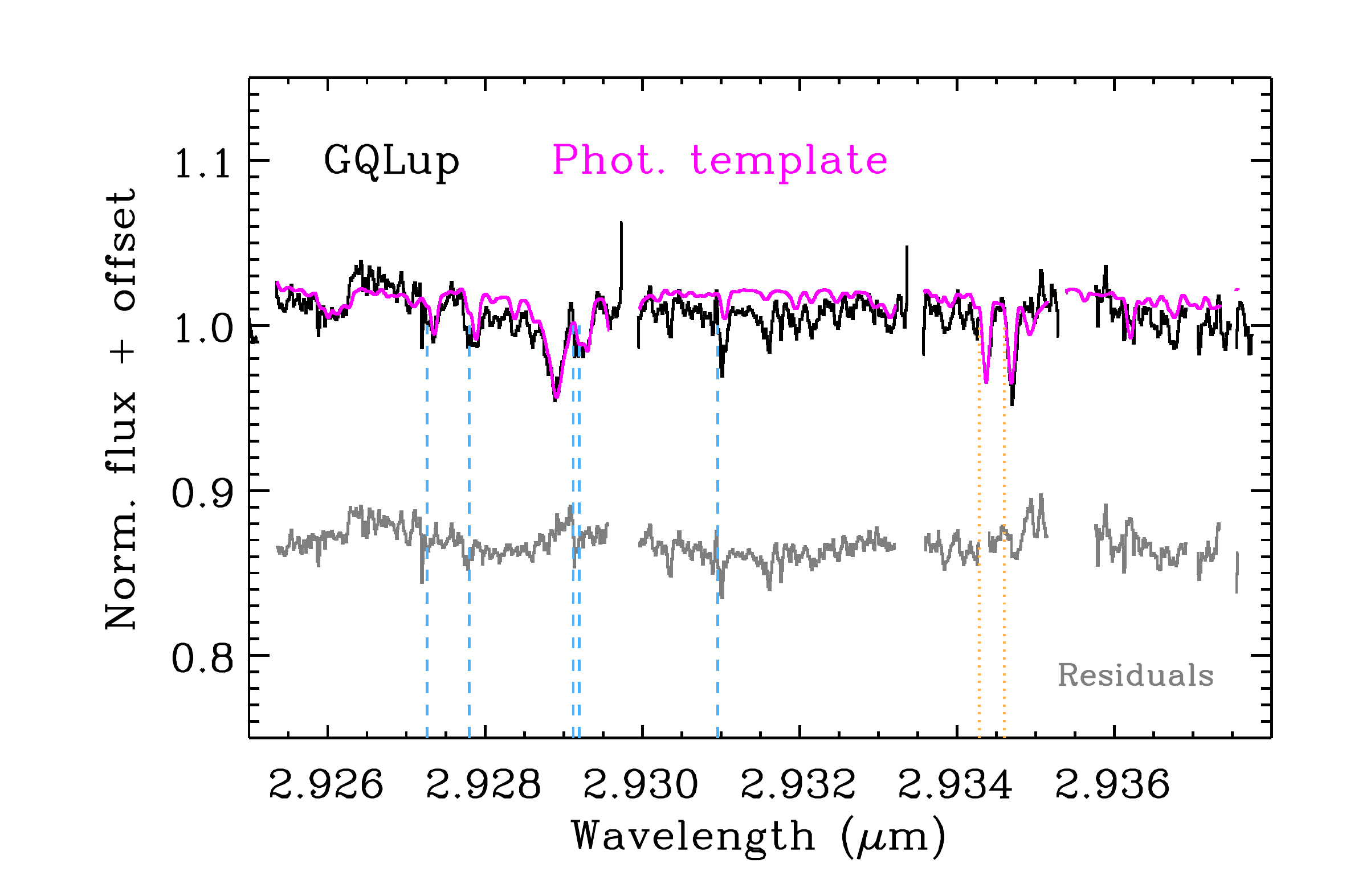} 
\includegraphics[width=0.5\textwidth]{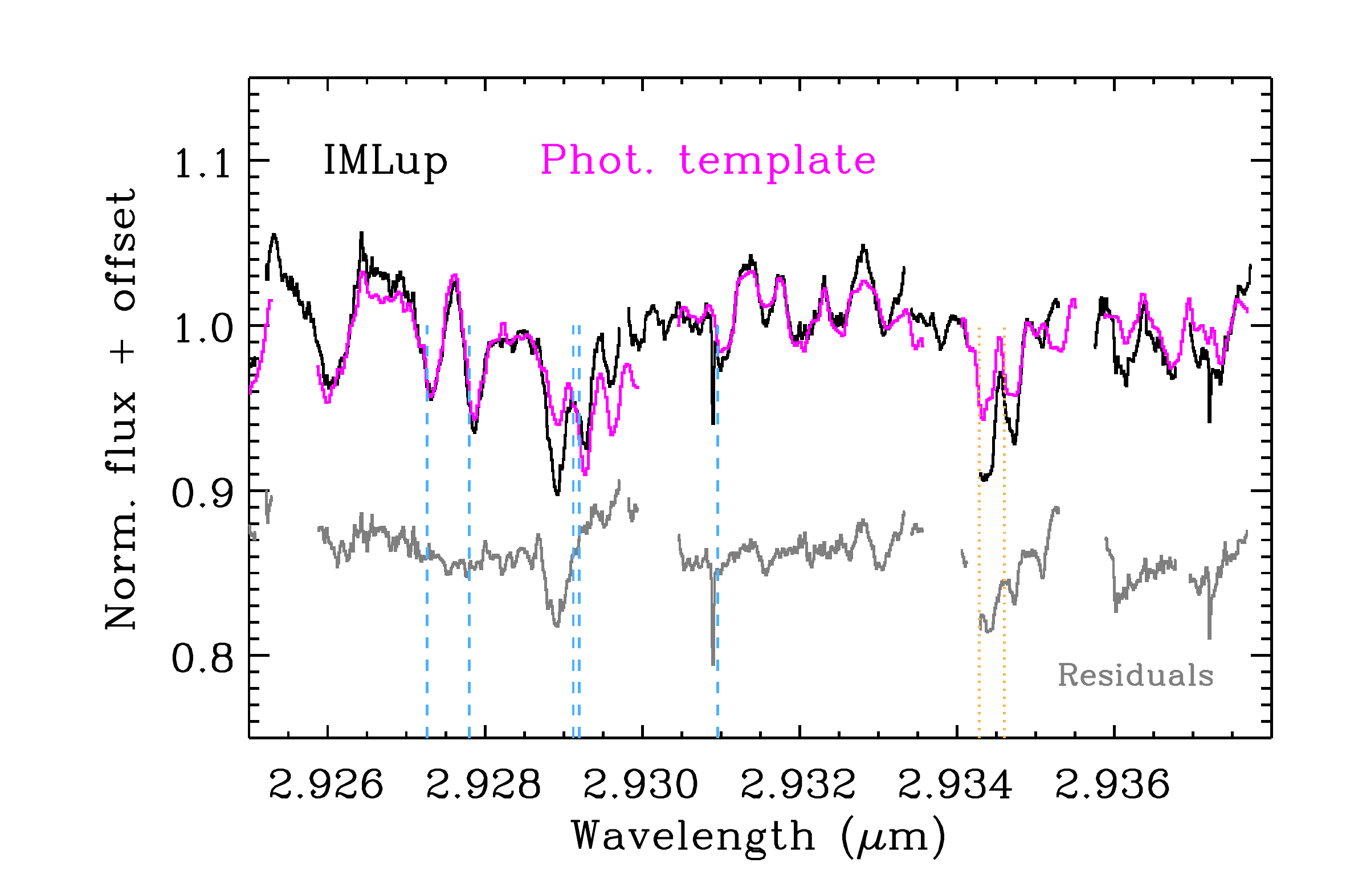} 
\caption{Photospheric correction of 2.9\,$\mu$m disk emission spectra observed by VLT-CRIRES.}
\label{fig: phot_corr_1}
\end{figure*}

\begin{figure*}
\includegraphics[width=0.5\textwidth]{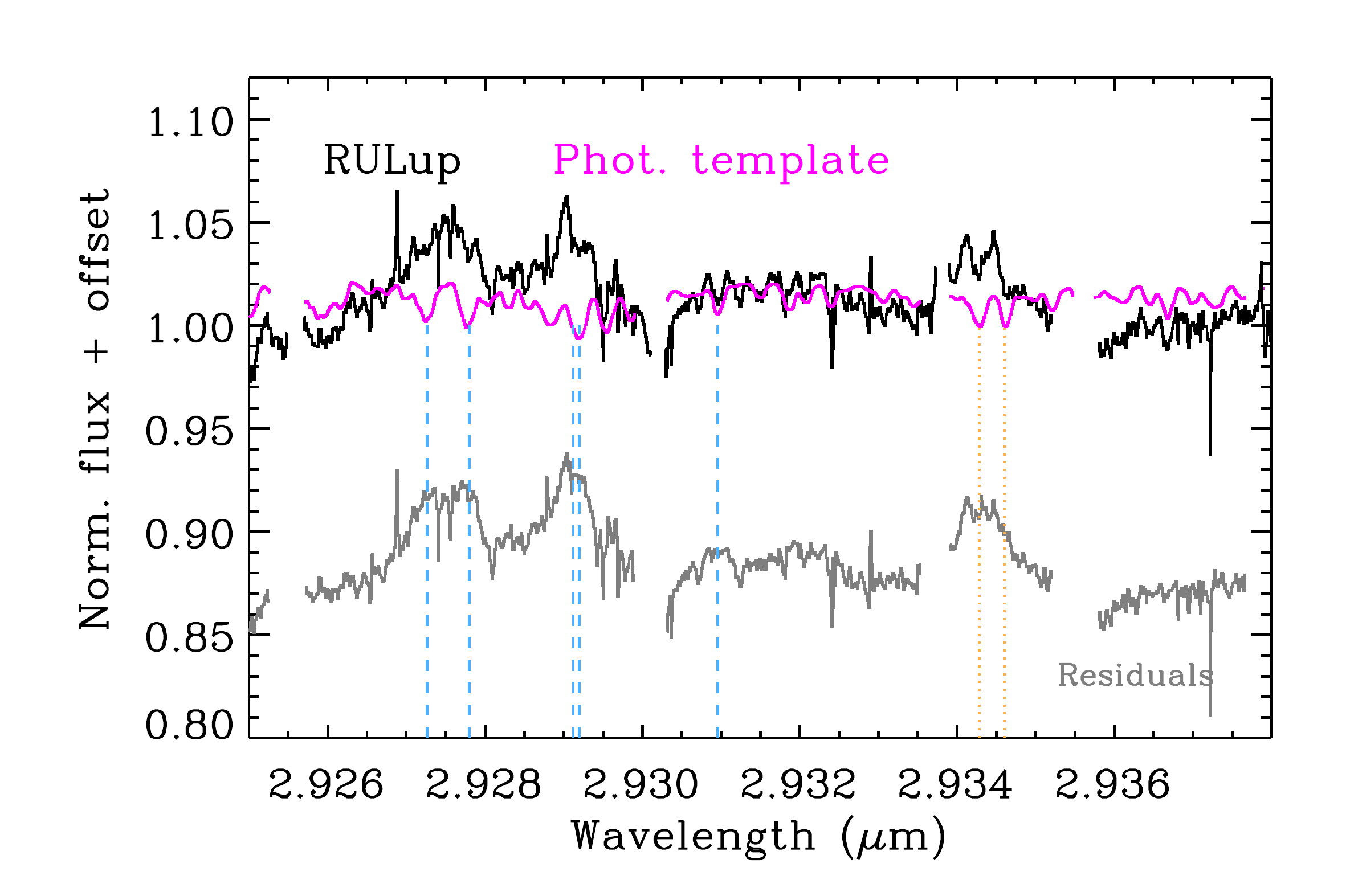} 
\includegraphics[width=0.5\textwidth]{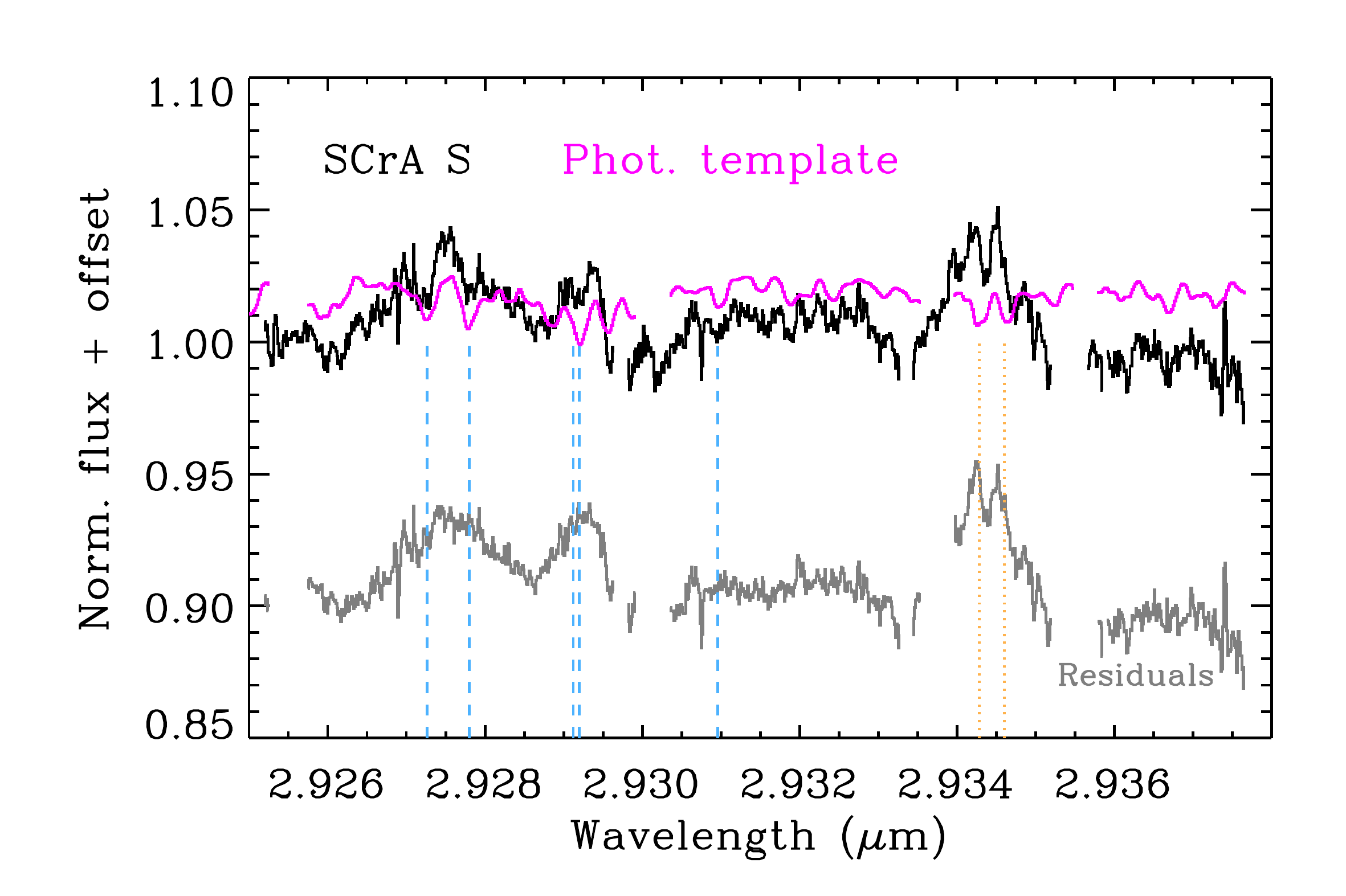} 
\includegraphics[width=0.5\textwidth]{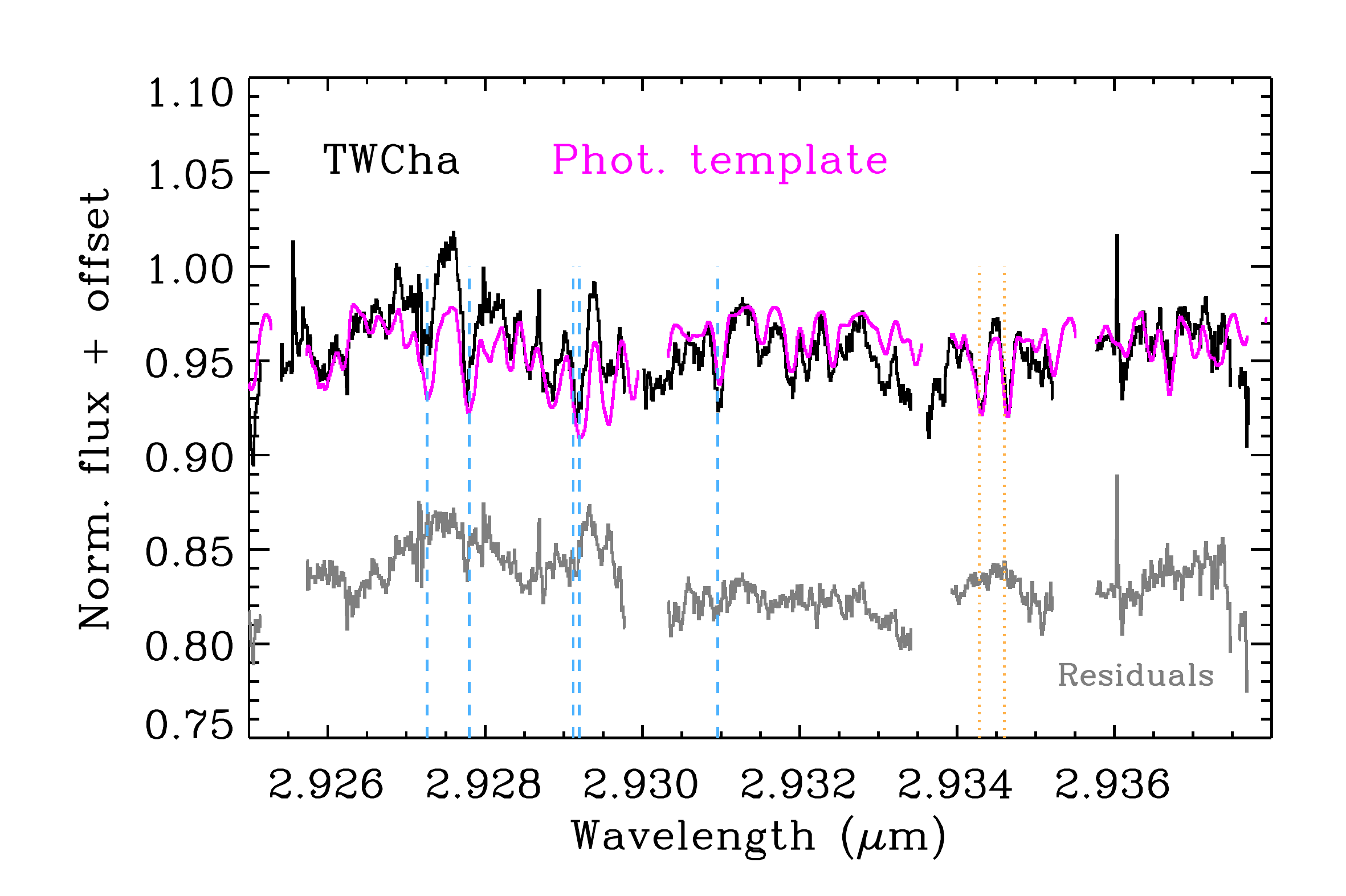} 
\includegraphics[width=0.5\textwidth]{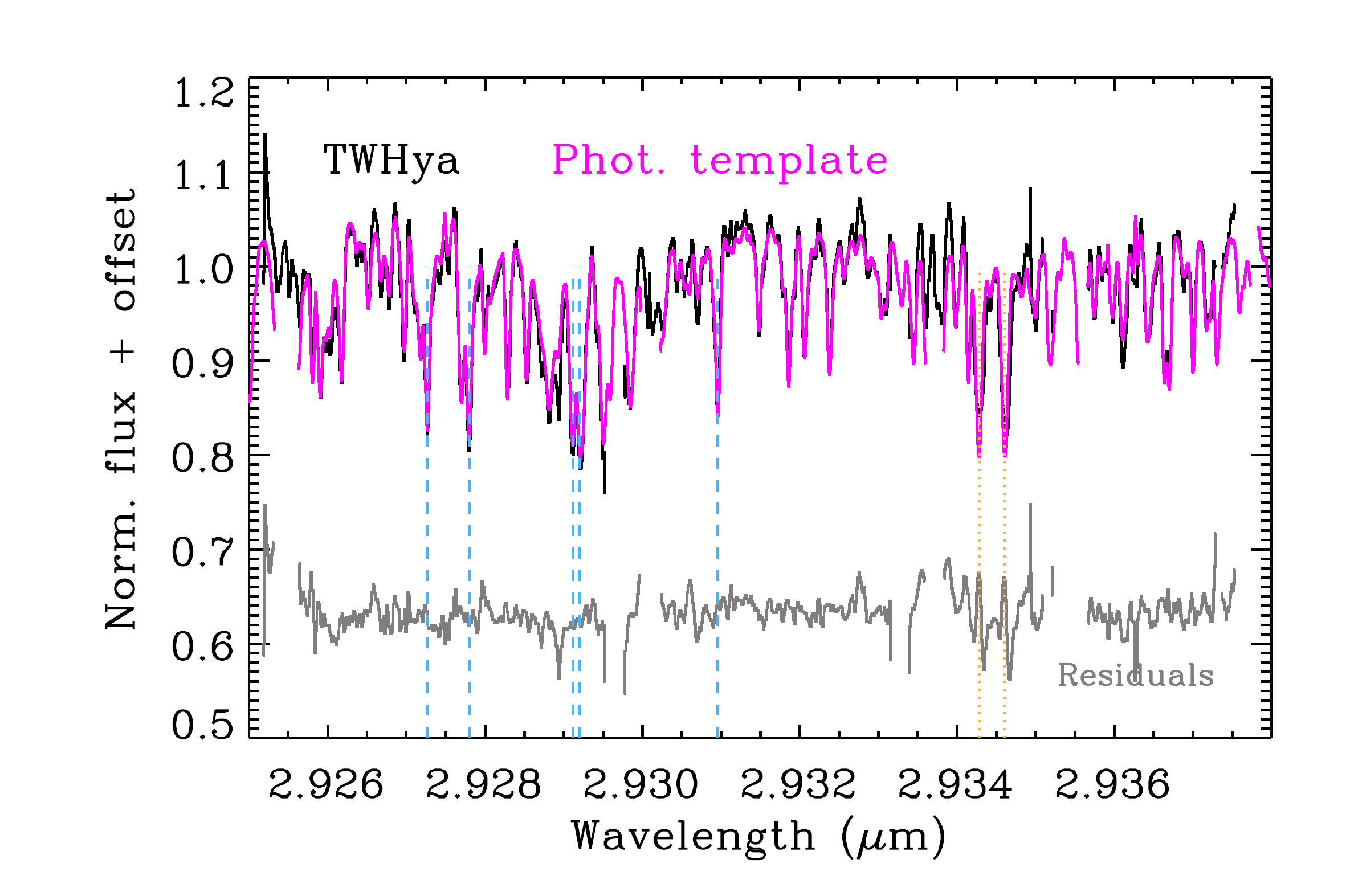} 
\includegraphics[width=0.5\textwidth]{VZCha_photcorr.pdf} 
\includegraphics[width=0.5\textwidth]{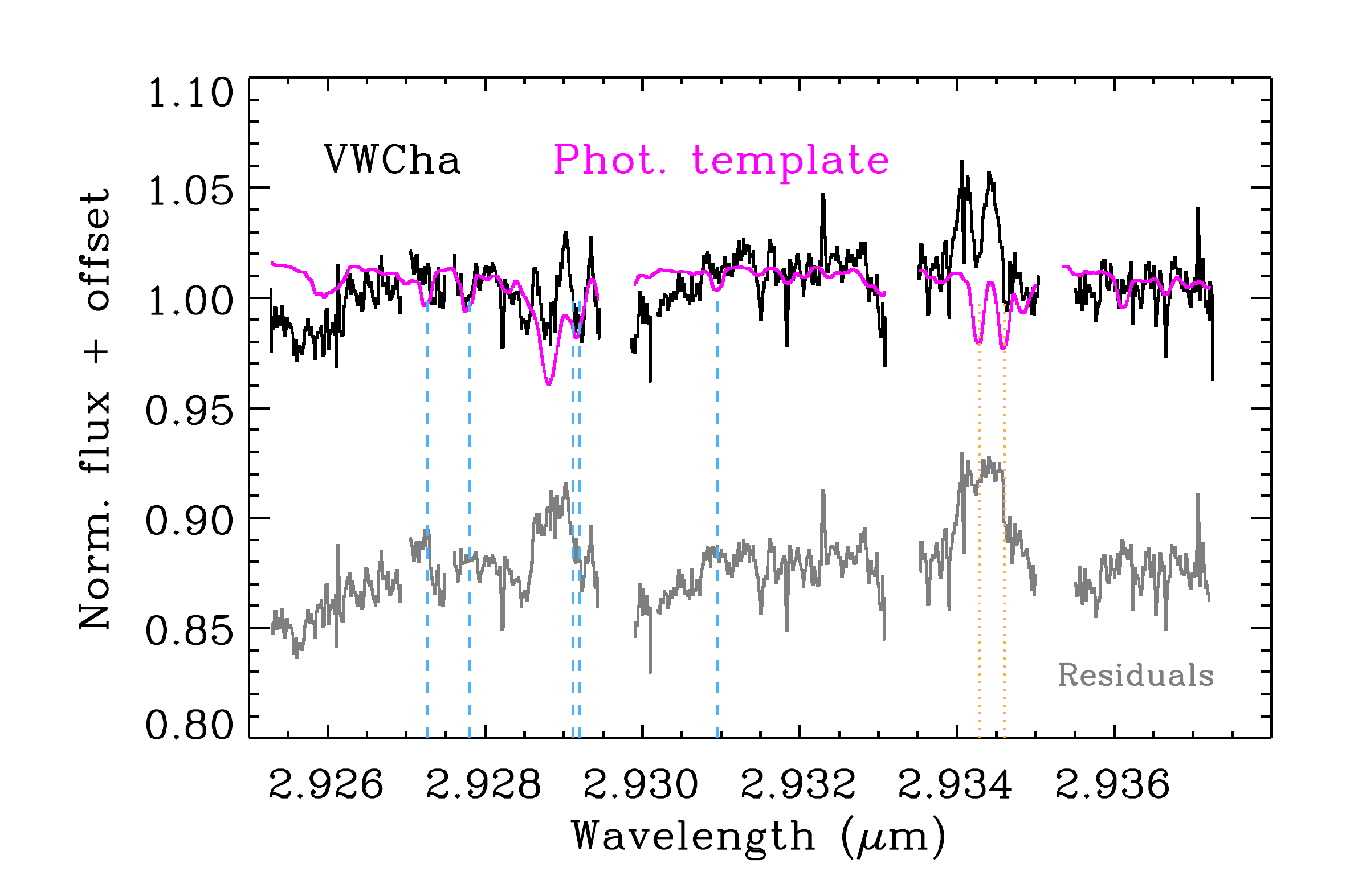} 
\includegraphics[width=0.5\textwidth]{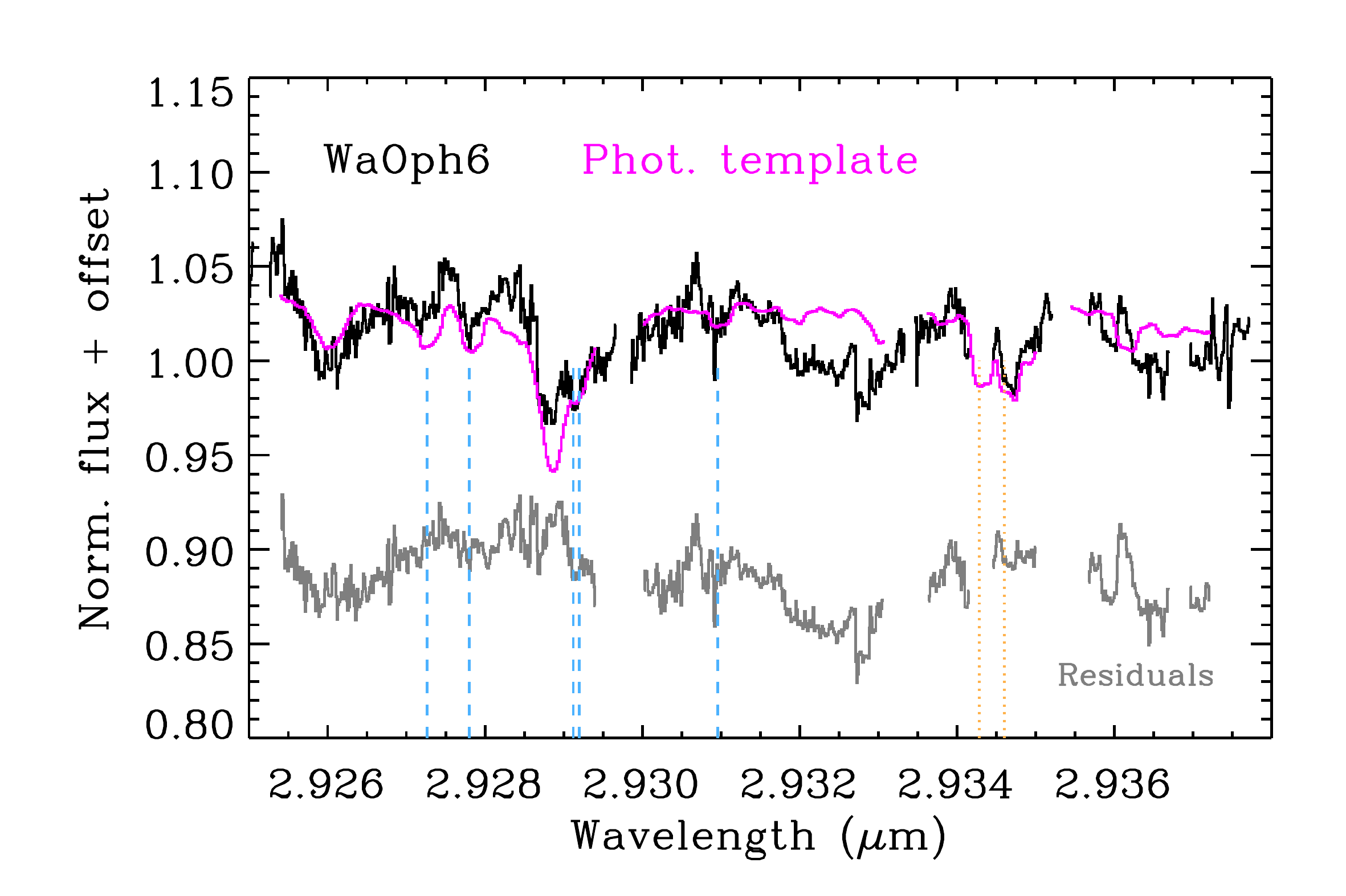} 
\includegraphics[width=0.5\textwidth]{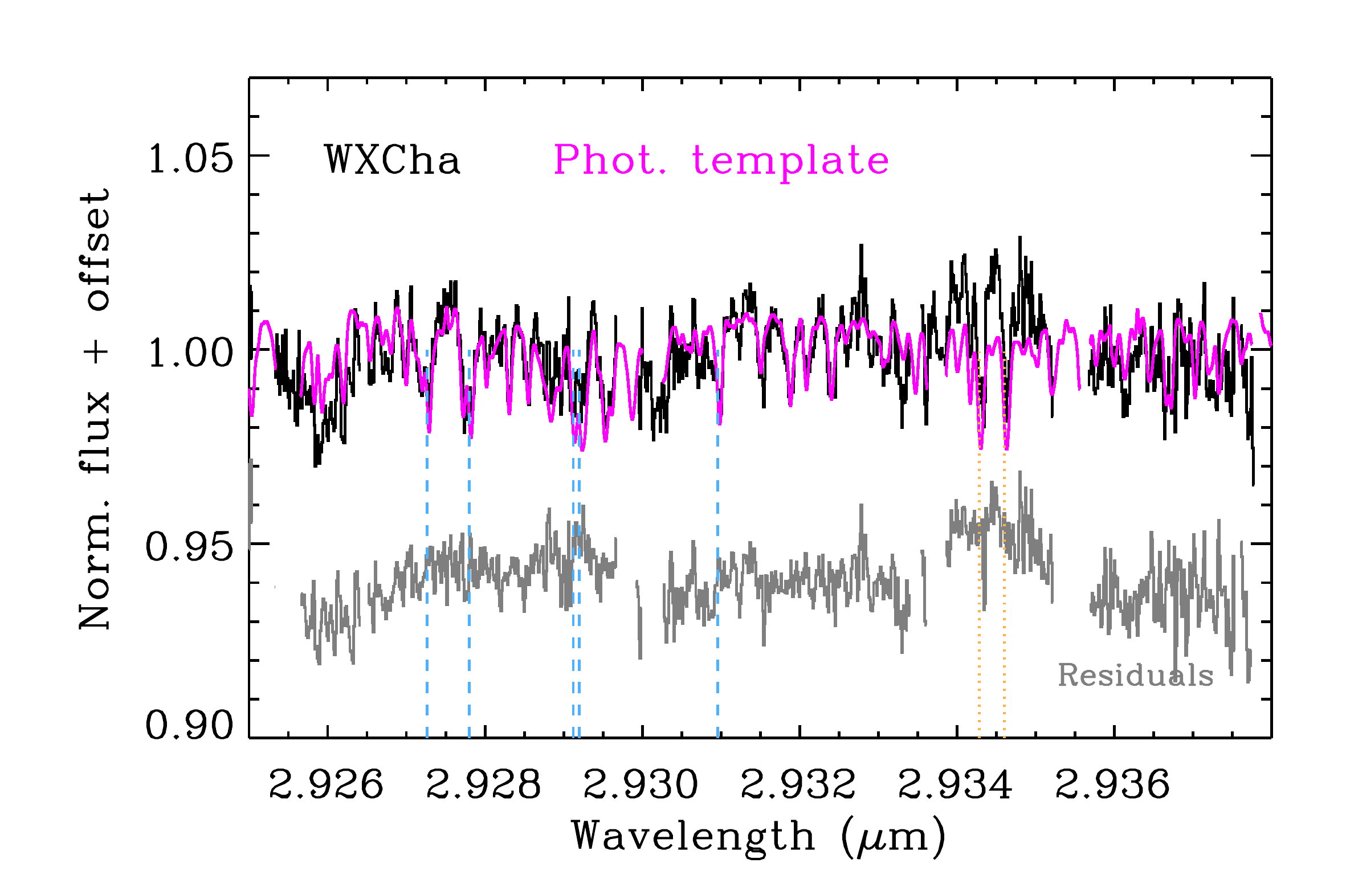} 
\caption{\textit{(Continued)} Photospheric correction of 2.9\,$\mu$m disk emission spectra observed by VLT-CRIRES.}
\label{fig: phot_corr_2}
\end{figure*}

\subsection{Fits to 2.9\,$\mu$m emission} \label{app:fit_plots}
\begin{figure*}
\includegraphics[width=0.5\textwidth]{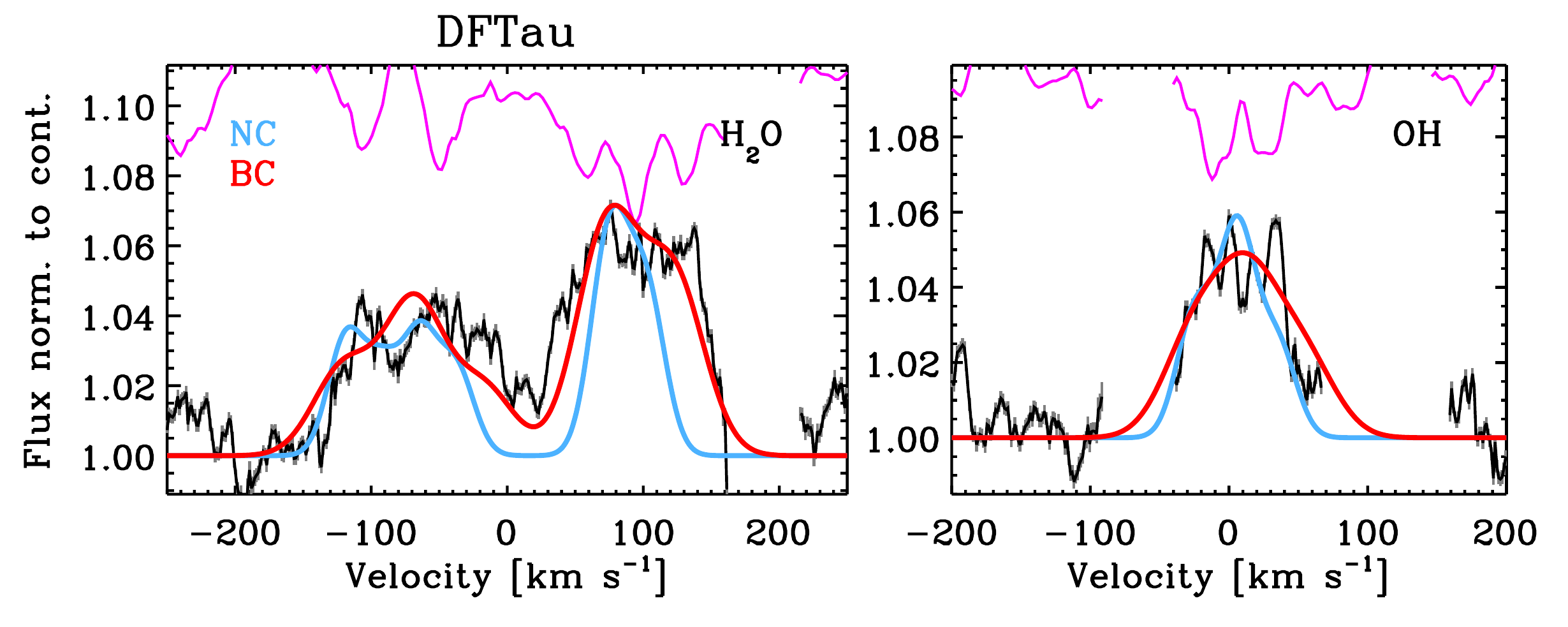} 
\includegraphics[width=0.5\textwidth]{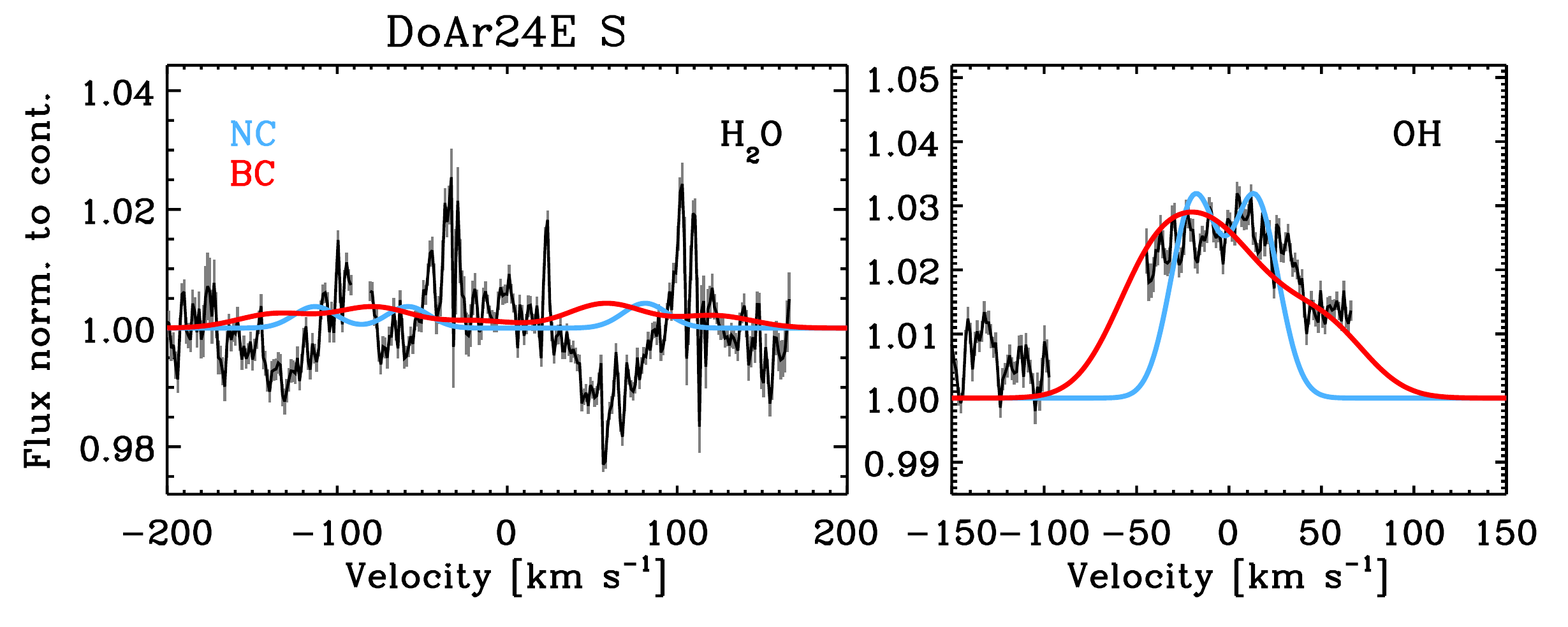} 
\includegraphics[width=0.5\textwidth]{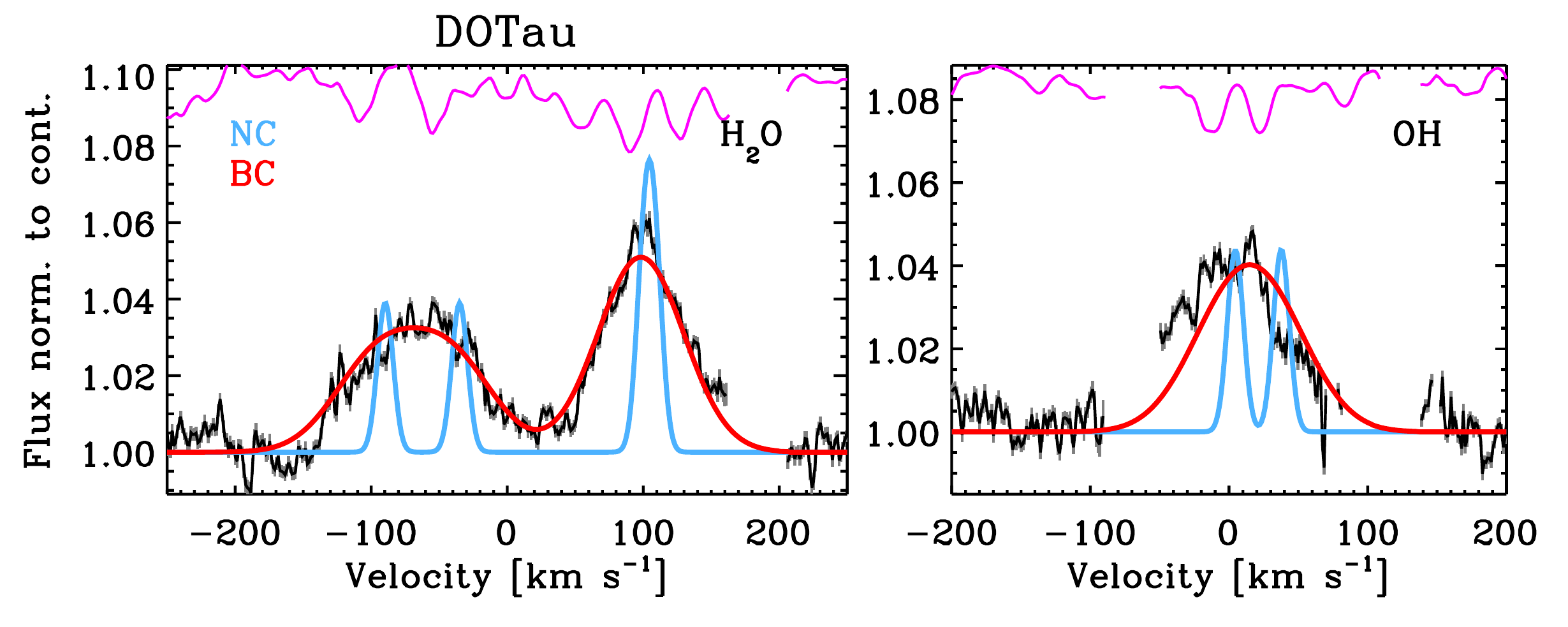} 
\includegraphics[width=0.5\textwidth]{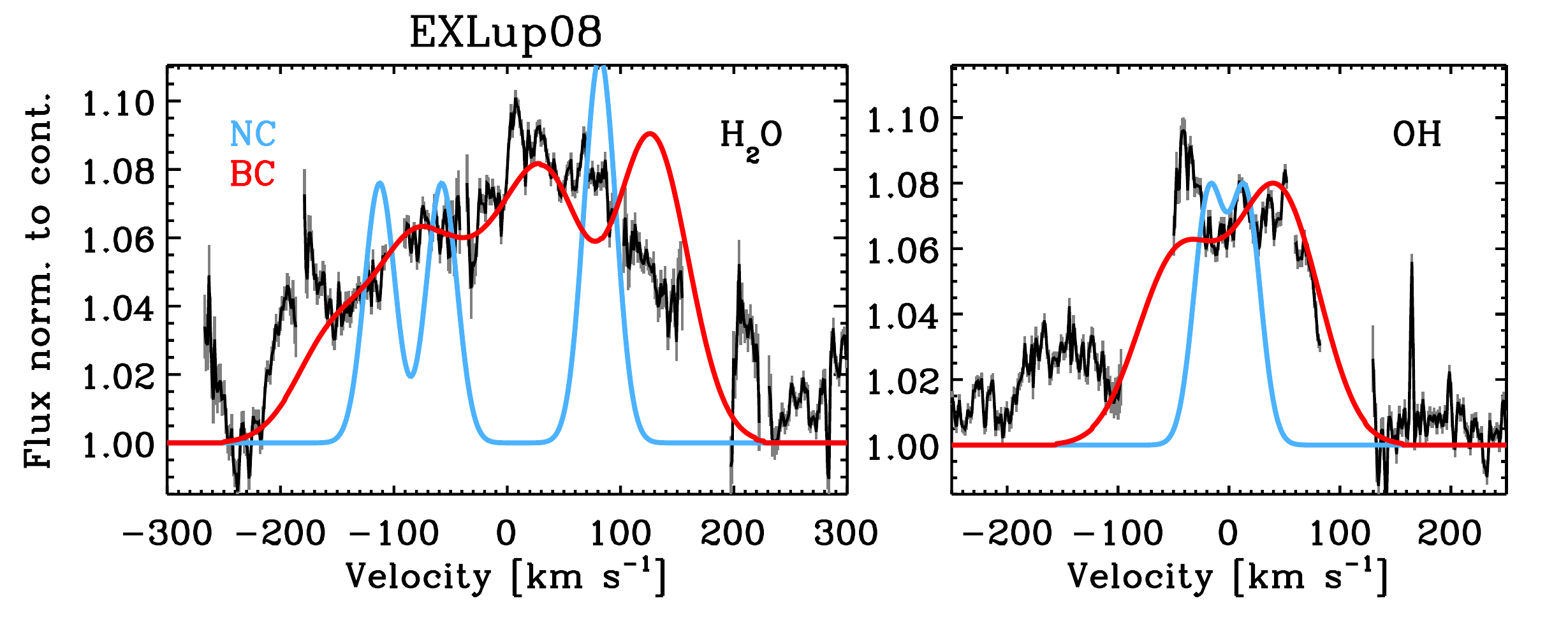} 
\includegraphics[width=0.5\textwidth]{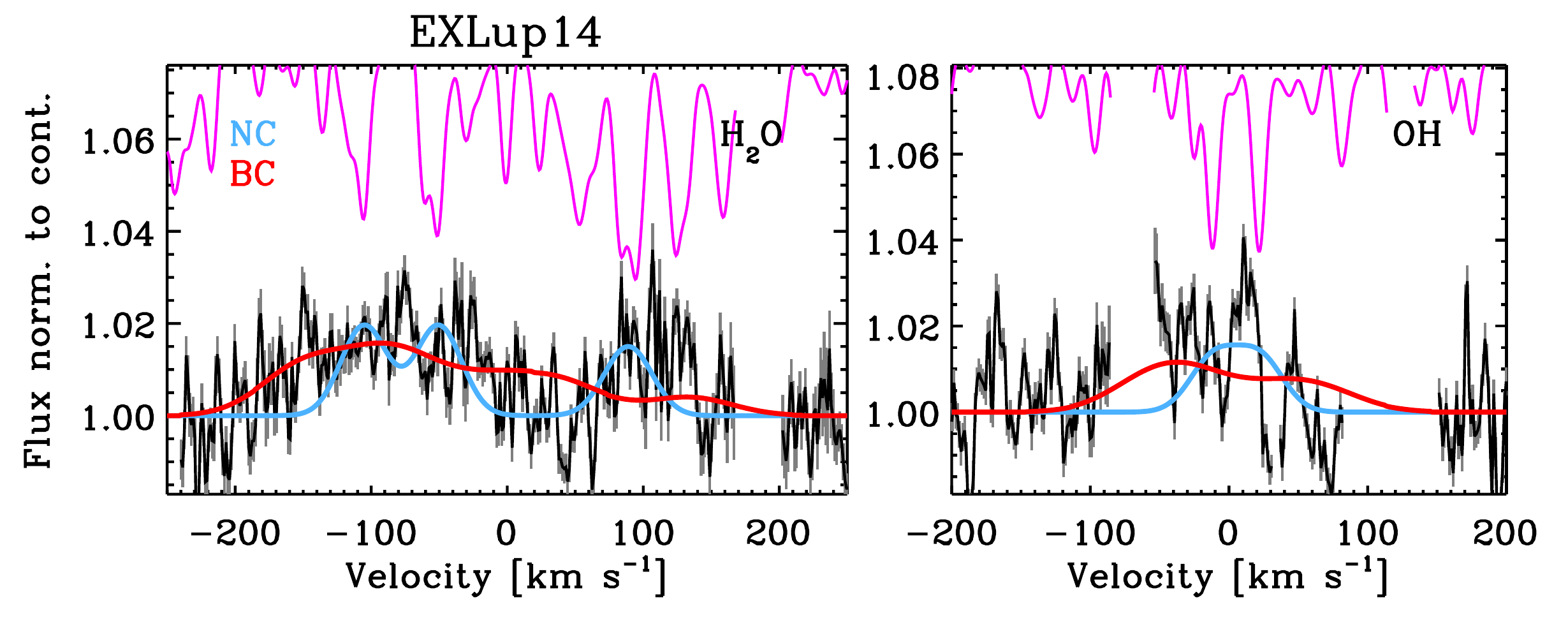} 
\includegraphics[width=0.5\textwidth]{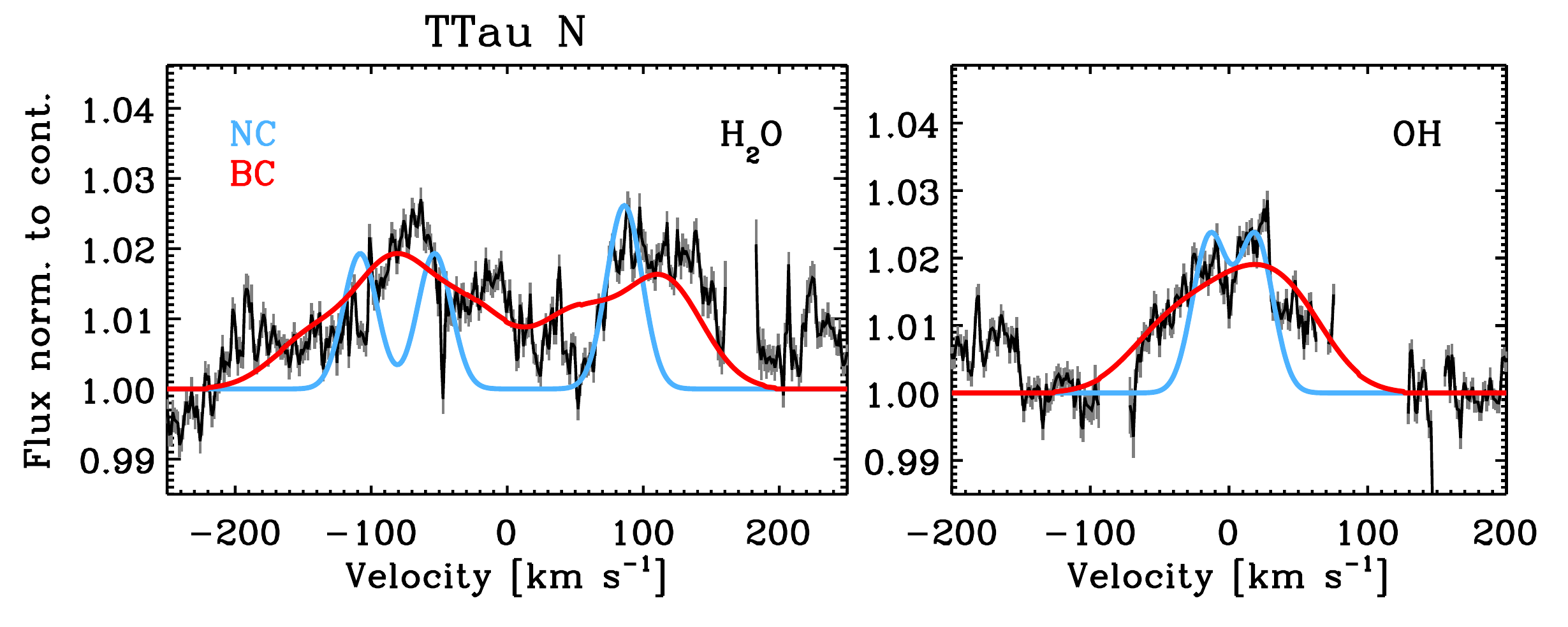} 
\includegraphics[width=0.5\textwidth]{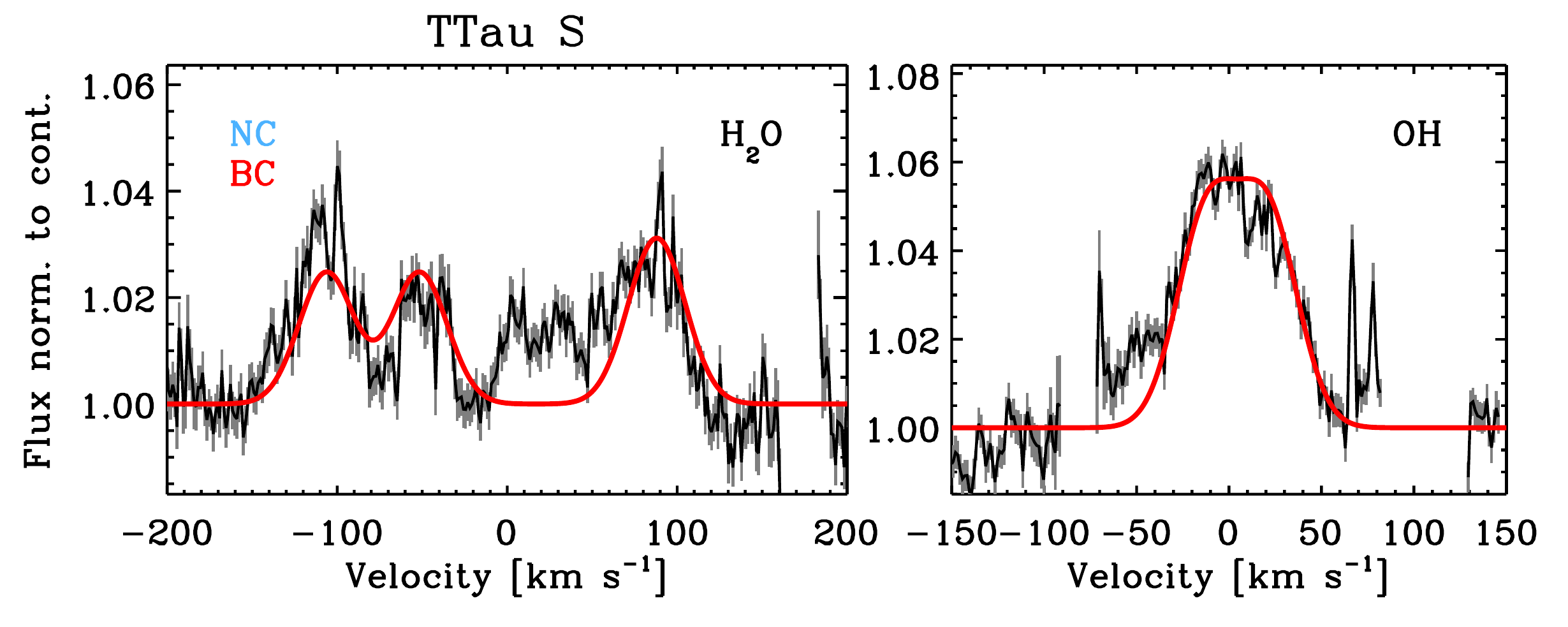} 
\includegraphics[width=0.5\textwidth]{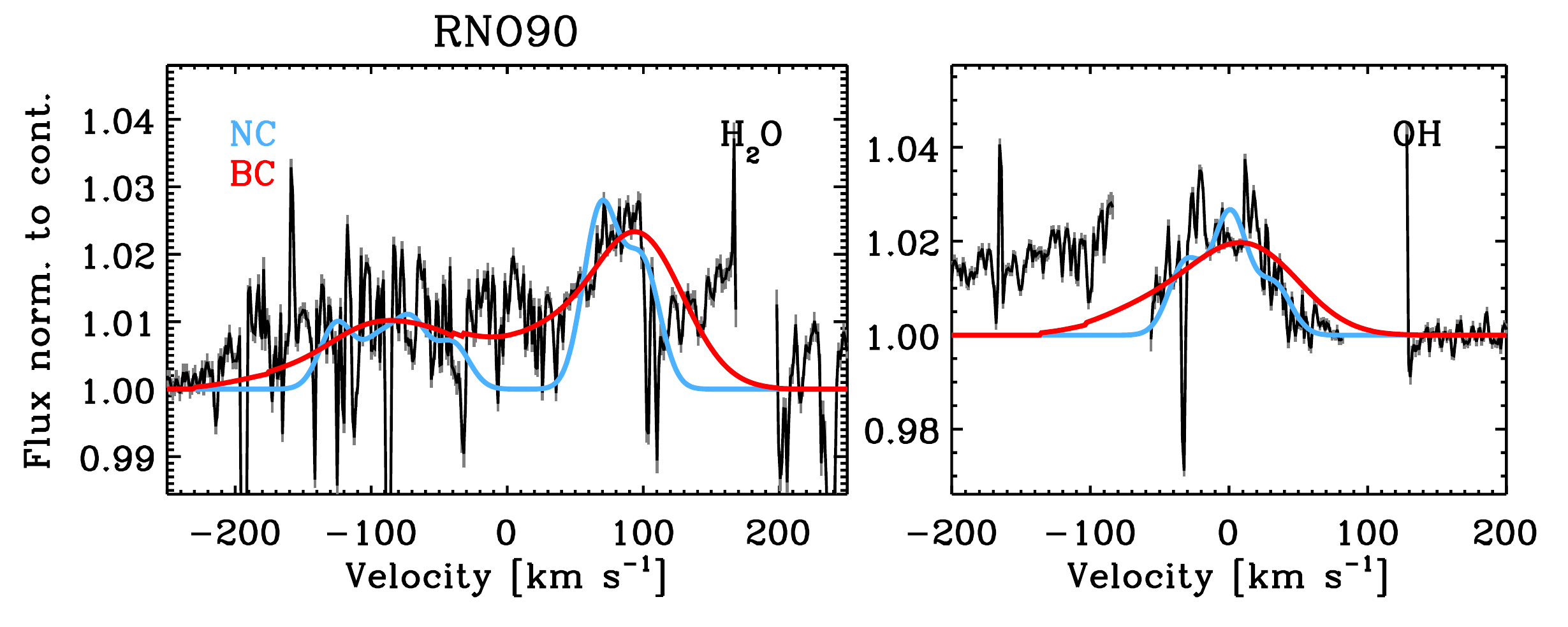} 
\includegraphics[width=0.5\textwidth]{VZCha_line_comp_mod.pdf} 
\includegraphics[width=0.5\textwidth]{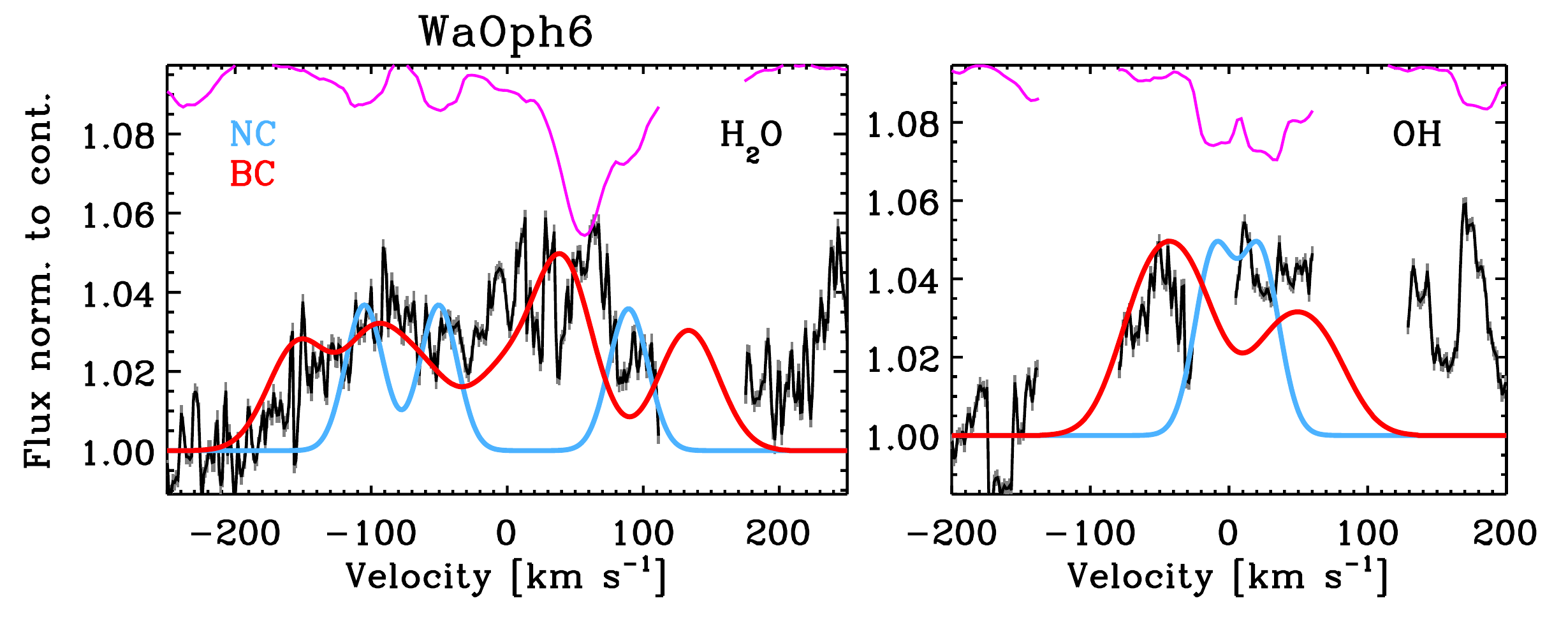} 
\includegraphics[width=0.5\textwidth]{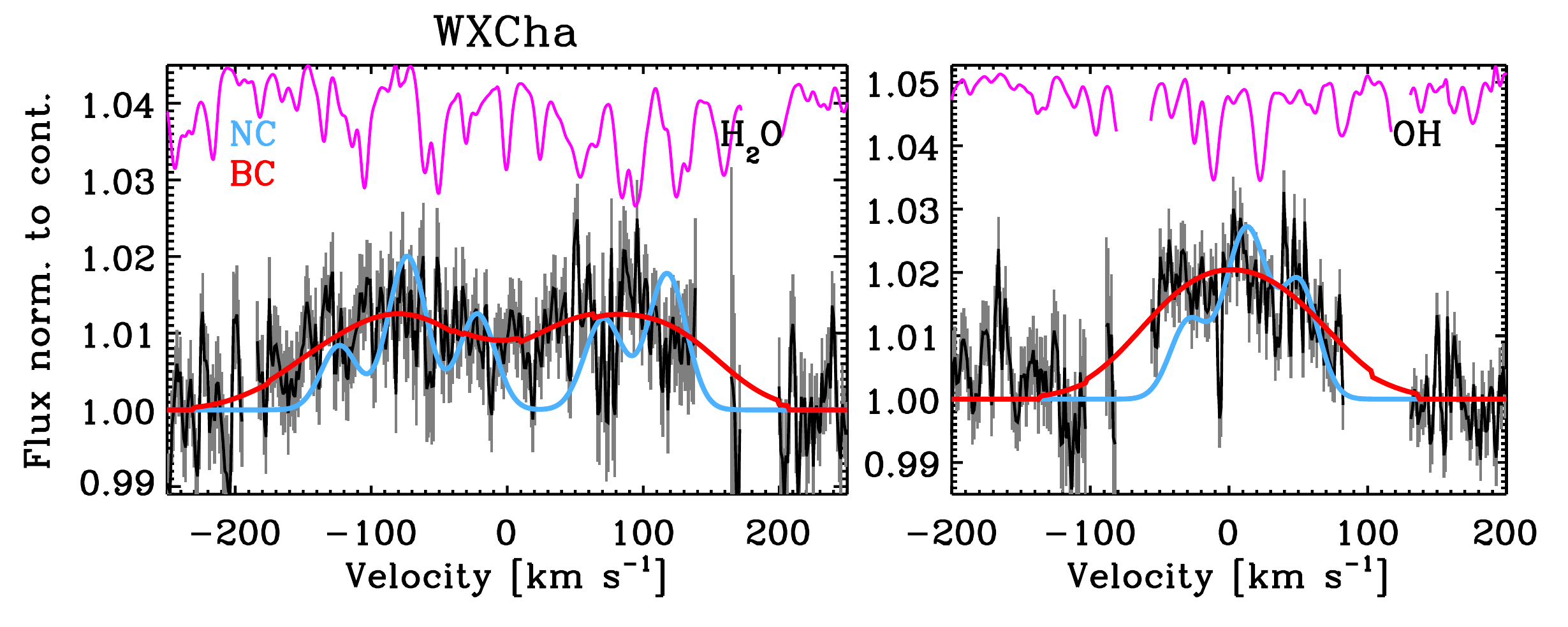} 
\includegraphics[width=0.5\textwidth]{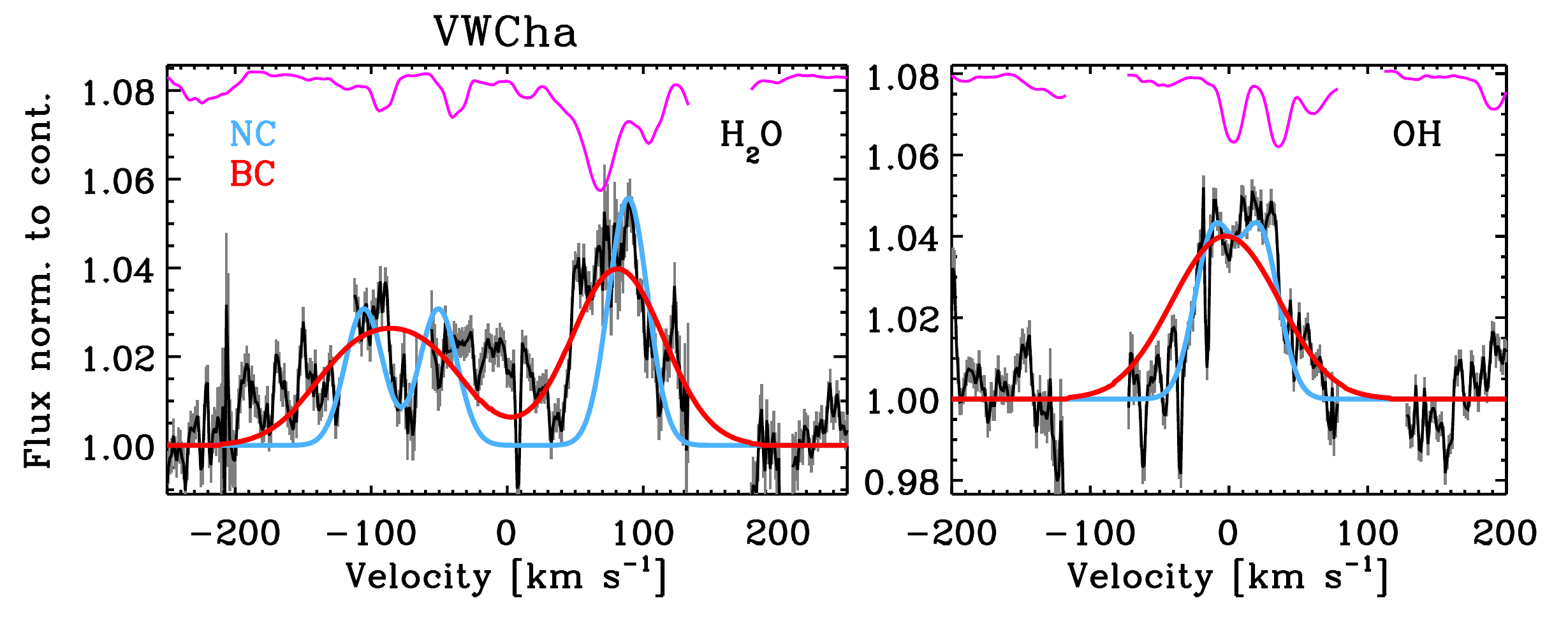} 
\caption{Fits to H$_{2}$O and OH line profiles at 2.9\,$\mu$m by using the BC (red) and NC (cyan) profiles from CO emission. Spectra where photospheric correction has been applied are shown with the photospheric template in magenta at the top, broadened and veiled. The NC in TTau\,S is not available due to a central broad absorption component \citep[see][]{brown13}.}
\label{fig: mol eq BC}
\end{figure*}

\begin{figure*}
\includegraphics[width=0.5\textwidth]{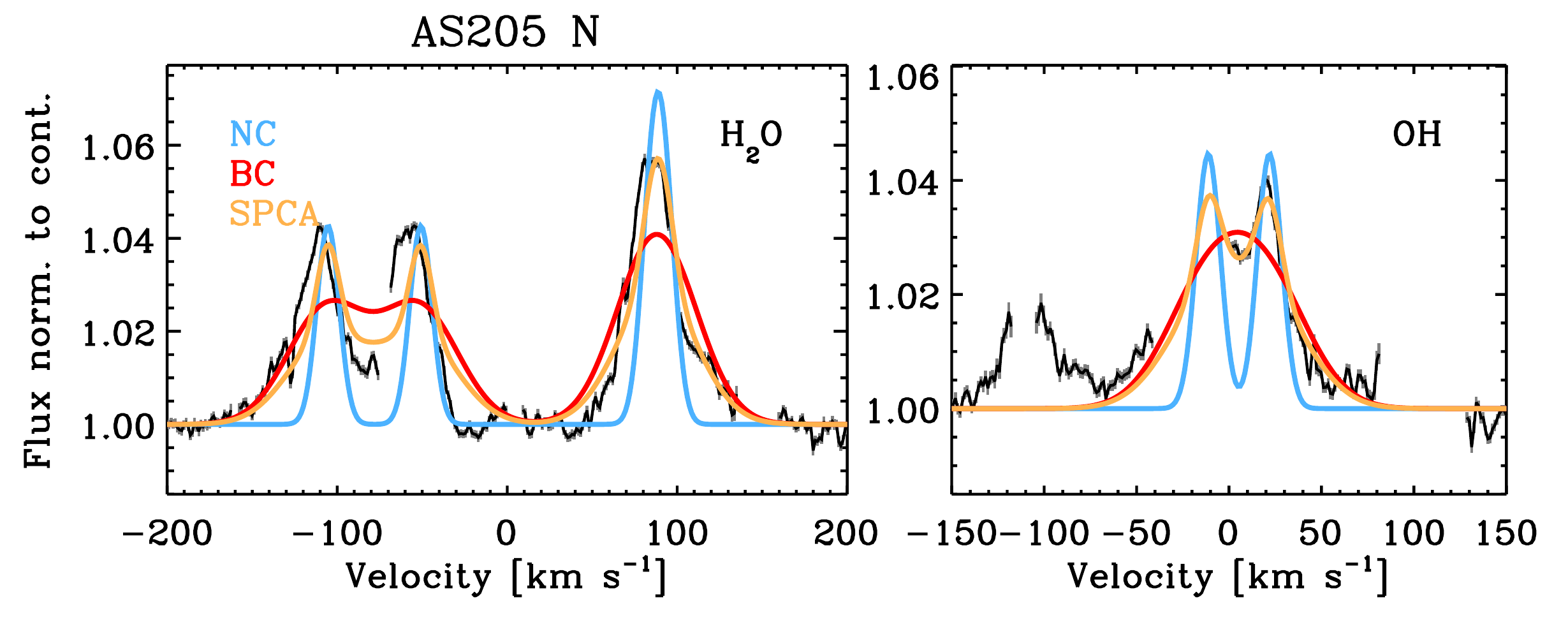} 
\includegraphics[width=0.5\textwidth]{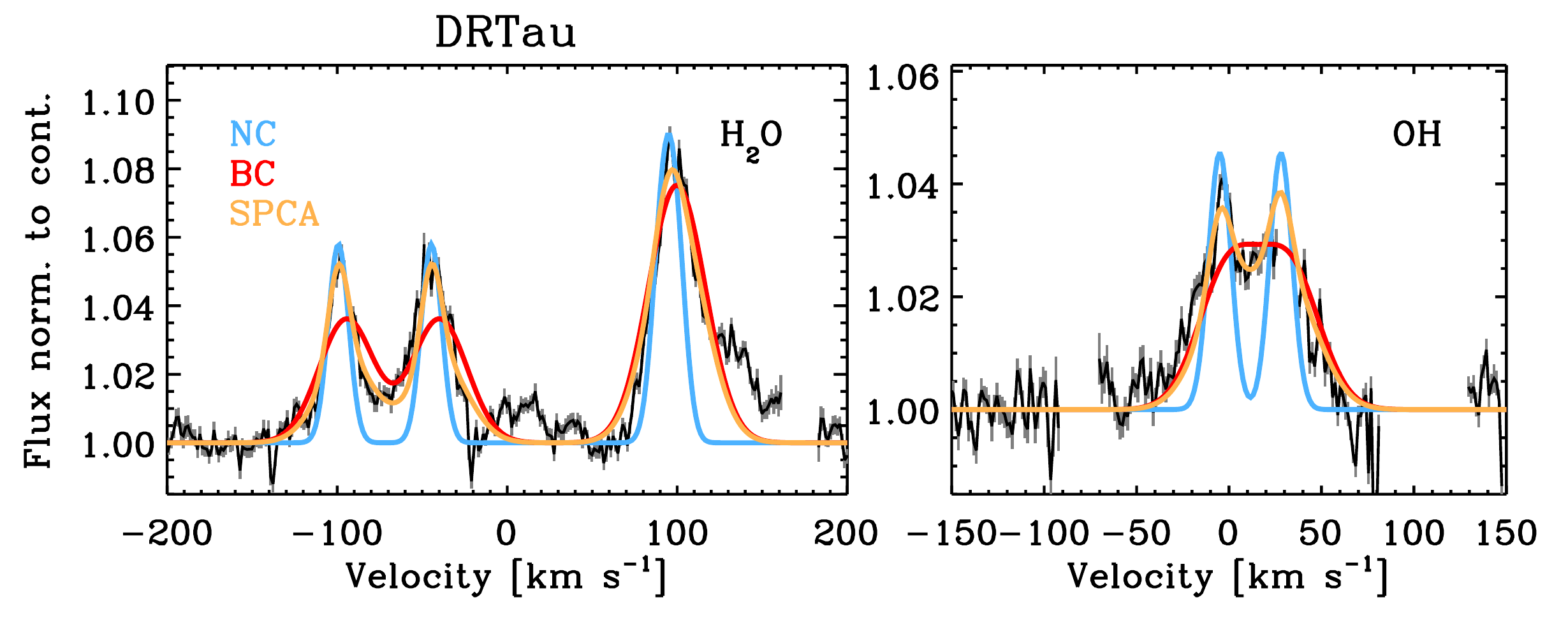} 
\includegraphics[width=0.5\textwidth]{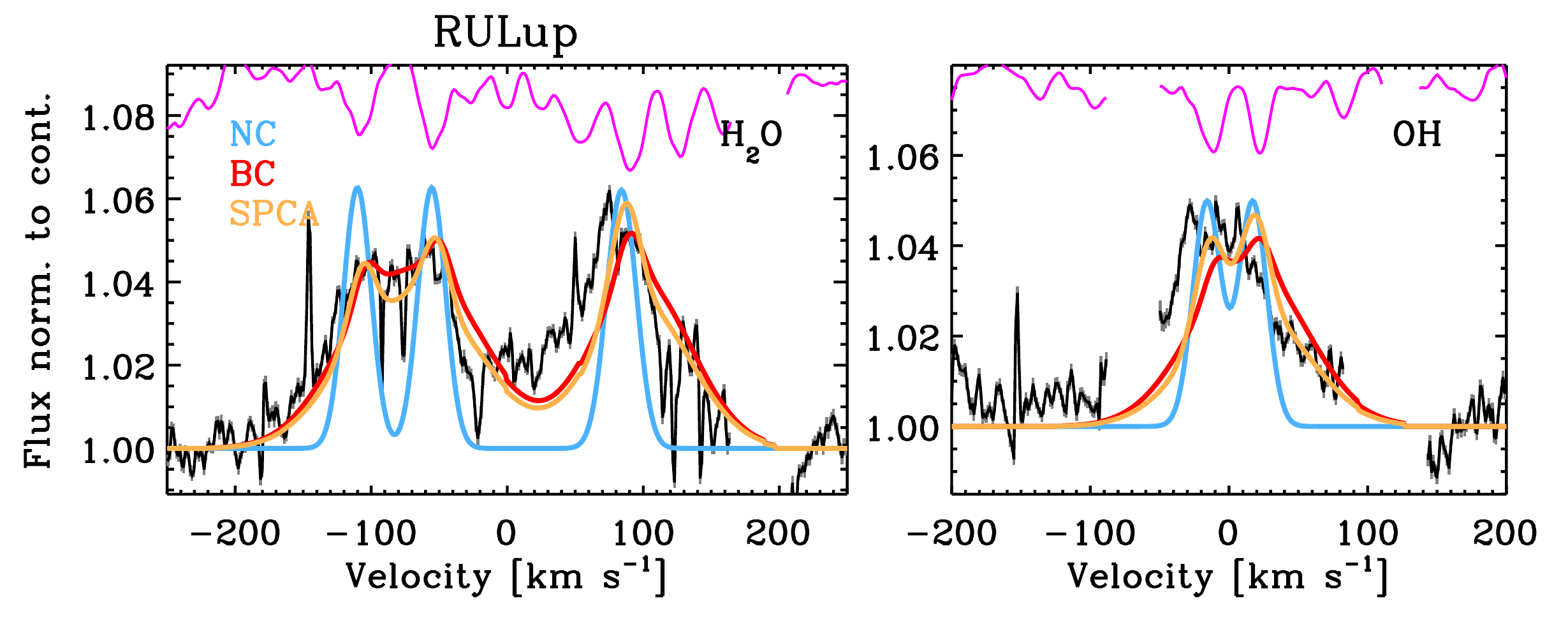} 
\includegraphics[width=0.5\textwidth]{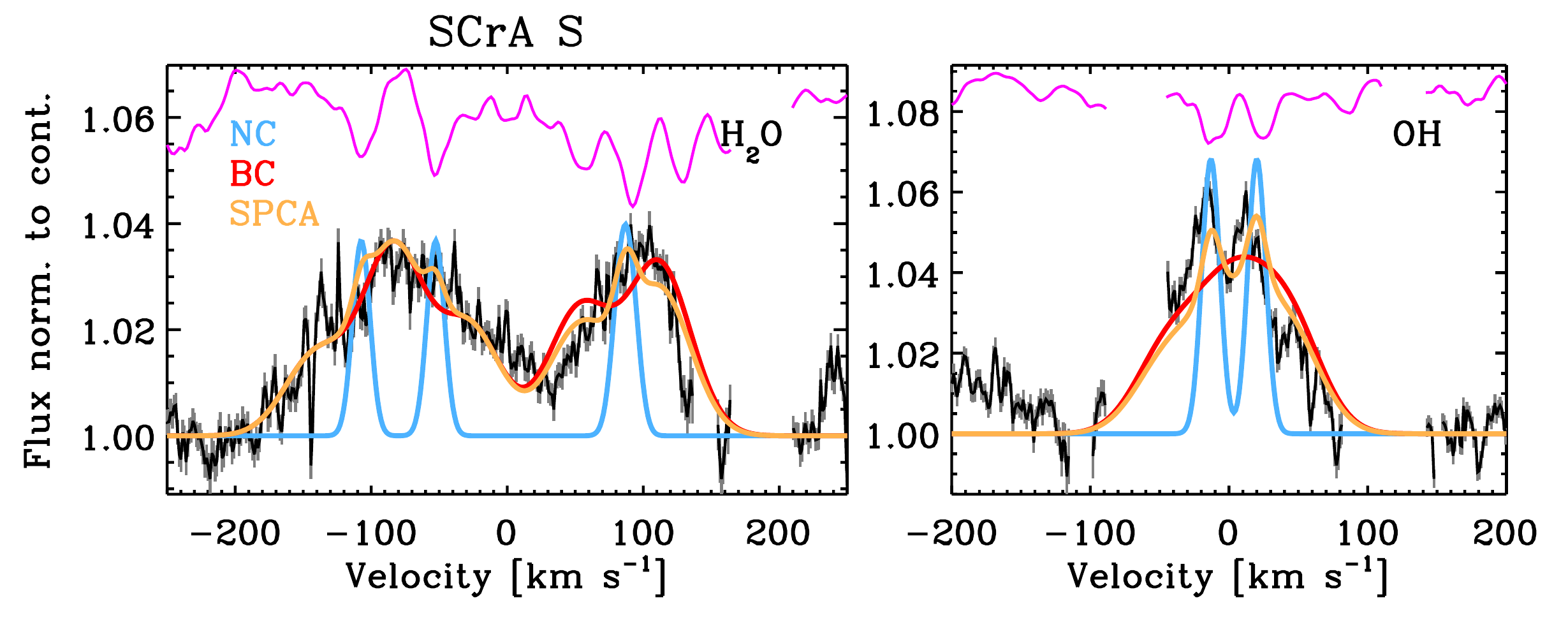} 
\includegraphics[width=0.5\textwidth]{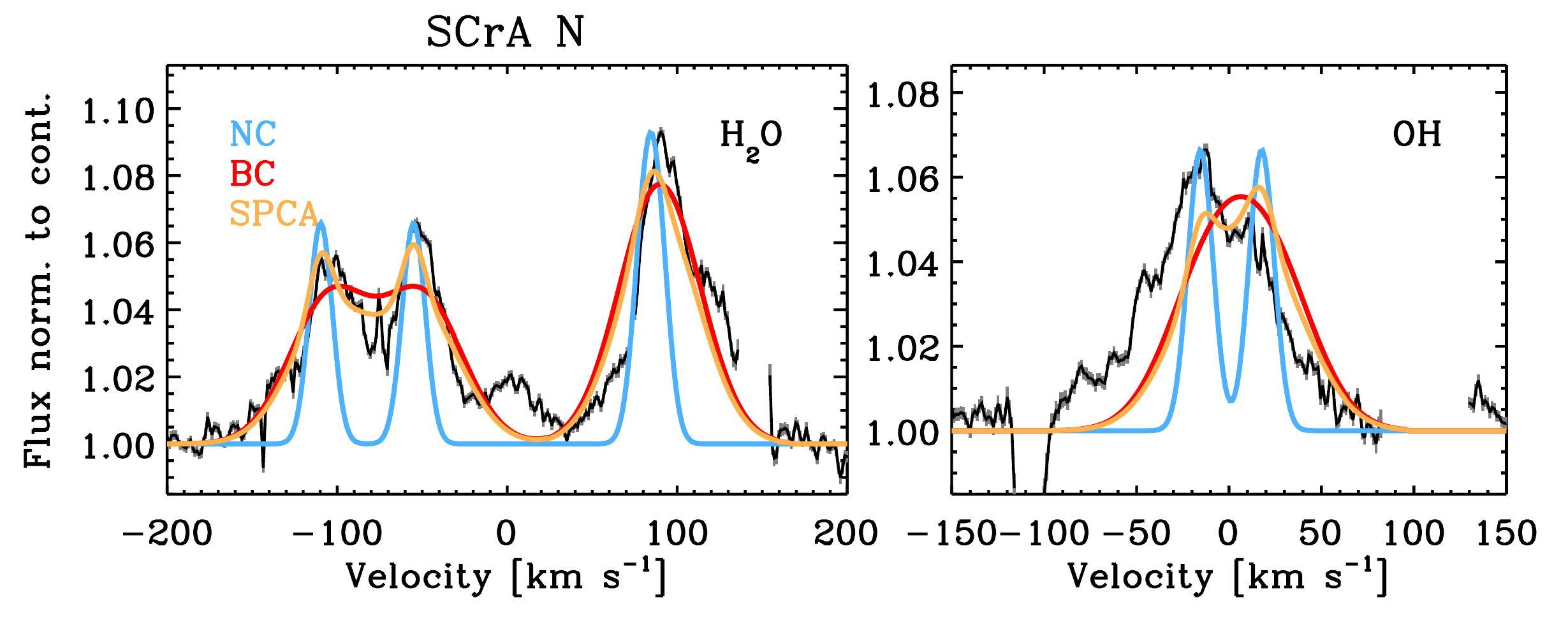} 
\includegraphics[width=0.5\textwidth]{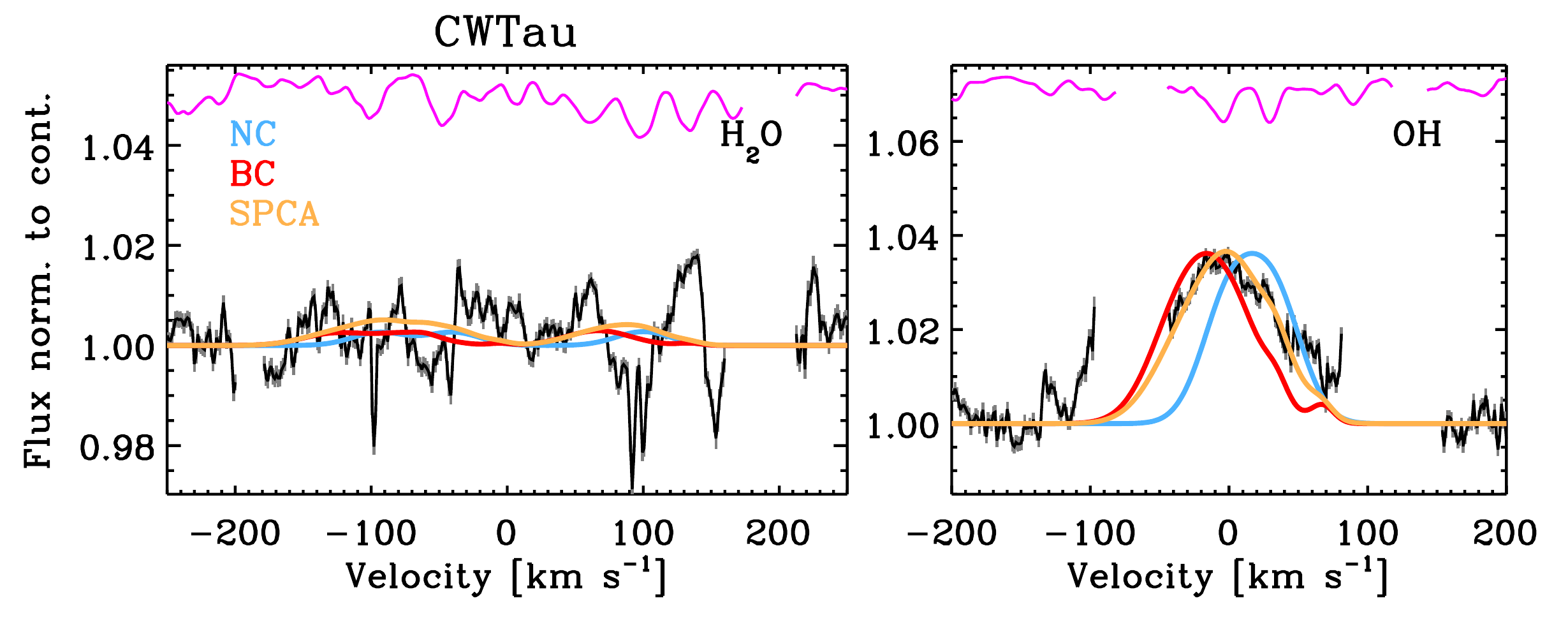} 
\caption{H$_{2}$O and OH lines at 2.9\,$\mu$m where a weak NC is detected on top of BC in H$_{2}$O and/or OH emission. In the spectrum of CW\,Tau, only OH is detected. The detection of NC is only tentative in the case of RU\,Lup and SCrA\,S, due to possible photospheric residuals. The photospheric template, where used for photospheric correction, is showed in magenta at the top, broadened and veiled. Line fits follow the same color code as in Figure \ref{fig: mol eq BC}, and the SPCA fit is shown in orange (see Section A\ref{app:fit_plots}).}
\label{fig: mol SPCA}
\end{figure*}

\begin{figure*}
\includegraphics[width=0.5\textwidth]{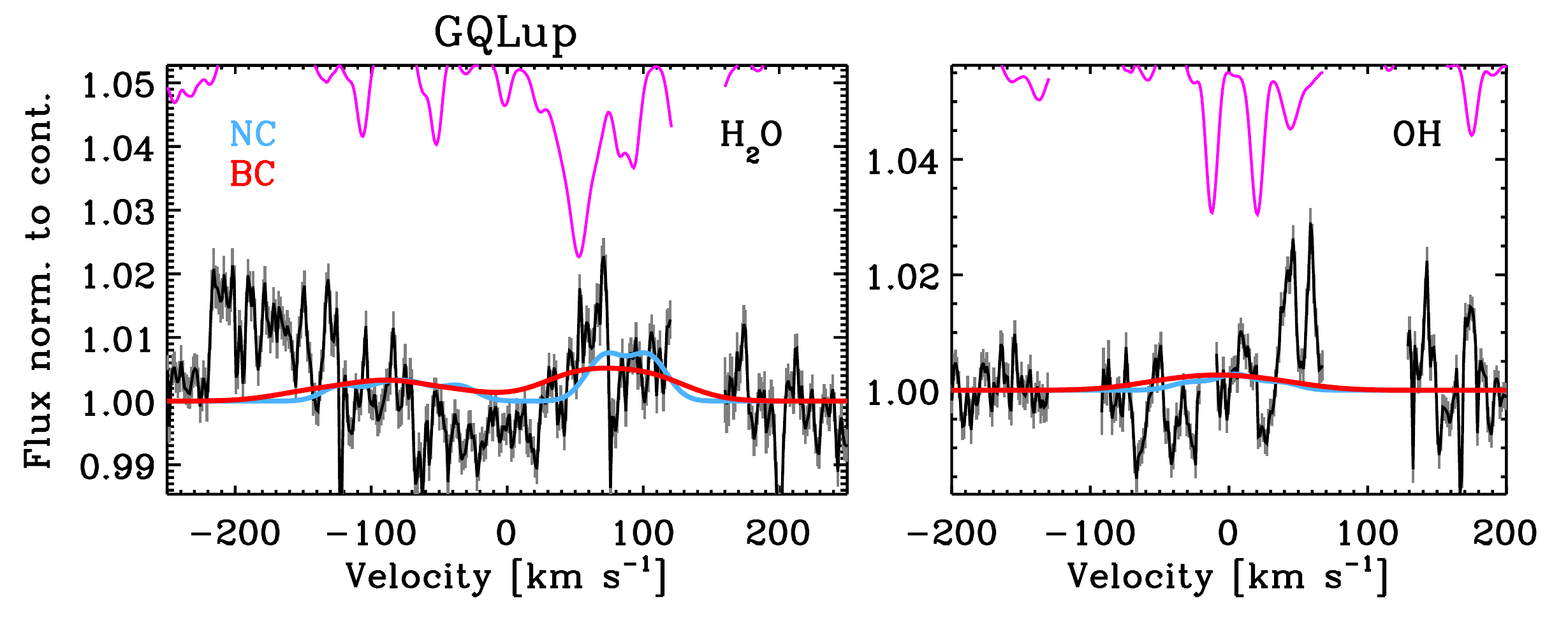} 
\includegraphics[width=0.5\textwidth]{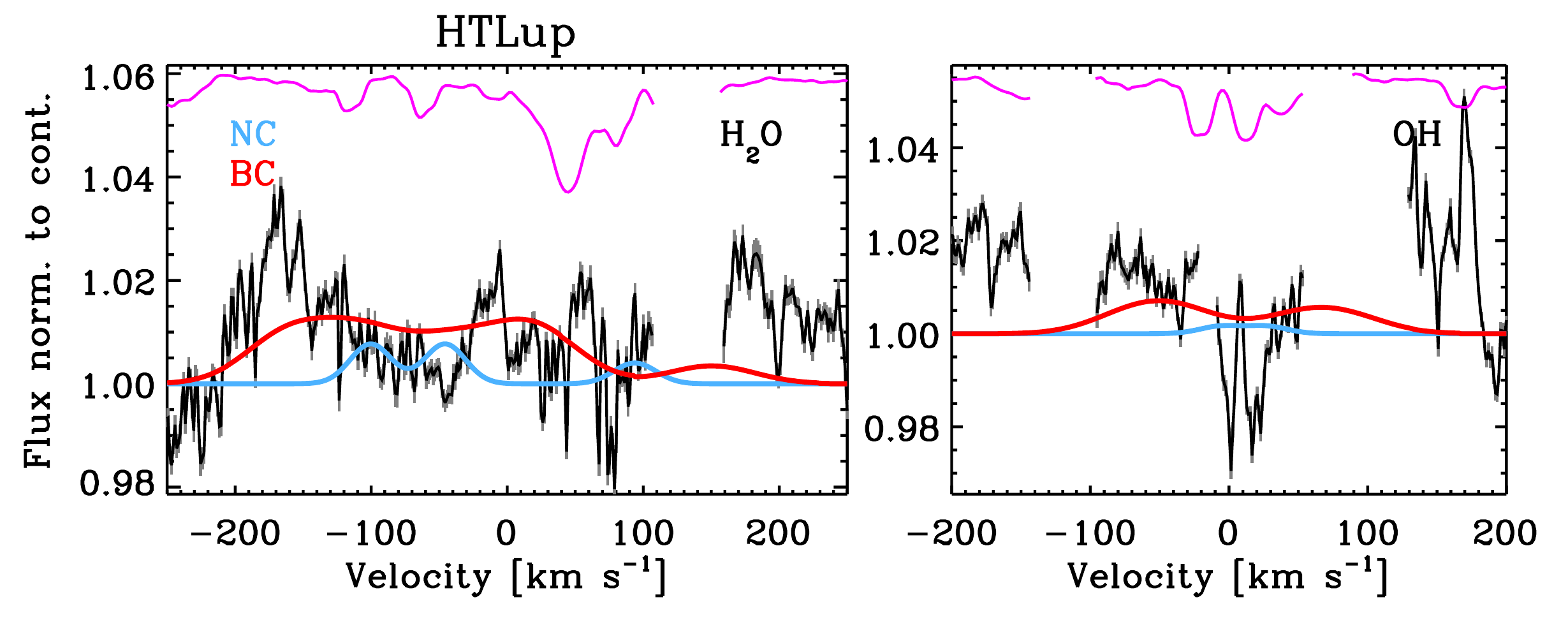} 
\includegraphics[width=0.5\textwidth]{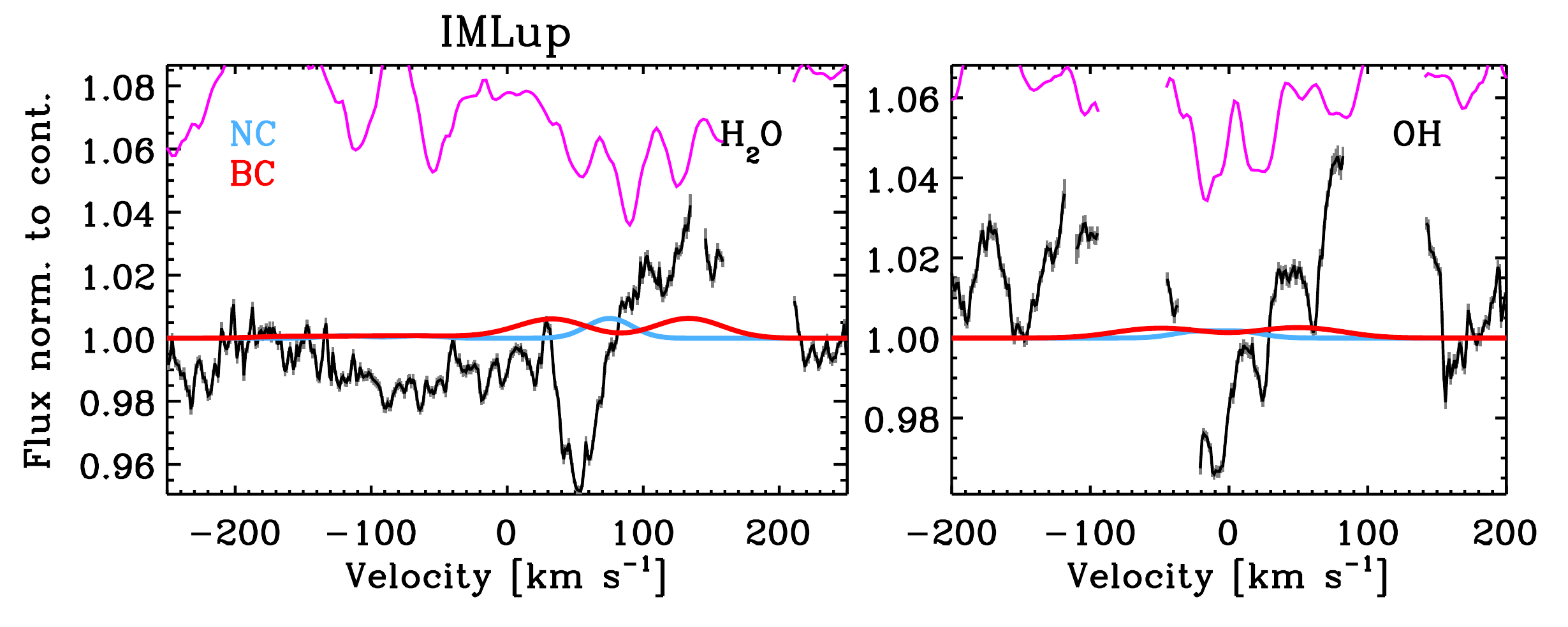} 
\caption{2.9\,$\mu$m spectra with photospheric residuals due to ambiguous/inappropriate match with the spectral type of the photospheric template. The photospheric template is showed in magenta at the top, broadened and veiled. Line fits follow the same color code as in Figure \ref{fig: mol eq BC}.}
\label{fig: mol unclear}
\end{figure*}

\begin{figure*}
\includegraphics[width=0.5\textwidth]{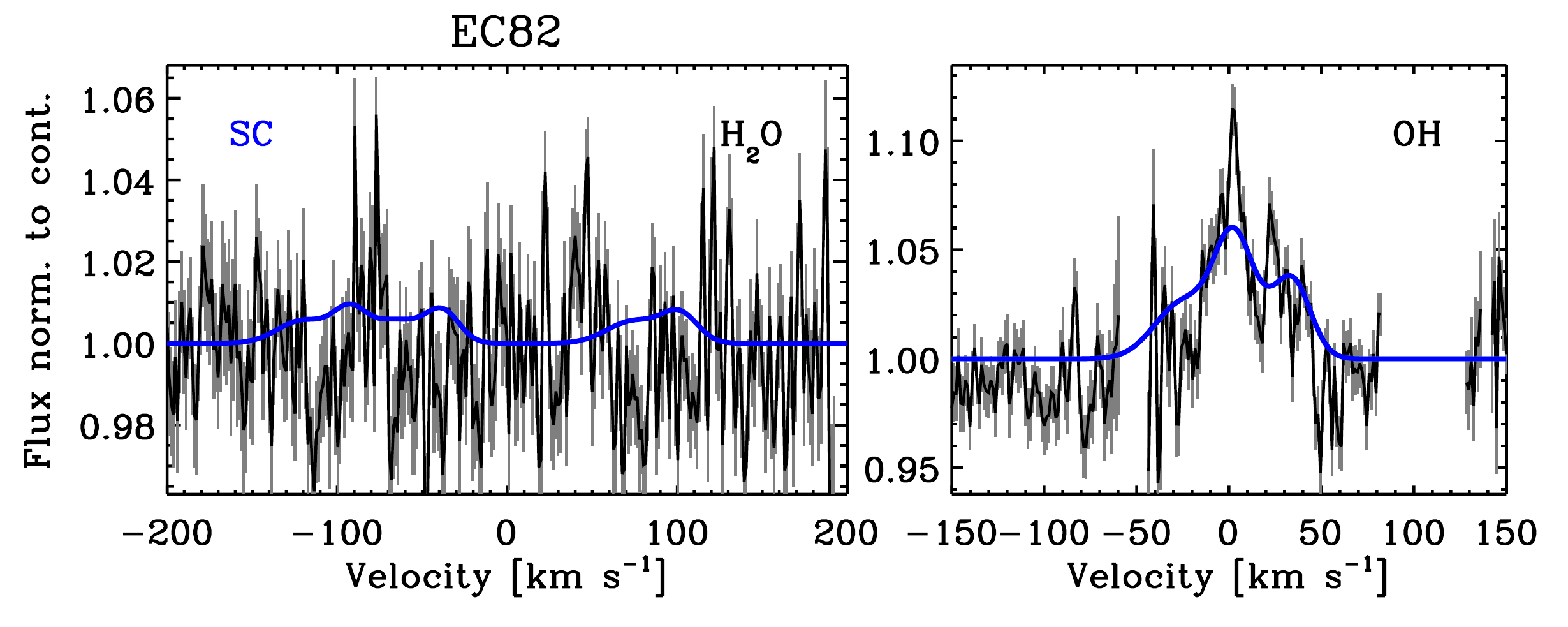} 
\includegraphics[width=0.5\textwidth]{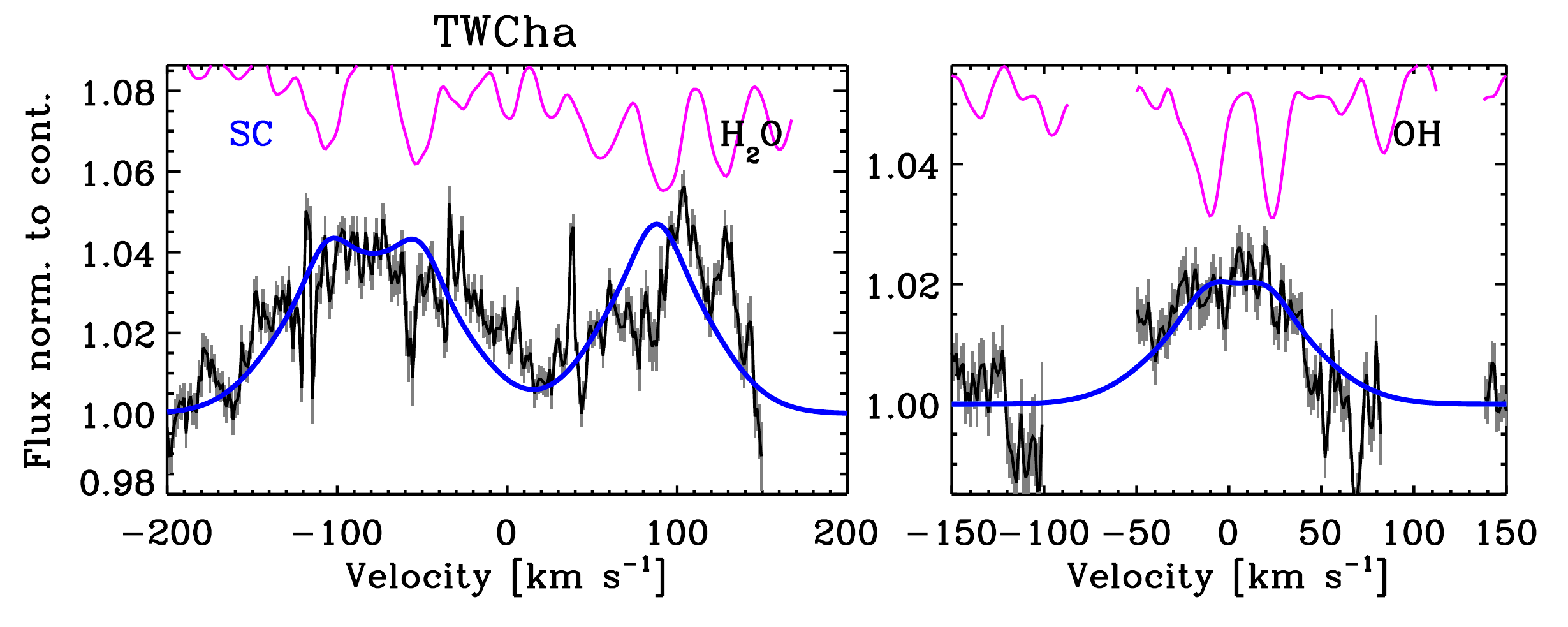} 
\caption{Single-component disks where H$_{2}$O and/or OH emission is detected at 2.9\,$\mu$m. Line fits include only the single CO component detected in the spectra (SC). The photospheric template, where used for photospheric correction, is showed in magenta at the top, broadened and veiled.}
\label{fig: mol eq CO1-0}
\end{figure*}

\begin{figure*}
\includegraphics[width=0.5\textwidth]{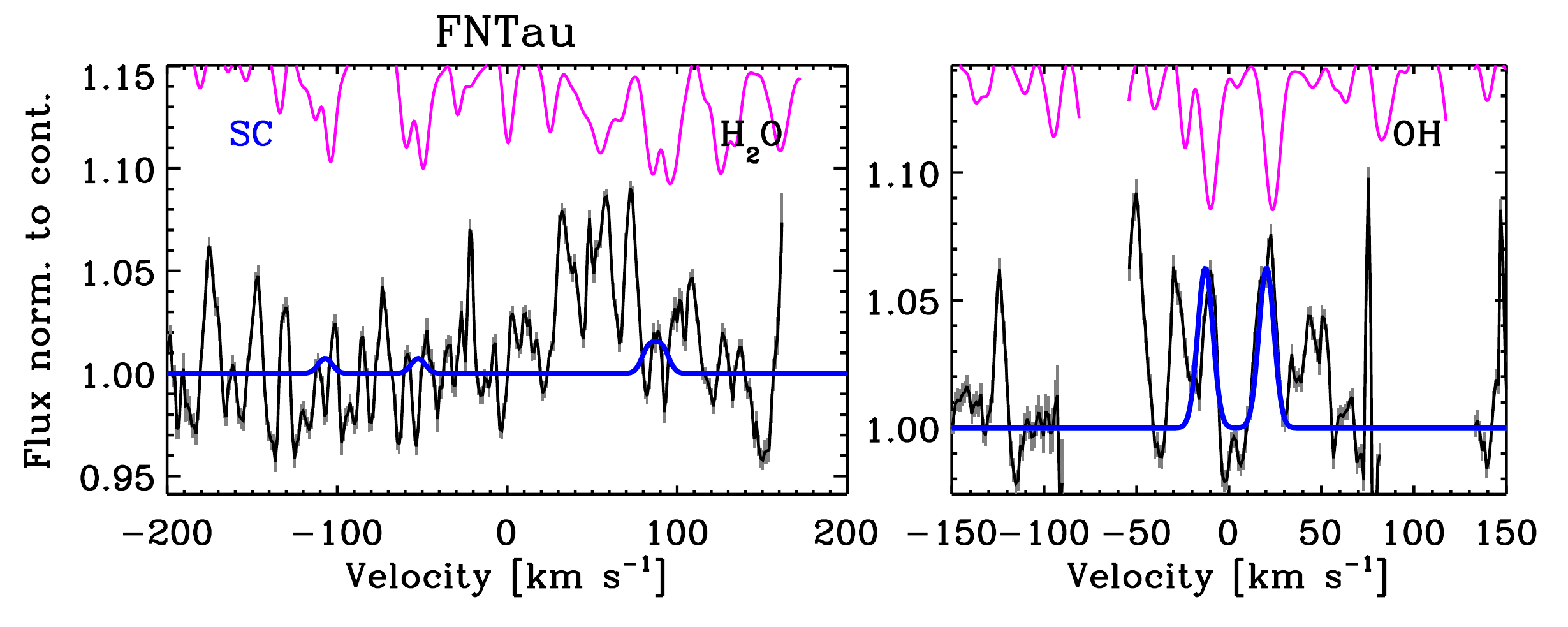} 
\includegraphics[width=0.5\textwidth]{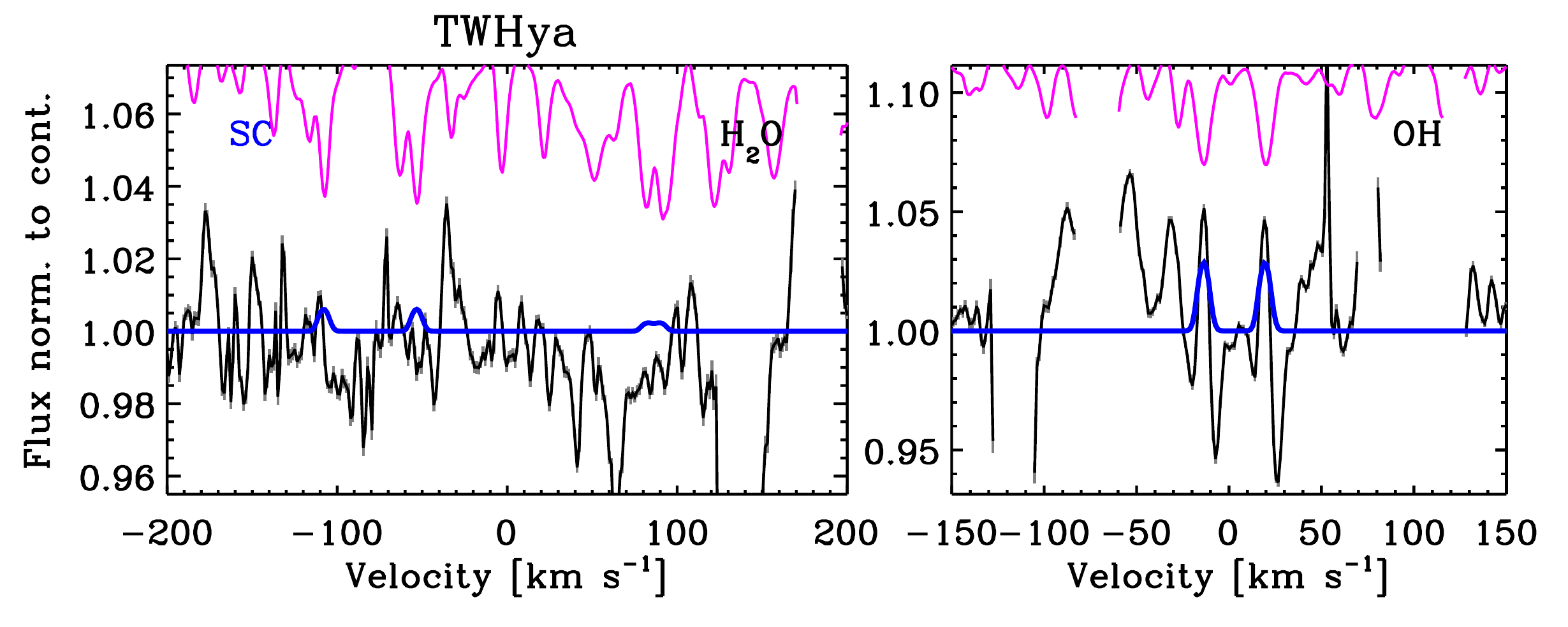} 
\caption{Uncertain OH detections in 2.9\,$\mu$m spectra where photospheric lines have similar FWHM to the potential OH emission lines. The photospheric template is showed in magenta at the top, broadened and veiled. OH is detected at longer wavelengths (10 to 30\,$\mu$m) in Spitzer spectra of TW\,Hya \citep[][and Table \ref{tab: fluxes}]{naji10}, and tentatively also in FN\,Tau.}
\label{fig: mol unclear1}
\end{figure*}

\begin{figure*}
\includegraphics[width=0.5\textwidth]{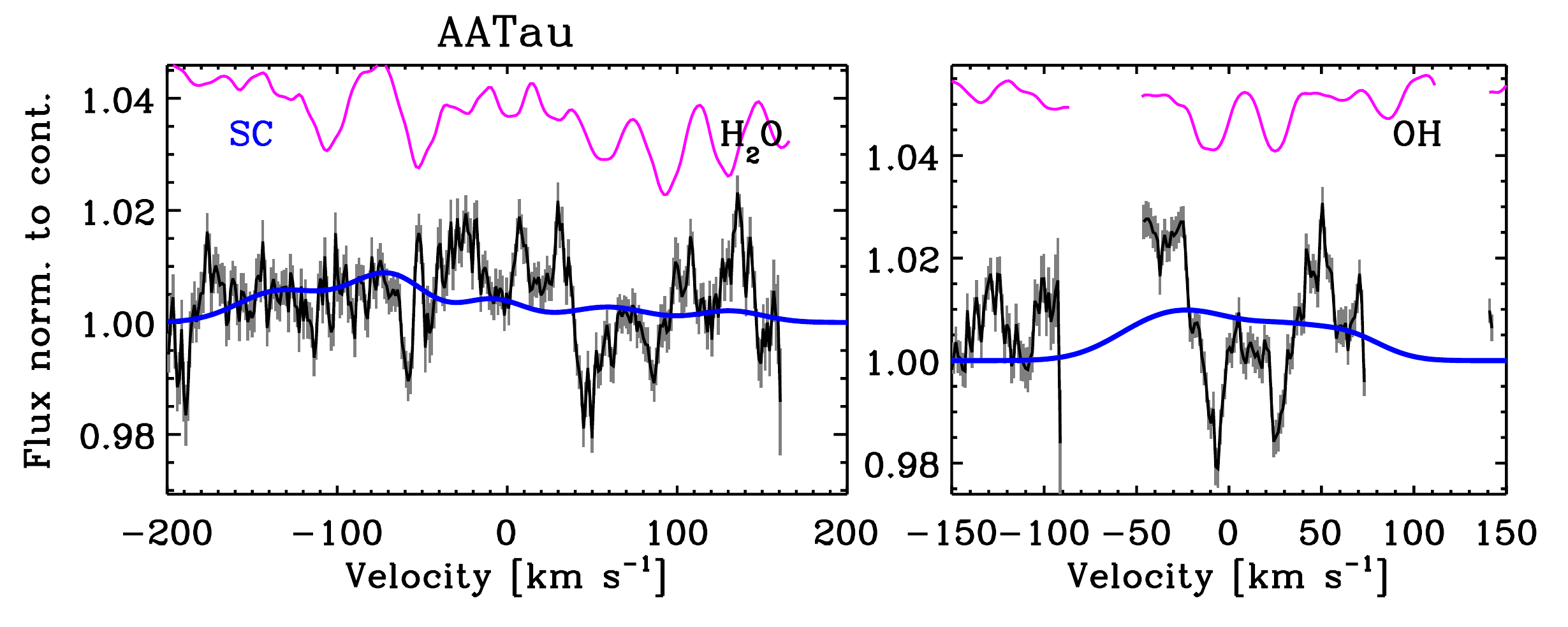} 
\includegraphics[width=0.5\textwidth]{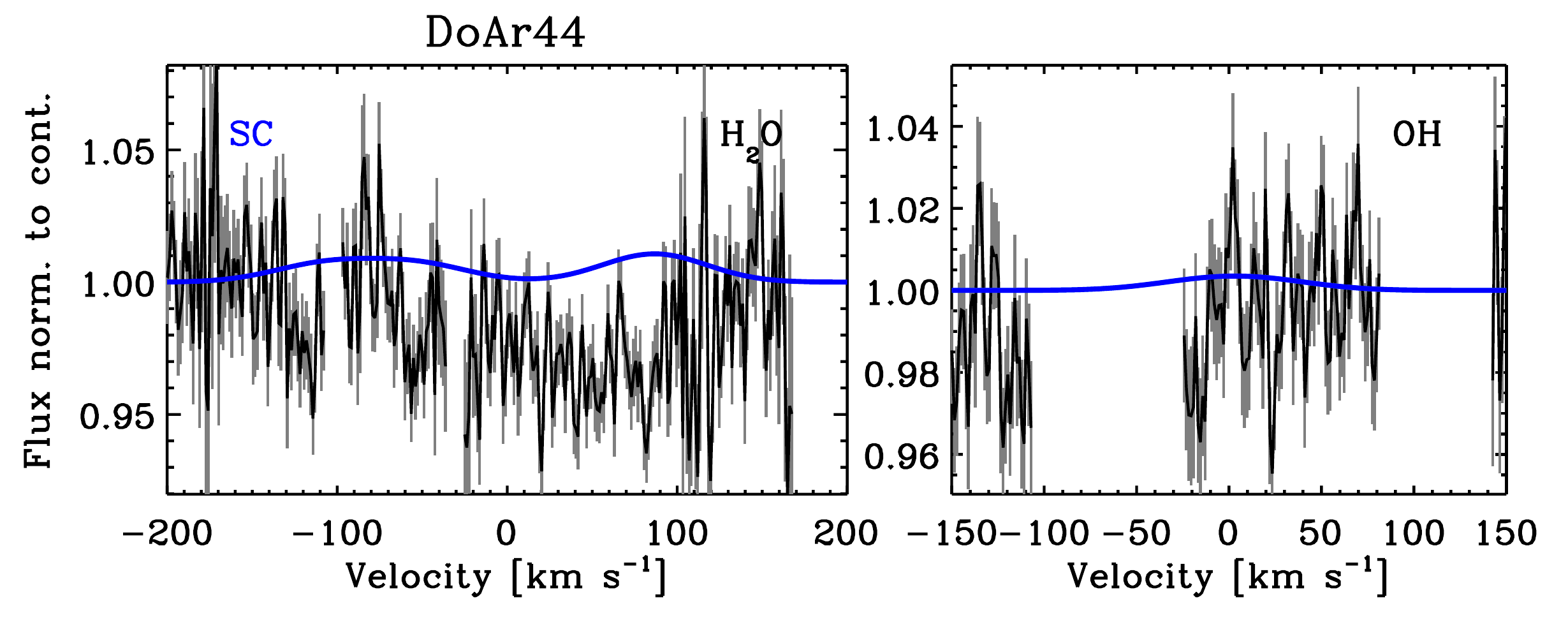} 
\includegraphics[width=0.5\textwidth]{SR21_line_comp_mod.pdf} 
\includegraphics[width=0.5\textwidth]{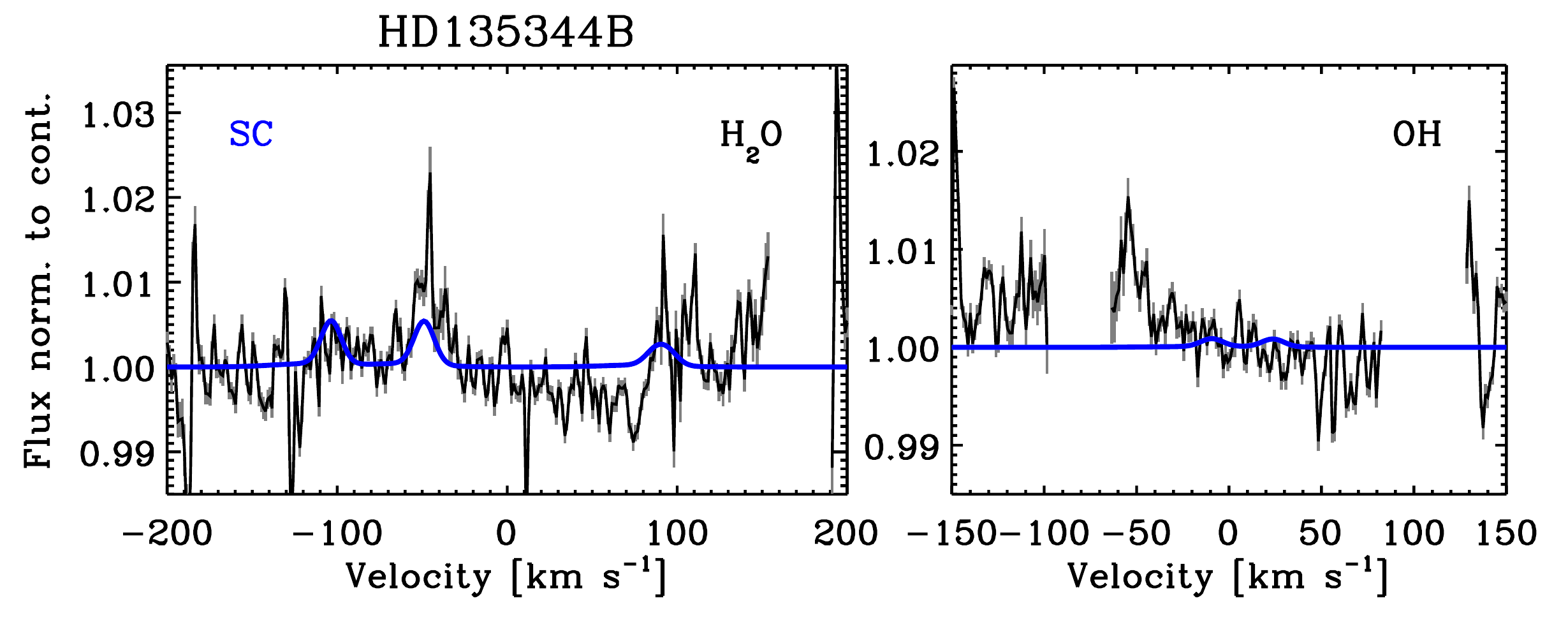} 
\caption{Single-component disks where H$_{2}$O and OH emission is not detected at 2.9\,$\mu$m. The photospheric template, where used for photospheric correction, is showed in magenta at the top, broadened and veiled.}
\label{fig: no h2o oh}
\end{figure*}

Figures \ref{fig: mol eq BC}, \ref{fig: mol SPCA}, \ref{fig: mol unclear}, \ref{fig: mol eq CO1-0}, \ref{fig: mol unclear1}, and \ref{fig: no h2o oh} show fits to CRIRES 2.9\,$\mu$m spectra made by using the CO line profiles as models for H$_{2}$O and OH emission, as explained in Section \ref{sec:ana4}. 
In addition to fitting the 2.9\,$\mu$m data with the NC or BC separately (in those disks that have two CO components), we also combine NC and BC profiles in different fractions, a technique that we call Spectral Principal Component Analysis (SPCA) as being based on principal spectral components given by NC and BC (Figure \ref{fig: mol SPCA}). In the double-component disks, we can therefore test whether H$_{2}$O and OH are composed of one rather than two components, and which one dominates. To determine which model best represents the data between NC, BC, and their combination, we use the Akaike information criterion \citep{akaike}, defined as AIC = $2\times$ the number of free model parameters + the chisquare value. The criterion penalizes models with a larger number of free parameters, so that a simpler model is preferred unless the more complex model decreases the AIC number significantly (we adopt the criterion of a $\Delta$AIC $> 14$, corresponding to a $10^4$ higher probability of minimizing the information loss). In most cases, the chisquare value is enough to identify the best model between BC and NC.

This procedure is generally robust against the photospheric correction performed on the 2.9\,$\mu$m velocity-resolved spectra obtained by CRIRES (Section \ref{sec:ana3}), because the disk emission lines are typically broader than photospheric lines from the same molecules (Figures \ref{fig: phot stand}, \ref{fig: mol comp}, \ref{fig: phot_corr_1}, \ref{fig: phot_corr_2}), and because the high resolving power of CRIRES allows to separate emission components with different velocities. In two disks (FN\,Tau, TW\,Hya), however, photospheric lines have similar width as the possibly detected OH emission lines, while H$_{2}$O is undetected; their OH line fluxes and profiles are therefore highly uncertain (see Figure \ref{fig: mol unclear1}).

Overall, both non-detections and detections of weak emission advocate for the predominance of BC over NC in H$_{2}$O and OH emission at 2.9\,$\mu$m. In a few cases, spectra are best fit with a combination of BC and NC: specifically AS\,205, DR\,Tau, and SCrA\,N (and only OH in CW\,Tau). These three disks are also the only ones where NC was detected in the CO $v=2-1$ lines, possibly due to the higher S/N and dust veiling in these spectra compared to other spectra in the survey \citep{bp15}. Two other disks show tentative evidence for a weak NC, but its presence can not be firmly established due to uncertainties from the photospheric correction: RU\,Lup and SCrA\,S. In all disks, the NC is about 2--10 times (or more) weaker in H$_{2}$O and OH than in CO, as compared to BC. In a few cases only, the OH profile is better matched by NC rather than by BC (DF\,Tau, RNO\,90, VW\,Cha). In single-component disks, OH is detected only in EC\,82 and TW\,Cha.

\subsection{Luminosity and accretion trends} \label{app:lum_trends}

\begin{figure*}[ht]
\includegraphics[width=1\textwidth]{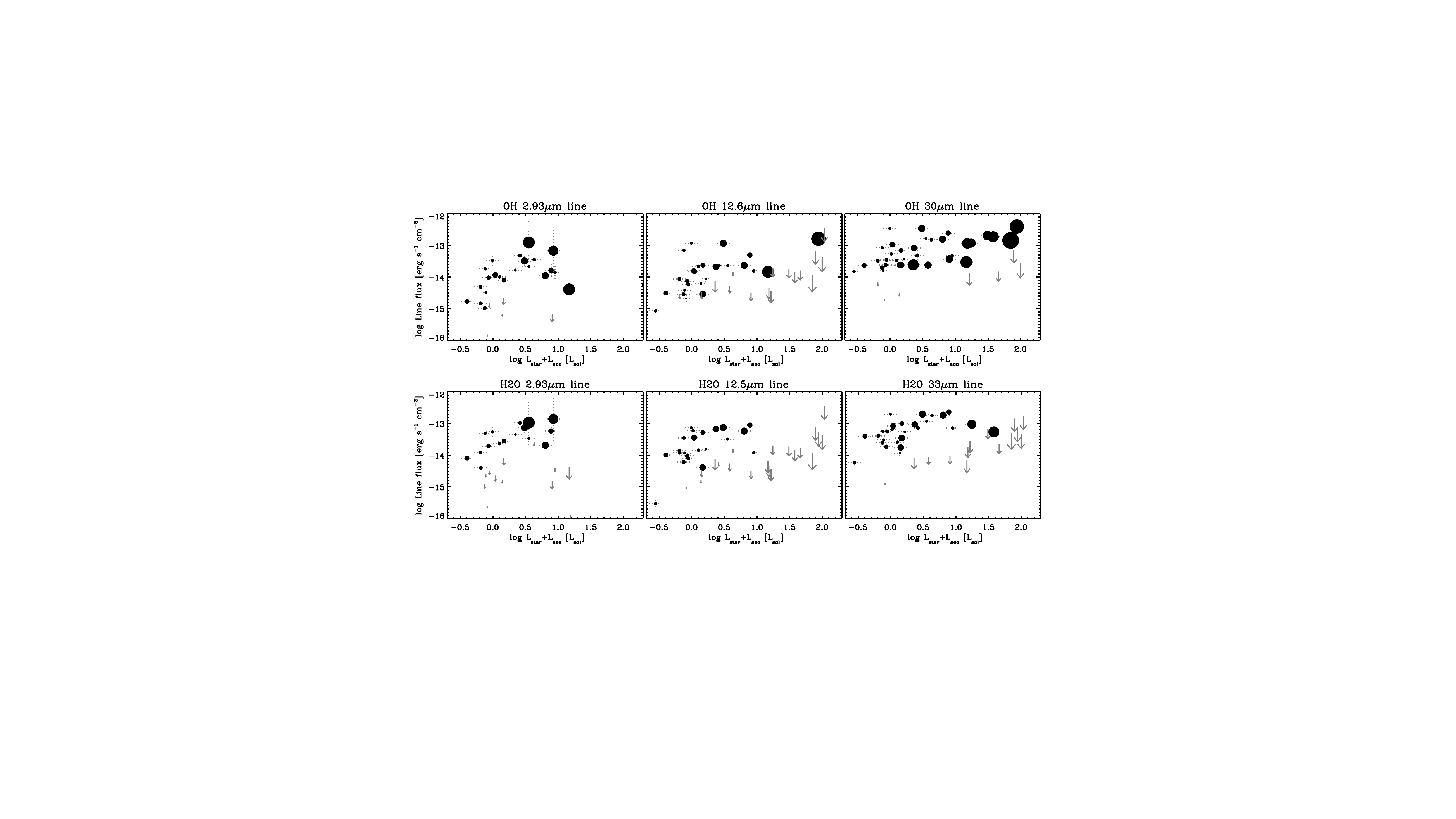} 
\caption{Trends between H$_{2}$O and OH line fluxes and the stellar+accretion luminosity. Symbol sizes are proportional to stellar masses as in Figure \ref{fig: TR_water}. All fluxes are normalized to a common distance of 140 pc.}
\label{fig: Lum_trends}
\end{figure*}

\begin{figure*}[ht]
\includegraphics[width=1\textwidth]{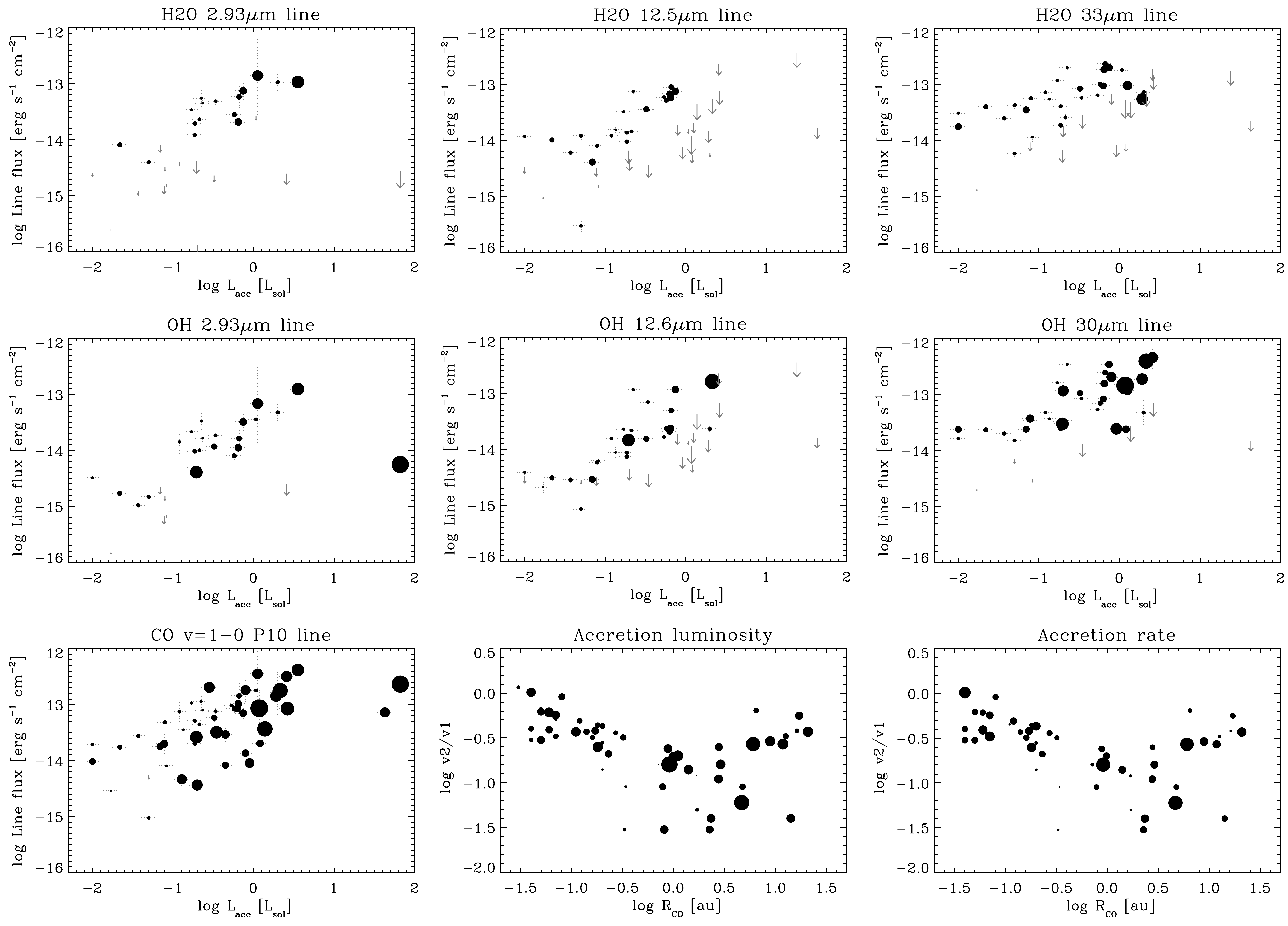} 
\caption{Dependence of measured line fluxes (normalized to 140 pc) on accretion luminosity. The symbol size is proportional to the stellar mass in panels showing line fluxes, while it is proportional to the logarithm of the accretion luminosity and rate in the last two panels showing the TR diagram in the bottom right, as indicated.}
\label{fig: accretion_figure}
\end{figure*}

Figure \ref{fig: Lum_trends} shows the measured H$_{2}$O and OH line fluxes plotted against the stellar+accretion luminosity, taken as a proxy for the irradiation and heating of inner disks. Only the OH line at 30\,$\mu$m shows a steady flux increase with increasing luminosity, consistent with an increasing area of the warm disk emitting region. Hotter OH lines show instead a turning point at a luminosity of $\sim3$\,L$_{\odot}$, beyond which line fluxes decrease (similarly to what seen in the H$_{2}$O at 33\,$\mu$m).
Figure \ref{fig: accretion_figure} shows the measured line fluxes plotted against the accretion luminosity alone, to highlight how line fluxes overall increase with accretion luminosity. The last two plots in the bottom right of the figure show the T-R diagram of CO emission, highlighting how there is not global trend between the sequence of CO gaps and the accretion luminosity and rate. 
OH emission is more frequently detected than water emission in disks, and is tentatively detected at 30\,$\mu$m out to R$_{\rm{co}}$/R$_{\rm{snow}} \sim 4$ (Figure \ref{fig: R_fl_corr}) and up to larger luminosities (Figure \ref{fig: Lum_trends}), possibly supporting a still active photo-dissociation of trace amounts of H$_{2}$O gas. OH formation through water photo-dissociation populates higher energy levels first \citep[e.g. the 12.6\,$\mu$m line, see][]{cn14}, though, which are less often detected in disks that have a large R$_{\rm{co}}$. It is therefore unclear whether a cold OH reservoir traced at 30\,$\mu$m may be formed by photo-dissociation of H$_{2}$O or by gas-phase chemistry, known to be efficient at temperatures of 100--300\,K \citep[100--300\,K, e.g.][]{vdish13}. Future observations of velocity-resolved H$_{2}$O and OH lines at wavelengths longer then 2.9\,$\mu$m will be able to test if two reservoirs of OH exist: one due to UV photo-dissociation of water (with similar line profiles for H$_{2}$O and OH, as found at 2.9\,$\mu$m), which dominates the hotter OH lines, and one due to OH formation in cold environments where water formation and/or release to the gas phase is less efficient (by finding larger line profiles for H$_{2}$O and smaller for OH).

\end{document}